\documentclass[twocolumn]{aastex631}

\usepackage{amssymb,amsmath,verbatim,mathtools,needspace,enumitem,etoolbox,graphicx,physics,microtype,afterpage,xspace,tabularx,lmodern,multirow,orcidlink}

\shorttitle{From NSCs to LRDs: BH growth, mergers, and TDEs}
\shortauthors{K.~Kritos \& J.~Silk}

\newcommand{\jhu}{\affiliation{William H. Miller III Department of Physics and Astronomy, Johns Hopkins University, 3400 North Charles
Street, Baltimore, Maryland 21218, USA}}

\newcommand{\iap}{\affiliation{Institut d’Astrophysique de Paris, UMR 7095 CNRS and UPMC, Sorbonne Universit\'e, F-75014 Paris, France}}

\newcommand{\bio}{\affiliation{Department of Physics, Beecroft Institute for Particle Astrophysics and Cosmology, University of Oxford, Oxford OX1 3RH, United Kingdom}}

\begin{document}

\title{From nuclear star clusters to Little Red Dots: black hole growth, mergers, and tidal disruptions}

\begin{abstract}
Little Red Dots, discovered by the James Webb Space Telescope, are hypothesized to be active galactic nuclei containing a supermassive black hole, possibly surrounded by a dense stellar cluster, large amounts of gas, and likely by a population of stellar-mass black holes. We develop a simple nuclear star cluster model to evolve the rapid mass growth of black hole seeds into the supermassive regime. The combined processes of tidal disruption events, black hole captures, and gas accretion are accounted for self-consistently in our model. Given the observed number density of Little Red Dots, and under reasonable assumptions, we predict at least a few tens of tidal disruption events and at least a few black hole captures at $z=4$--$6$, with a tidal disruption event rate an order of magnitude larger than the black hole capture rate. We also estimate the uncertainties in these estimates. Finally, we comment on the low x-ray luminosity of Little Red Dots.
\vspace{1cm}
\end{abstract}

\author{Konstantinos Kritos$\,$\orcidlink{0000-0002-0212-3472}}
\email{kkritos1@jhu.edu}
\jhu

\author{Joseph Silk$\,$\orcidlink{0000-0002-1566-8148}}
\jhu
\iap
\bio

\date{\today}

\keywords{Supermassive black holes (1663) --- Star clusters (1567) --- Tidal disruption (1696) --- Black hole physics (159) --- High-redshift galaxies (734) --- Gravitational wave sources (677)}

\section{Introduction}
\label{sec:Introduction}

Nuclear star clusters (NSCs) are the densest systems observed in the local Universe with masses up to $10^8\,M_\odot$~\citep{Neumayer:2020gno}.
Studies have verified the black hole (BH) mass-stellar velocity dispersion relation up to a redshift of $z\sim9$, while the BH mass-to-stellar mass ratio is higher beyond $z\gtrsim2$ than in the local Universe by a factor of $\sim10$--$100$~\citep{2025arXiv250403551J,2025arXiv251007376J}. Sometimes, the supermassive BH is heavier than the stellar component of the galaxy~\citep{Maiolino:2025tih}. This likely implies that the majority of the missing dynamical mass in star-depleted dark halos is in the form of baryonic gas, and that some massive BHs form early in gas-dominated dark halos~\citep{2025arXiv250613852M}.

Strongly gravitationally lensed, high-redshift, dense star clusters have also been detected with the James Webb Space Telescope (JWST), with masses $\sim10^6\,M_\odot$ and effective radii $<1\,\rm pc$~\citep{2023ApJ...945...53V,2024Natur.632..513A,2024Natur.636..332M} and reproduced in simulations~\citep{2024A&A...690A..94P,2024MNRAS.529.4104V,2025ApJ...981L..28M}.
These clusters hierarchically merge to form heavier NSCs in the centers of their dwarf galaxies~\citep{2022A&A...667A.101F}.
Simulations carried out by~\cite{2025arXiv250321879R} and~\cite{2025arXiv250418620L} have demonstrated that intermediate-mass black hole (IMBH) seeds rapidly form in the centers of gas-rich star cluster complexes that resemble the conditions in some high-redshift regions.
Runaway stellar collisions rapidly form a very massive star within $\sim1\,\rm Myr$ which can collapse into an IMBH seed in metal-poor environments~\citep{2025arXiv250814260V}. Cluster density controls the growth rate, while cluster mass controls the asymptotic mass value of the IMBH.

The diverse population of Little Red Dots (LRDs) discovered by the JWST at $z=4$--$6$ suggests the high abundance of $10^7$--$10^8\,M_\odot$ BHs at those redshifts and measured assuming the local virial relations that connect the width of broad-lines to BH mass~\citep{Matthee:2023utn,Maiolino:2023bpi,Greene:2024phl}. Some low-redshift LRD candidates have also been suggested with lower number densities in the local Universe~\citep{Lin:2025pnq,2025arXiv250612141L}. The absence of x-rays in all but a few LRDs is puzzling. It may either indicate high gas column densities near the accreting BH~\citep{Maiolino:2024uon,2025ApJ...980L..27I,Naidu:2025rpo,2025arXiv250316600D,2025arXiv250113082J}, or lower BH masses~\citep{Ananna:2024jug}, or else that some LRDs could be in the form of accreting supermassive stars~\citep{Begelman:2025upi,Zwick:2025eik}, or even be an extremely dense purely stellar system~\citep{2024ApJ...977L..13B}. The high BH mass of $>10^7\,M_\odot$ is verified from resolved narrow line kinematics by direct BH mass measurement in a strongly lensed LRD~\citep{2025arXiv250821748J}. Furthermore, \cite{Pacucci:2025ojp} suggest that the purely stellar interpretation for LRDs is problematic since the extremely high densities would inevitably lead to massive BH formation from collisional runaways. It is also indicated that a combined model of an accreting BH, a stellar population, and dense gas seems to better fit the spectral properties of LRDs~\citep{2025arXiv250818358W,2025arXiv250919422I}.
LRDs are likely the places where SMBHs grew for the first time~\citep{Inayoshi:2025isg}.

The presence of supermassive black holes (SMBHs) in the centers of LRDs may induce a cuspy distribution of stars around them, leading to increased tidal disruption events (TDEs)~\citep{Rozner:2025iec}.
\cite{2025arXiv250308779G} have also argued that high-redshift NSC hierarchical formation may seed SMBHs and evolve into LRDs during the BH accretion phase.

We are not currently in a position to fully constrain the seeding mechanism of massive BHs at high redshift. However, the observation of BHs with a mass of up to $\sim10^8\,M_\odot$ at redshift $z\sim10$~\citep{Kovacs:2024zfh,Maiolino:2023zdu}, and of inactive SMBHs~\citep{2024Natur.636..594J}, suggest that BHs likely grow through repeated bursts of super-Eddington episodes.
This view is also supported by theoretical studies of BH growth models that compare with observational data from JWST and Pulsar Timing Arrays~\citep{2025arXiv250912325B,Zana:2025uuk}.

We revisit the hypothesis that NSCs evolve dynamically into the observed high-redshift population of LRDs, assumed here to be highly compact and gas-rich galaxies containing massive central BHs~\citep{2023Natur.616..266L}. A semi-analytical framework is developed that includes loss cone feeding, stellar evaporation, BH mergers, and gas accretion to form intermediate and even supermassive central BHs from plausible initial conditions.

Our key results are that early-forming NSCs, as motivated by theory (merging tree studies to redshifts of $\sim 10$ or higher; low metallicity star formation studies that favor clusters of $\sim 1000\,M_\odot$ stars) and observations of gravitationally lensed high-redshift compact star clusters, can provide a favorable environment for forming massive BHs at high redshift. 

We predict enhanced and correlated rates of tidal disruptions in parallel with BH mergers and extreme-mass ratio inspiral (EMRI) captures. Our estimated central massive  BHs extend in mass to $\sim 10^7\, M_\odot$
in the first $\sim 100\,\rm Myr$ of the Universe, as well as at much later epochs in less extreme NSCs.

In what follows, we develop a model of NSC evolution and runaway BH formation, inspired by our earlier work~\citep{Kritos:2022non,Kritos:2024upo}. Section~\ref{sec:CMZ} contains a description of the best nearby proxy for LRDs, the Central Molecular Zone of the Milky Way galaxy (MWG) as a gas reservoir for the nuclear star cluster and SMBH at
the Galactic Center.  We develop the physics of NSC evolution and BH growth in Section~\ref{sec:Model} using a semi-analytic approach that rests on a relatively subjective choice of parameters. Our intent is to demonstrate what is possible in the context of conservative and more speculative assumptions, rather than to make precise predictions. 
We appeal to confirmation or otherwise of our inferred NSC evolution (Sec.~\ref{sec:Solutions}), as well as our multimessenger predictions of TDEs and captured EMRIs (Sec.~\ref{sec:Rates}), which combine simple physics prescriptions with astrophysical uncertainties to justify the more extreme limits of our modeling. The x-ray predictions are described in Sec.~\ref{sec:x-rays}, and we finally conclude in Sec.~\ref{sec:Conclusions}.

\section{The CMZ and  nuclear star cluster as a proxy for LRD models}
\label{sec:CMZ}

Our  LRD modelling combines the stellar environment modelled by the  massive equivalent of an  NSC along with a substantial gas reservoir. To illustrate these initial conditions we use as a nearby proxy the MWG NSC and the surrounding CMZ as a less extreme counterpart. The local environment involves black hole mergers along with gas and stellar accretion, all crucial ingredients for our LRD model.

First,  some definitions. The Bondi radius is defined as $GM_{\rm BH}/v_s^2$  where $v_s$  is a generalized sound speed that includes turbulence, both supersonic and  Alfvenic. 
The BH sphere of influence within which the potential is dominated by that of the central SMBH is $R_{\rm infl} = GM_{\rm BH}\sigma_\star^{-2}$. We use this definition for $R_{\rm infl}$ in this section as an order-of-magnitude estimate for the influence radius and in subsequent sections define it more accurately as the radius that contains a stellar mass of $2M_{\rm BH}$.
We define $R_{\rm CMZ}$ as the extent of the gas reservoir in the central molecular zone  (CMZ) of the MWG, and use the inner bulge of the MWG as a local template for an LRD. We use CMZ parameters taken from a recent review~\citep{2023ASPC..534...83H}.

At high $z$, there is more gas. We expect in generic early CMZ regions that $M_{\rm gas}/M_\star\sim 0.1$--$1$ whereas $M_{\rm gas}/M_\star \sim  0.03$ for the MWG CMZ. 

\subsection{CMZ properties}
The properties of the CMZ are:  
$1\,\rm kpc$ radius, $300\,\rm pc$ scale height, gas density $n= 150\,{\rm cm}^{-3} = 1\, M_\odot\,\rm pc^{-3}$;  $M_{\rm gas}\sim3\times10^7\,M_\odot$; $\sigma_{\rm los}= 10\,\rm km\,s^{-1}$.  The inflow rate from the bar is $\sim 0.1\,M_\odot\,\rm yr^{-1}$. For comparison, the MWG stellar bulge is nearly spherical with radius $\sim2\,\rm kpc$ and mass $2\times10^{10}\,M_\odot$.  

\subsection{CMZ components}
The CMZ consists of, cumulatively, the ``nuclear bulge'' in the MWG that has the following components: 
\begin{itemize}
\item SMBH:  $4\times10^6 M_\odot$, 
sphere of influence of SagA* is about $1\,\rm pc$, 
\item nuclear star cluster, of 
mass $2.5\times10^7\,M_\odot,$ size $1\,\rm pc$,  
sphere of gravitational influence of the NSC, about $30\,\rm pc$,
\item CND: Circumnuclear stellar disk.  
Gas mass $M_{\rm gas}= 3\times10^4\, M_\odot$. Size from 1--7$\,\rm pc$. Gas density $n_H= 10^{5-7}\,\rm cm^{-3}$. The line width is $\sigma_{\rm gas}= 10^4\,\rm km\, s^{-1}$. This contains the closest gas reservoir to SagA*.
\item NSD: Nuclear stellar disk.
Mass is $10^9\,M_\odot$. Sphere of influence $300\,\rm pc$. NSD is cospatial with CMZ.
\item Galactic stellar bar.
Mass $1.9\times 10^{10}\, M_\odot$. Sphere of influence $4\,\rm kpc$. Found in 2/3 of nearby disk galaxies. Transport matter to CMZ by rotational torquing.
\end{itemize}
\subsection {Current star formation rate and gas flows in the CMZ} 
The star formation rate is $\sim 0.07\,M_\odot\,\rm yr^{-1}$. 
The star formation rate is bursty, separated by periods of 20--100$\,\rm Myr$, cf. Fermi bubbles. It is generated by a central explosion  $\sim 50$ Myr ago, and initiated by central BH outflow and/or multiple supernovae. 
The gas inflow rate is driven by a bar and fuels the  CMZ at $\sim  0.8\, M_\odot\,\rm yr^{-1}$. 
Further in, the inflow from the  CMZ to the nucleus is $\sim 0.03\, M_\odot\,\rm yr^{-1}$. 
The observed gas linewidth due to turbulence is $12\,\rm km\,s^{-1}$. 
The inferred CMZ outflow rate, responsible for driving the Fermi bubbles, is  
$\sim 0.8\,M_\odot\,\rm yr^{-1}$.  
The ratio of gas outflow rate to star formation rate is $\sim 10$. 
The accompanying supernova rate is 2--$15\times10^{-4}\,\rm yr^{-1}$. 

\subsection{CMZs in distant galaxies}  
CMZs are observed out to  cosmic noon at $z\sim 2$. 
The  gas velocity dispersion is $\sigma_{\rm gas}=20$--$70\, \rm km\,s^{-1}$. 
The star formation rate is bursty. The inferred CMZ gas mass is of order the stellar mass or $\sim 10^9 \,M_\odot$.

\subsection{Schematic CMZ-based model for LRD} 
Consider a scaled-up early gas-rich version of the CMZ as a precursor of an LRD. It is compact due to strong gas cooling and turbulent transport of angular momentum. We expect that $M_{\rm gas} \sim M_\star$.  We assume that the LRD is a very massive NSC with a growing central massive BH via BH mergers, TDEs, and gas accretion. The model assumes an early phase of super-Eddington accretion fed by Bondi accretion. Within the CMZ, we use dust-dominated cooling. The corresponding infall time-scale is 
$t_{\rm acc} = GM_{\rm BH}/\sigma_\star^3$  within the  influence radius. We use the observed local scaling law: 
$M_{\rm BH}= \alpha \sigma_{200}^5\times10^8\,M_\odot$,
where the stellar velocity dispersion is $\sigma_{200} = \sigma/(200\,\rm km\,s^{-1})$, and the normalization constant $\alpha$ is of order unity.
The gas reservoir is contained by the current Bondi radius, $R_B=GM_{\rm BH}/v_s^2.$  We equate this with $R_{\rm CMZ}$. Gas self-gravity traps gas within the influence radius, followed by free-fall once  $M_{\rm gas} \gtrsim M_{\rm BH}$. 
We use  
$M_8= \alpha (\sigma_{200})^5 $
to obtain   
$t_{\rm acc} = GM_{\rm BH}\sigma_\star^{-3}= 3\times10^5\ \alpha^{3/5}M_8^{2/5}\,\rm yr$ where $M_8=M_{\rm BH}/(10^8\,M_\odot)$. 
We define $M_g$ as the gas mass within the Bondi radius: 
$M_g=R_B\sigma_\star^2/G=M_{\rm BH}(\sigma_\star/v_s)^2$ 
and the Bondi time-scale  
$t_{\rm Bondi} = Gv_s^3M_{\rm BH}^{-1}M_g^2v_s^3\sigma_\star^{-6}. $ 
The Bondi radius is identified with $R_{\rm CMZ}$ and we use  
$\rho_g =\sigma_\star^6 G^{-3}M_g^{-2}$
to obtain  
 $t_{\rm Bondi} = G M_{\rm BH}^{-1}M_{\rm gas}^2v_s^3/\sigma_{\star}^6  \propto 1/M_{\rm BH}$.

The duty cycle is $t_{\rm Bondi}/ t_{\rm acc}= (v_s/\sigma_{\star})^3(M_g/M_{\rm BH})^2 $ with $M_{\rm gas}$ identified as the mass of the CMZ. 
The Eddington ratio is $L_{\rm acc}/L_{\rm Edd}$ and is proportional to $M_{\rm BH}^{-1}. $ Hence $L_{\rm Edd}$ is very large and $t_{\rm Bondi}$ is very short at early phases, so that BH growth is rapid.

\section{Semi-analytical model for LRDs}
\label{sec:Model}

We model galactic nuclei as a double power-law density profile for both stars with mass $m_1$ and stellar-mass black hole remnants with mass $m_2$, as a two-mass model. This choice is partly made based on observations of the central regions of galaxies and partly because, as we show later, it leads to an analytical description of velocity and potential profiles, thus minimizing the use of numerical integration as much as possible.
These other past works employ a semianalytical framework similar to ours~\citep{1977ApJ...217..281S,1990Ap&SS.168..233Y,1991MNRAS.251..564D,Kaur:2024ofj}, but our model differs by including the stellar-mass BH population and captured EMRIs together with the stellar TDEs and gas accretion episodes onto the central massive BH.

\subsection{Stellar and black-hole mass distributions}

We denote by $R_1$ and $R_2$ the break radii of the stellar and black hole density profile, and by $n_1$ and $n_2$ the value of the number density at the break radii, respectively. Moreover, $\alpha_1$ and $\beta_1$ are the indices of the stellar power law density for radii $r<R_1$ and $r>R_1$, respectively. Similarly, we define $\alpha_2$ and $\beta_2$. We thus have
\begin{align}
    n_1(r)=\begin{cases}
        n_1\left({r\over R_1}\right)^{\alpha_1},\ r<R_1\\
        n_1\left({r\over R_1}\right)^{\beta_1},\ r\ge R_1
    \end{cases}
\end{align}
and a similar expression for the stellar-mass black hole remnant density law $n_2(r)$. We assume $\beta_1,\beta_2<0$ and $\alpha_1,\alpha_2\le0$ so that densities vanish as $r\to\infty$, but allow the possibility of a flat density within the break radius.
Such a broken-power-law two-mass model is a highly flexible and simple model that can describe a range of possible profile configurations, including both cored and cuspy nuclei. A broken power law is similar to the Nuker law, which adequately describes the central regions within $\sim300\,\rm pc$ of early-type galaxies observed at low redshift~\citep{1995AJ....110.2622L}.
In particular, it is observed that the central density transitions from a steeper outer to a shallower inner profile at the break radius. The NSC of the Milky Way is also fitted well by a broken power-law~\citep{2014A&A...566A..47S}.

The total number of stars enclosed within radius $r$ can be computed by integrating $4\pi x^2n_1(x)$ from $0$ to $r$ and is
\begin{align}
    N_1(r)={4\pi n_1R_1^3\over\alpha_1+3}\times\begin{cases}
        \left({r\over R_1}\right)^{\alpha_1+3},\ r<R_1\\
        1 + {\alpha_1+3\over\beta_1+3}\left[\left({r\over R_1}\right)^{\beta_1+3}-1\right],\ r\ge R_1
    \end{cases}
\end{align}
and similarly for $N_2(r)$. Moreover, the total stellar and black hole remnant mass contained within radius $r$ is $M_1(r)=m_1N_1(r)$ and $M_2(r)=m_2N_2(r)$, respectively. Convergence of the total mass in the system as $r\to\infty$ is achieved only when $\beta_1+3<0$ and the asymptotic value becomes $M_1(r\to\infty)\to4\pi n_1 m_1R_1^3(\beta_1-\alpha_1)(\alpha_1+3)^{-1}(\beta_1+3)^{-1}$. Similarly, we require $\beta_2+3<0$ such that $M_2(r\to\infty)<\infty$ for the stellar-mass black hole subsystem.

\subsection{Velocity dispersion profiles}

If $M_{\rm BH}$ is the mass of the central massive black hole in the system, then the total mass enclosed within radius $r$ is $M(r)=M_{\rm BH}+M_2(r)+M_1(r)$. The one-dimensional stellar velocity dispersion $\sigma_1(r)$ can be computed by integrating Jeans' equation [see Eq.~(3.58) in~\cite{Merritt:2013book}]
\begin{subequations}
\begin{align}
    n_1(r)\sigma_1(r)^2 &= \int_r^{\infty} {G\over x^2}M(x)n_1(x)dx\\&= I_{B1}(r) + I_{11}(r) + I_{21}(r)
    \label{eq:Jeans-eq}
\end{align}
\end{subequations}
Notice that this equation neglects the collision term in the Fokker-Planck equation; we assume the system is collisionless on the crossing timescale, which is orders of magnitude smaller than the relaxation time (but see~\cite{2023ApJ...955...30R,2024ApJ...963L..17R} for the effect of stellar collisions in NSC environments).
To compute the integral in Eq.~\eqref{eq:Jeans-eq}, we decompose it into three terms associated with the contribution of the central massive black hole $I_{B1}(r)$, the self-gravity of the stars $I_{11}(r)$, and the contribution from the stellar-mass black holes $I_{21}(r)$.
These terms are computed in closed-form expressions as
\begin{widetext}
\begin{subequations}
\begin{align}
    I_{\rm B1}(r) &= \begin{cases}
        I_{B1}(R_1) + {GM_{\rm BH}n_1\over R_1^{\alpha_1}}\times\begin{cases}
            \ln{R_1\over r},\ \alpha_1-1=0\\
            {R_1^{\alpha_1-1}-r^{\alpha_1-1}\over \alpha_1-1},\ \alpha_1-1\ne0
        \end{cases},\ r<R_1\\
        -{GM_{\rm BH}n_1\over (\beta_1-1)R_1}\left({r\over R_1}\right)^{\beta_1-1},\ r\ge R_1
    \end{cases}\\
    I_{11}(r) &= \begin{cases}
        I_{11}(R_1) + {4\pi Gn_1^2m_1\over(\alpha_1+3)R_1^{2\alpha_1}}\times\begin{cases}
            \ln{R_1\over r},\ \alpha_1+1=0\\
            {R_1^{2(\alpha_1+1)} - r^{2(\alpha_1+1)}\over2(\alpha_1+1)},\ \alpha_1+1\ne0
        \end{cases},\ r<R_1\\
        {4\pi Gn_1^2m_1R_1^2\over(\alpha_1+3)(\beta_1+3)}\left[{\beta_1-\alpha_1\over\beta_1-1}\left({r\over R_1}\right)^{\beta_1-1} + {\alpha_1+3\over2(\beta_1+1)}\left({r\over R_1}\right)^{2(\beta_1+1)}\right],\ r\ge R_1
    \end{cases}
\end{align}
\end{subequations}
\end{widetext}
and the equations for $I_{B2}(r)$ and $I_{22}(r)$ can be computed by replacing all indices ``$1$'' by ``$2$'' above.
Moreover, the cross term is computed analytically as
\begin{widetext}
\begin{subequations}
\begin{align}
    I_{21}(r) &= \begin{cases}
        I_{21}(R_{\rm min}) + {4\pi Gn_1n_2m_2\over(\alpha_2+3)R_1^{\alpha_1}R_2^{\alpha_2}}\times\begin{cases}
            \ln{R_{\rm min}\over r},\ \alpha_1+\alpha_2+2=0\\
            {R_{\rm min}^{\alpha_1+\alpha_2+2}-r^{\alpha_1+\alpha_2+2}\over \alpha_1+\alpha_2+2},\ \alpha_1+\alpha_2+2\ne0
        \end{cases},\ r<R_{\rm min}\\
        I_{21}(R_{\rm max}) + {4\pi Gn_1n_2m_2\over\alpha_2+3}\times{\cal I}_{21}(r),\ R_{\rm min}\le r<R_{\max}\\
        -{4\pi Gn_1n_2m_2R_2^3\over(\alpha_2+3)(\beta_2+3)R_1}\left[{\beta_2-\alpha_2\over\beta_1-1}\left({r\over R_1}\right)^{\beta_1-1} + {\alpha_2+3\over\beta_1+\beta_2+2}{r^{\beta_1+\beta_2+2}\over R_1^{\beta_1-1}R_2^{\beta_2+3}}\right],\ r\ge R_{\rm max}\\
    \end{cases}\\
    {\cal I}_{21}(r)&=\begin{cases}
        {R_2^3\over(\beta_2+3)R_1^{\alpha_1}}\times\left[ (\beta_2-\alpha_2)\times\begin{cases}
                \ln{R_1\over r},\ \alpha_1=1\\
                {R_1^{\alpha_1-1}-r^{\alpha_1-1}\over\alpha_1-1},\ \alpha_1\ne1
            \end{cases} + {\alpha_2+3\over R_2^{\beta_2+3}}\times\begin{cases}
                \ln{R_1\over r},\ \alpha_1+\beta_2+2=0\\
                {R_1^{\alpha_1+\beta_2+2}-r^{\alpha_1+\beta_2+2}\over\alpha_1+\beta_2+2},\ \alpha_1+\beta_2+2\ne0
            \end{cases} \right],\ R_1\ge R_2\\
            {1\over R_1^{\beta_1}R_2^{\alpha_2}}\times\begin{cases}
                \ln{R_2\over r},\ \beta_1+\alpha_2+2=0\\
                {R_2^{\beta_1+\alpha_2+2}-r^{\beta_1+\alpha_2+2}\over\beta_1+\alpha_2+2},\ \beta_1+\alpha_2+2\ne0
            \end{cases},\ R_1<R_2
        \end{cases}
\end{align}
\end{subequations}
\end{widetext}
where $R_{\rm max}\equiv\max(R_1,R_2)$ and $R_{\rm min}\equiv\min(R_1,R_2)$.
Finally, $\sigma_1(r) = \sqrt{[I_{B1}(r)+I_{11}(r)+I_{21}(r)]/n_1(r)}$. In case $n_1=0$, we then write $\sigma_1(r)=0$.
We can also write a similar set of equations for $\sigma_2(r)$, the one-dimensional velocity dispersion of stellar-mass black holes.
Convergence of all $I_{...}(r\to\infty)$ requires $\beta_1+1<0$ and $\beta_2+1<0$.

\subsection{Gravitational potential}

The total gravitational potential is $\Phi(r)=\Phi_\infty+\Phi_{\rm BH}(r)+\Phi_2(r)+\Phi_1(r)$ where $\Phi_\infty$ is its value as $r\to\infty$. It can be computed as a superposition of three contributions, due to the potential of the central massive black hole $\Phi_{\rm BH}(r)$, the stellar-mass black hole remnants, $\Phi_2(r)$, and the stars, $\Phi_1(r)$. 
The former contribution is written as
\begin{align}
    \Phi_{\rm BH}(r)=-{GM_{\rm BH}\over r}.
\end{align}
The stellar contribution can be written as
\begin{widetext}
\begin{align}
    \Phi_1(r)=-{4\pi Gm_1n_1R_1^3\over\alpha_1+3}\times\begin{cases}
        {1\over R_1^{\alpha_1+3}}\times\begin{cases}
            {R_1^{\alpha_1+2}-r^{\alpha_1+2}\over\alpha_1+2},\ \alpha_1+2\ne0\\
            \ln(R_1/r),\ \alpha_1+2=0
        \end{cases},\ r<R_1\\
        {1\over\beta_1+3}\left[ {\beta_1-\alpha_1\over r} - {\alpha_1+3\over \beta_1+2}{1\over R_1}\left({r\over R_1}\right)^{\beta_1+2} \right],\ r\ge R_1
    \end{cases}
    \label{eq:Potential}
\end{align}
\end{widetext}
and a similar expression for $\Phi_2(r)$. We may set $\Phi_\infty=0$ as long as $\beta_1+2<0$ and $\beta_2+2<0$.

\subsection{Particle evaporation rates}

The specific energy of a particle in the cluster is $\varepsilon(r)={1\over2}v^2+\Phi(r)$ and the escape velocity at radius $r$ is thus $v_{\rm e}(r)=\sqrt{-2\Phi(r)}$ determined by the condition that a marginally escaping particle has zero energy. Particles with a velocity that exceeds this escape speed at radius $r$ are removed from the system in one crossing time, which is given by $t_{\rm D}(r)=r/\sigma(r)$. Thus, if $p(v|\sigma(r))$ is the velocity distribution at radius $r$, assumed here to be a Maxwell-Boltzmann, then the fraction of stars with $v>v_{\rm e}(r)$ is
\begin{align}
    f_{\rm ev}(r) &= \int_{v_{\rm e}(r)}^{\infty} p(v|\sigma(r))dv \nonumber\\&= 1 - {\rm erf}\left[{v_{\rm e}(r)\over\sqrt{2}\sigma(r)}\right] + \sqrt{2\over\pi} {v_{\rm e}(r)\over\sigma(r)}\exp\left[-{v_{\rm e}(r)\over2\sigma(r)^2}\right]
    \label{eq:f_ev}
\end{align}
where $\rm erf$ and $\rm exp$ are the error and exponential functions, respectively.
If we use $\sigma_1(r)$ ($\sigma_2(r)$) in Eq.~\eqref{eq:f_ev}, then we obtain the fraction of stars (stellar-mass black hole remnants) that are moving at a velocity that exceeds the local escape speed at radius $r$.
We further assume that the high-velocity tail ($v>v_{\rm e}$) of the velocity distribution is refilled via two-body relaxation as the primary mechanism. The two-body relaxation timescale for stars is given by
\begin{align}
    t_{R,1}(r) = {0.34 \sigma_1(r)^3\over G^2 n_1(r)m_1^2\ln\Lambda_1}
\end{align}
where $\ln\Lambda_1=\ln(0.5N_1(\infty))$ is the Coulomb logarithm. 
Thus, the evaporation rate of stars at radius $r$ per unit star is given by
\begin{align}
    F_{\rm ev,1}(r)\equiv{d\dot{N}_{\rm ev,1}(r)\over dN_1} = {f_{\rm ev,1}(r)\over t_{\rm R,1}(r)}.
    \label{eq:F_ev1}
\end{align}
We can obtain the total evaporation rate of stars from the system by integrating Eq.~\eqref{eq:F_ev1} over all radii and noting that $dN_1(r) = 4\pi r^2 n_1(r) dr$,
\begin{align}
    \dot{N}_{\rm ev,1} = \int_{r_{\rm min,1}}^{\infty} F_{\rm ev,1}(r)4\pi r^2 n_1(r) dr
\end{align}
where $r_{\rm min,1}$ is the radius that contains a single star, i.e., determined by the condition $N_1(r_{\rm min,1})=1$.

Each time a particle is removed from the cluster, an amount of energy equal to the particle's energy is lost from the system. The fraction of energy per unit mass that is lost each relaxation time from radius $r$ is
\begin{align}
    g_{\rm ev,1}(r) &= \int_{v_{\rm e}(r)}^\infty p(v|\sigma_1(r))\left[{1\over2}v^2 + \Phi(r)\right]dv\nonumber\\ &= \left( {3\over2}\sigma_1(r)^2 + \Phi(r) \right)f_{\rm ev,1}(r) \nonumber\\ &+ {v_{\rm e}(r)^3\over\sqrt{2\pi \sigma_1(r)^2}}\exp\left[-{v_{\rm e}(r)^2\over2\sigma_1(r)^2}\right].
\end{align}
We compute the total energy loss rate at radius $r$ per unit star as $G_{\rm ev,1}(r)\equiv m_1g_{\rm ev,1}(r)/t_{\rm R,1}(r)$, and the total energy loss rate from the stellar system due to evaporation is
\begin{align}
    \dot{E}_{\rm ev,1} = \int_{r_{\rm min,1}}^\infty G_{\rm ev,1}(r)4\pi r^2n_1(r)dr.
\end{align}
We implement a similar set of expressions to compute the stellar-mass BH number and energy evaporation rates, $\dot{N}_{\rm ev,2}$ and $\dot{E}_{\rm ev,2}$, respectively, where we replace all indices ``1'' by ``2'' in the equations above.
Since stars and stellar-mass BHs have different velocity dispersions due to their distinct density profiles and masses, their evaporation rates are thus distinct.

\subsection{Central BH Brownian motion}

The massive BH in the center will not be sitting motionless at $r=0$, but will execute Brownian motion due to interactions with stellar-mass BHs and stars. It will thus have a velocity dispersion due to its wandering near the center, whose value is denoted by $\langle v_{\rm BH}^2\rangle$. Depending on whether it interacts more frequently with stars or stellar-mass BHs, the massive BH will be in equipartition with one or the other mass component, and thus $M_{\rm BH}\langle v_{\rm BH}^2\rangle$ will be equal to $3m_2\sigma_2^2$ or $3m_1\sigma_1^2$, respectively. We compute the rate of strong interactions (those that result in significant velocity change during the encounter) of the massive BH with stars and stellar-mass BHs and take their ratio $\Gamma_1/\Gamma_2\approx(n_1/n_2)(\sigma_2/\sigma_1)$ where densities and velocities are computed at their minimum radius because we assume the massive BH moves near the central region.
Therefore, we write for the root-mean-square velocity of the massive BH,
\begin{align}
    \langle v_{\rm BH}^2\rangle^{1/2}=\sqrt{3\over M_{\rm BH}}\times\begin{cases}
        \sigma_2(r_{\rm min,2})\sqrt{m_2},\ \Gamma_2>\Gamma_1\\
        \sigma_1(r_{\rm min,1})\sqrt{m_1},\ \Gamma_1>\Gamma_2
    \end{cases}
\end{align}
as follows from kinetic energy equipartition [see Eq.~(5.127) in~\cite{Merritt:2013book}].
In particular, if stellar-mass BHs dominate the central region, then the wandering radius may be larger than if Brownian motion into the massive BH was caused solely by stars because typically $m_2\sigma_2^2>m_1\sigma_1^2$~\citep{Perets:2006bz}.

\subsection{Accretion onto the central BH}

The presence of the massive BH in the center provides a sink for stars and BHs, which, if they approach close enough with a pericenter distance below a threshold value, are swallowed by the massive BH and are thus lost from the system. In contrast, evaporating particles are lost to infinity.
We compute the tidal disruption event rate following the loss-cone formalism, as presented by~\cite{Syer:1998xh}.
As for stellar-mass BHs captured by the central massive BH, we use the same loss-cone theory and substitute the tidal radius with the critical pericenter for gravitational-wave capture [see Eq.~(11) in~\cite{1989ApJ...343..725Q}]. Since the tidal and capture radii increase non-linearly with BH mass, we set the loss-cone radius to be either the tidal/capture critical pericenter distance or the horizon radius of the massive BH. We call these stellar-mass BHs captured by the central massive BH through the mechanism described as ``captured EMRIs''.
These tend to have high eccentricities at formation and excite higher gravitational-wave harmonics~\citep{OLeary:2008myb}.

We denote by $N_1\equiv N_1(\infty)$ and $N_2\equiv N_2(\infty)$ the total number of stars and stellar-mass BHs, respectively. Moreover, $M_1=m_1N_1$ and $M_2=m_2N_2$ are correspondingly the total masses. These numbers vary over time due to evaporation and loss cone effects. We may write the time evolution of the total numbers as follows,
\begin{subequations}
    \begin{align}
    \dot{N}_1 &= - \dot{N}_{\rm ev,1} - \dot{N}_{\rm lc,1}\\
    \dot{N}_2 &= - \dot{N}_{\rm ev,2} - \dot{N}_{\rm lc,2}
    \end{align}
\end{subequations}
As a consequence of the loss cone effects associated with the consumption of stars and BHs, the mass of the central massive BH grows at a rate
\begin{align}
    \dot{M}_{\rm BH,lc} = f_{\rm lc,1}m_1\dot{N}_{\rm lc,1} + m_2\dot{N}_{\rm lc,2}
    \label{eq:dMBHdtlc}
\end{align}
where $f_{\rm lc, 1}$ is the fraction of the mass of the star accreted during a tidal disruption event, typically assumed to be $50\%$~\citep{1988Natur.333..523R}.

Gas may be transported into the central $\rm pc$ region of the galaxy from the larger $\rm kpc$ scales. The process occurs in two steps~\citep{2023MNRAS.523.2918S}: first, from the galactic disc ($\sim\rm kpc$ scales) into the central molecular zone ($\sim\rm 100\,\rm pc$) through dust lanes via which the gas is accreted, and then funneled into the NSC. Inflow mechanisms that can drive gas into the nuclear region of the galaxy involve, among others, stellar feedback, magnetohydrodynamic turbulence, the presence of a bar structure, and extragalactic perturbations, such as those from satellites.
Such a hierarchical inflow of gas towards the center through successive dynamical instability episodes (the ``bars-within-bars'' channel) is one viable mechanism for fueling active galactic nuclei~\citep{1989Natur.338...45S,1990Natur.345..679S}.
More recently, multi-scale simulations adopting the cyclic zoom-in method have demonstrated the feasibility for gas to be rapidly driven from galactic to horizon scales~\citep{Guo:2025sjb}.

We expect the gas to be in a plasma state, due to stellar and radiation feedback from the central accreting BH. Furthermore, we take the gas uniformly distributed within the break radius of stars and compute the mean gas density as $\rho_{\rm g}=3M_{\rm g}/(4\pi R_1^3)$.

The inflow rate can be computed from the free fall mass rate as $f_{\rm g}V^3/G$, where $f_{\rm g}$ is the gas mass fraction of the galaxy and $V$ the circular velocity~\citep{Inayoshi:2019fun}. In turn, gas is accreted into the galaxy from the intergalactic medium.
The gas must first accumulate into the Bondi sphere at the Bondi rate, become self-gravitating, and then collapse within a dynamical timescale [$(c_s^2 + \langle v_{\rm BH}^2\rangle)^{3/2}G^{-1}$] to be accreted by the BH finally. Hyperaccretion is episodic; gas is subdominant in the dynamics of the cluster $M_{\rm g}\ll M_1,M_2$, and we ignore its contribution to the potential. We cap the gas accretion rate at a fixed Eddington ratio parametrized by the radiation efficiency $f_{\rm rad}$. The gas accretion rate contribution to the central BH's mass is written as
\begin{align}
    \dot{M}_{\rm BH, g} = \min & \left[{4\pi\rho_{\rm g}(GM_{\rm BH})^2\over(c_s^2 + \langle v_{\rm BH}^2\rangle)^{3/2}}, {(c_s^2+\langle v_{\rm BH}^2\rangle)^{3/2}\over G},\right. \nonumber\\ & \left. {0.1\over f_{\rm rad}}{M_{\rm BH}\over t_{\rm Sal}}\right]
\end{align}
where $t_{\rm Sal}\approx45.2\,\rm Myr$ is the Salpeter timescale corresponding to a radiative efficiency of $10\%$. The radiative efficiency has a very low value within the photon trapping radius during super-Eddington accretion that can reach below $1\%$ for Eddington ratios that exceed $\sim10$~\citep{Poutanen:2006uc}.
According to radiation-dominated magnetohydrodynamic simulations, the super-Eddington accretion phase can be sustained for several $\sim10^4\times GM_{\rm BH}/c^3$ and drives power outflows~\citep{Zhang:2025uug}.

We write the mass growth equation of the BH mass as the sum of the gas accretion and loss cone contributions:
\begin{align}
    \dot{M}_{\rm BH} = \dot{M}_{\rm BH,lc} + \dot{M}_{\rm BH,g}.
    \label{eq:dMBHdt}
\end{align}
Notice that the loss-cone growth term contains both accretion of stars and stellar-mass BHs (cf.~\eqref{eq:dMBHdtlc}).

\begin{figure*}
    \centering
    \includegraphics[width=0.329\linewidth]{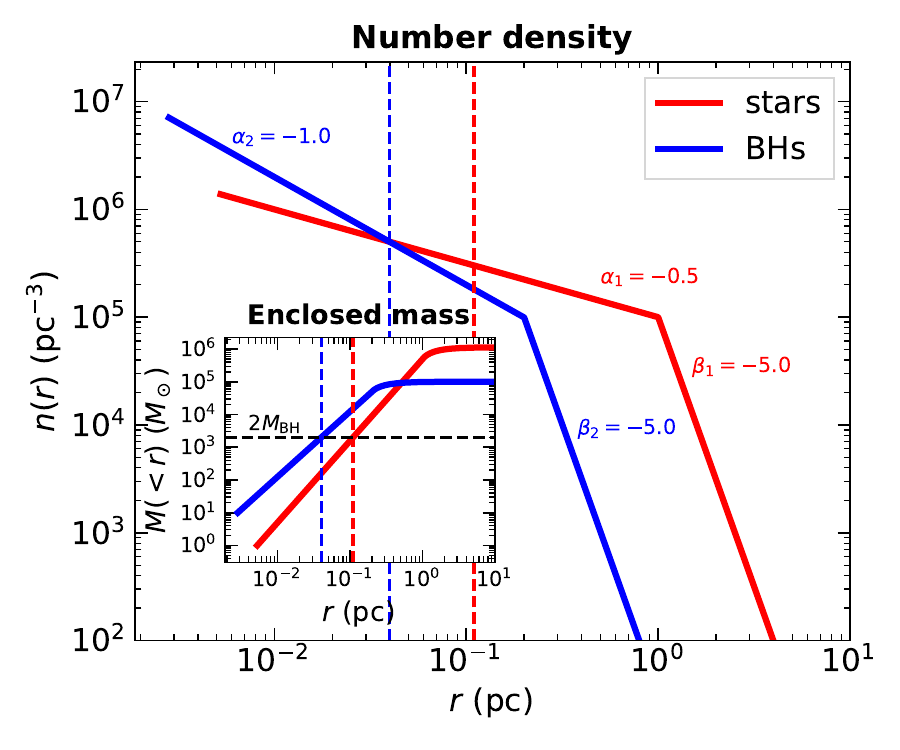}
    \includegraphics[width=0.329\linewidth]{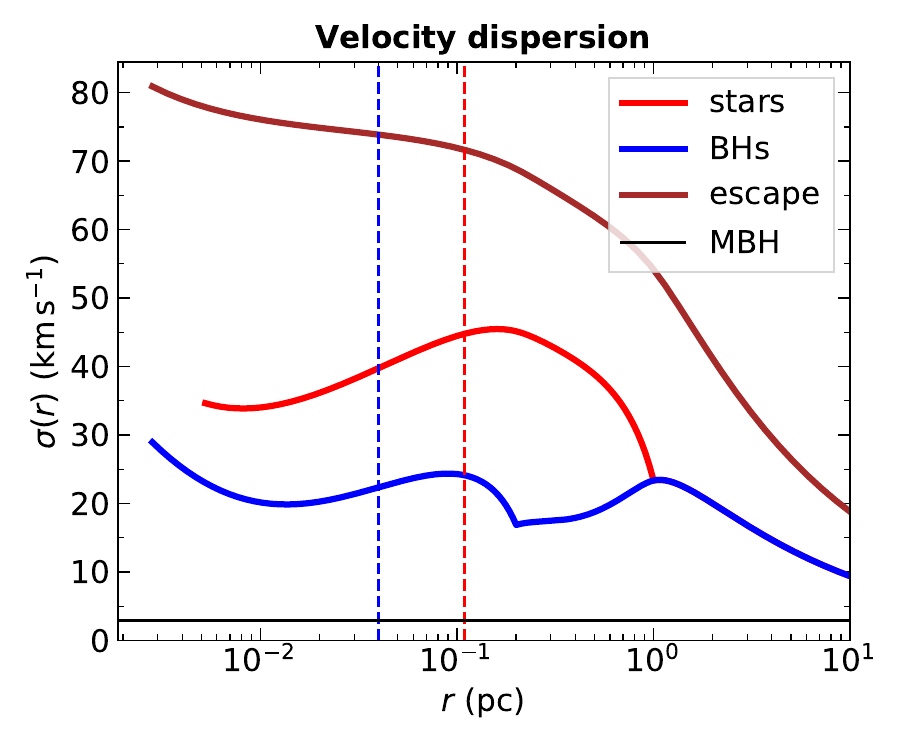}
    \includegraphics[width=0.329\linewidth]{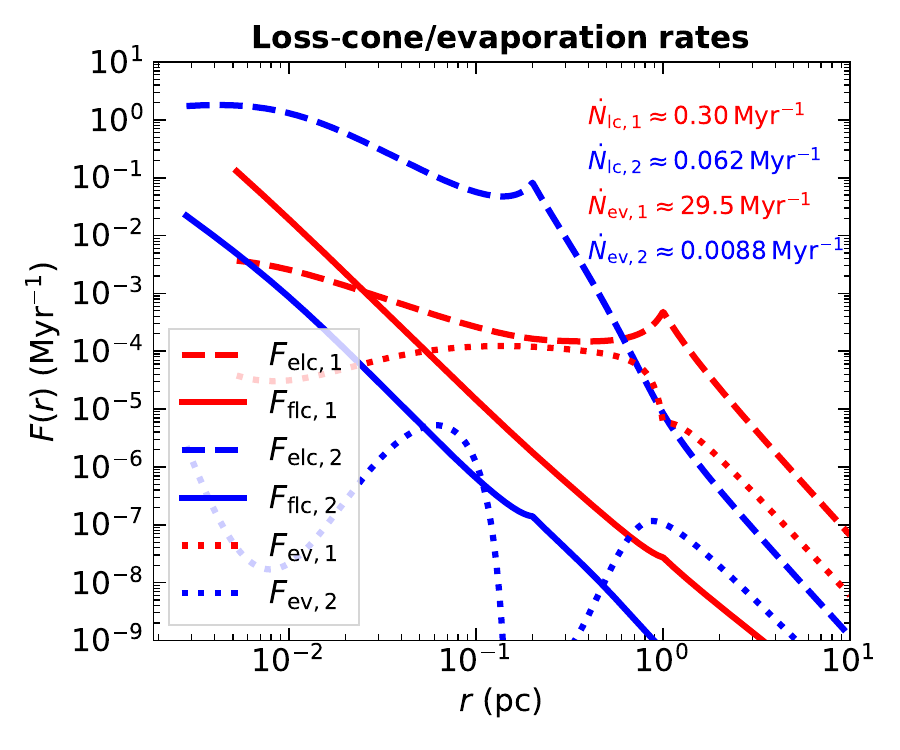}
    \caption{Radial profiles of number densities, total enclosed masses, velocity dispersions, and loss-cone and evaporation fluxes. {\it Left}: number density of stellar-mass BHs (blue) and stars (red). A massive BH with mass $M_{\rm BH}=10^3\,M_\odot$ is placed in the center of the system. The blue and red vertical dashed lines correspond to the influence radii of BHs and stars, respectively. The inset shows the total mass of BHs (blue) and stars (red) enclosed within radius $r$. The horizontal black dashed line at $M(r)=2M_{\rm BH}$ defines the influence radii. {\it Middle}: One-dimensional velocity dispersion for BHs (blue) and stars (red), as well as the escape velocity at radius $r$ (brown dash-dotted). The horizontal black line represents the root-mean-square wandering velocity of the central massive BH. {\it Right}: Full and empty loss-cone fluxes of BHs (blue-solid and blue-dashed) and stars (red-solid and red-dashed), as well as the evaporation fluxes of BHs (blue-dotted) and stars (red-dotted). Also shown in the upper right corner are the volume-integrated loss-cone and evaporation fluxes.}
    \label{fig:initial-snapshot}
\end{figure*}

\subsection{Energy conservation conditions}

Let $E_1$ and $E_2$ be the total energies of the stellar and stellar-mass BH populations within their break radii, respectively. We compute the total stellar energy as
\begin{align}
    E_1 = \int_{r_{\rm min,1}}^{\infty}\left[{3\over2}\sigma_1(r)^2+\Phi(r)\right]4\pi m_1n_1(r)r^2dr
\end{align}
and similarly for $E_2$.
We compute the total rate of change of $E_1$ and $E_2$ by accounting for both heating rates from evaporation and loss cone consumption. In addition, we assume that stellar-mass BHs serve as an energy source for the stars, as they have a higher temperature. The equations become
\begin{subequations}
\begin{align}
    \dot{E}_{1} &= \dot{E}_{\rm ev,1} + \dot{E}_{\rm lc,1}\\
    \dot{E}_{2} &= \dot{E}_{\rm ev,2} + \dot{E}_{\rm lc,2}
\end{align}
\end{subequations}
Applying the chain rule, we also write these energy production rates as
\begin{subequations}
\begin{align}
    \dot{E}_1 &= {\partial E_1\over \partial R_1}\dot{R}_1 + {\partial E_1\over \partial R_2}\dot{R}_2 + {\partial E_1\over \partial n_1}\dot{n}_1 + {\partial E_1\over \partial n_2}\dot{n}_2 + {\partial E_1\over \partial M_{\rm BH}}\dot{M}_{\rm BH}\\
    \dot{E}_2 &= {\partial E_2\over \partial R_1}\dot{R}_1 + {\partial E_2\over \partial R_2}\dot{R}_2 + {\partial E_2\over \partial n_1}\dot{n}_1 + {\partial E_2\over \partial n_2}\dot{n}_2 + {\partial E_2\over \partial M_{\rm BH}}\dot{M}_{\rm BH}
\end{align}
\end{subequations}
Moreover, 
\begin{align}
    \dot{N}_1 = {4\pi R_1^3\over\alpha_1+3}\dot{n}_1 + {12\pi n_1 R_1^2\over \alpha_1+3}\dot{R}_1
    \label{eq:dN1dt}
\end{align}
and similarly for $\dot{N}_2$. We can combine all equations above to write a differential equation for $\dot{R}_1$ and $\dot{R}_2$. These equations become
\begin{subequations}
    \begin{align}
        \dot{R}_1 &= \begin{cases}
            {bf-ec\over D},\ D\ne0\\
            \begin{cases}
                0,\ N_1\le2\\
                {c\over a},\ N_1>2
            \end{cases},\ D=0
        \end{cases}\\
        \dot{R}_2 &= \begin{cases}
            {cd-af\over D},\ D\ne0\\
            \begin{cases}
                0,\ N_1\le2\\
                {c\over b},\ N_1>2
            \end{cases},\ D=0
        \end{cases}
    \end{align}
    \label{eq:dRdt}
\end{subequations}
where we define
\begin{subequations}
\begin{align}
    a &\equiv {\partial E_1\over\partial R_1} - {3n_1\over R_1}{\partial E_1\over \partial n_1}\\
    b &\equiv {\partial E_1\over\partial R_2} - {3n_2\over R_2}{\partial E_1\over \partial n_2}\\
    c &\equiv \dot{E}_1 - {\partial E_1\over\partial n_1}{\dot{N}_1\over N_1} - {\partial E_1\over\partial n_2}{\dot{N}_2\over N_2}\\
    d &\equiv {\partial E_2\over\partial R_1} - {3n_1\over R_1}{\partial E_2\over \partial n_1}\\
    e &\equiv {\partial E_2\over\partial R_2} - {3n_2\over R_2}{\partial E_2\over \partial n_2}\\
    f &\equiv \dot{E}_2 - {\partial E_2\over\partial n_1}{\dot{N}_1\over N_1} - {\partial E_2\over\partial n_2}{\dot{N}_2\over N_2}\\
    D &\equiv bd - ae
\end{align}
\end{subequations}

\subsection{Stellar-mass BH binary formation}

We consider stellar-mass binary BH formation mechanisms related to 2-body and 3-body interactions. The former channel involves the strong interaction between two single stellar-mass BHs and the formation of a bound state following the efficient emission of gravitational waves along the hyperbolic encounter. The cross section for this process $\Sigma_{\rm 2b}$ is given by Eq.~(4) in~\cite{Mouri:2002mc} in which we set $m_i=m_j=m_2$. The volumetric rate density of 2-body binary formation is given by $\gamma_{\rm 2b}=n_2\langle\Sigma_{\rm 2b}v_{\rm rel}\rangle$ where $\langle...\rangle$ denote average over the velocity distribution (assumed to be the Maxwell-Boltzmann) and $\langle v_{\rm rel}^2\rangle=2\sigma_2^2$. The total 2-body (2b) binary formation rate is:
\begin{align}
    \Gamma_{\rm 2b} = \int_{r_{\rm min,2}}^\infty 12.1cn_2(r)^2\left({Gm_2\over c^2}\right)^2 \left({c\over\sigma_2(r)}\right)^{11/7} 4\pi r^2 dr
\end{align}
The interaction of three single BHs can also lead to the formation of a bound state, and the volumetric formation rate density of surviving 3-body binaries is given by $\gamma_{\rm 3b}=0.75n_2^3(Gm_2)^5/\sigma_2^9$~\citep{1993ApJ...403..271G}. The total 3-body (3b) binary formation rate is obtained by integrating over the volume of the system:
\begin{align}
    \Gamma_{\rm 3b} = \int_{r_{\rm min,2}}^\infty 0.75 cn_2(r)^3 \left({Gm_2\over c^2}\right)^5\left({c\over \sigma_2(r)}\right)^9 4\pi r^2dr.
\end{align}
The total binary formation rate is then the sum of these two components, $\Gamma_{\rm 2b}+\Gamma_{\rm 3b}$.

\section{LRD time evolution}
\label{sec:Solutions}

Below we simulate two individual NSC examples with growing SMBHs. First, we show the initial conditions (Sec.~\ref{sec:ICs}) and time evolution of a typical NSC (Sec.~\ref{sec:evolution-of-typical-NSC}) and later we consider a more compact and massive system (Sec.~\ref{sec:evolution-of-extreme-NSC}). For each system, we compare its evolution and the growth of the central BH with and without gas accretion episodes.
We compare the dependence of our results on the slope of the inner cusp in the Appendix.

\begin{figure*}
    \centering
    \includegraphics[width=0.329\linewidth]{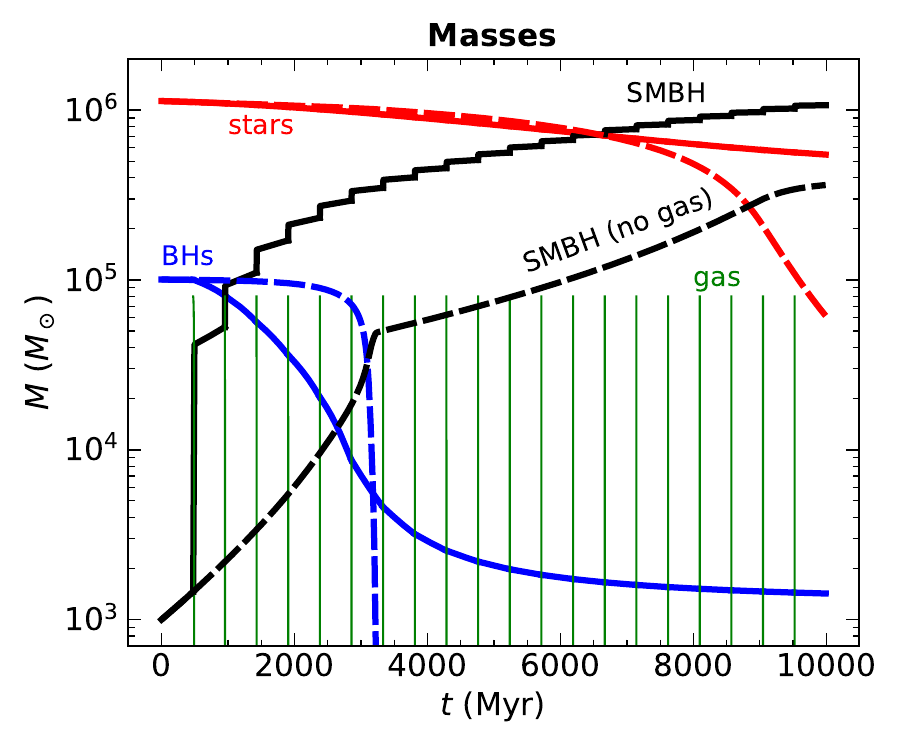}
    \includegraphics[width=0.329\linewidth]{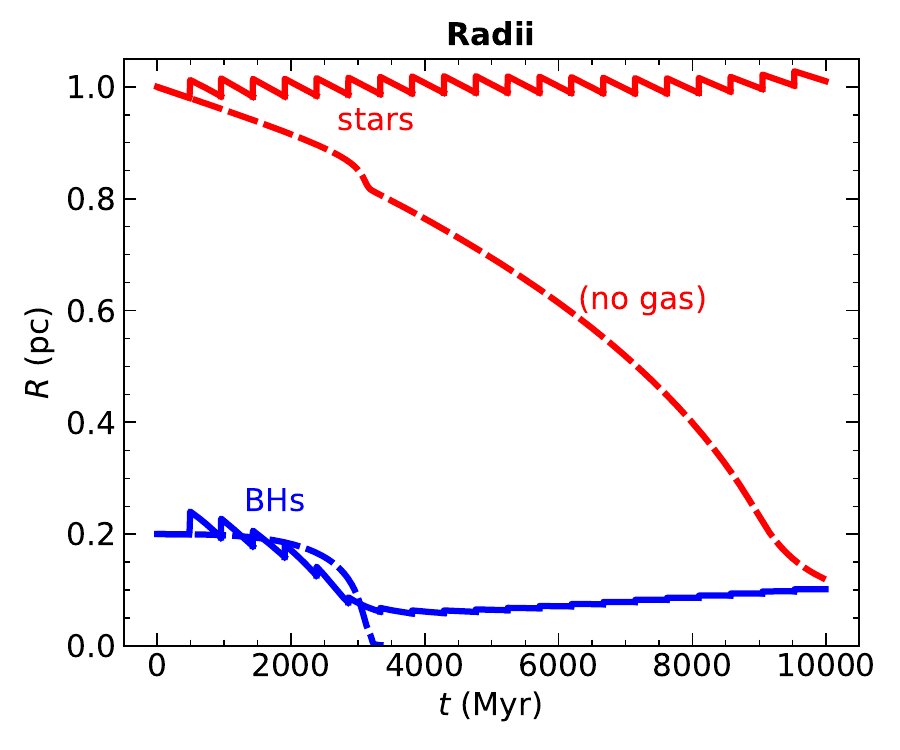}
    \includegraphics[width=0.329\linewidth]{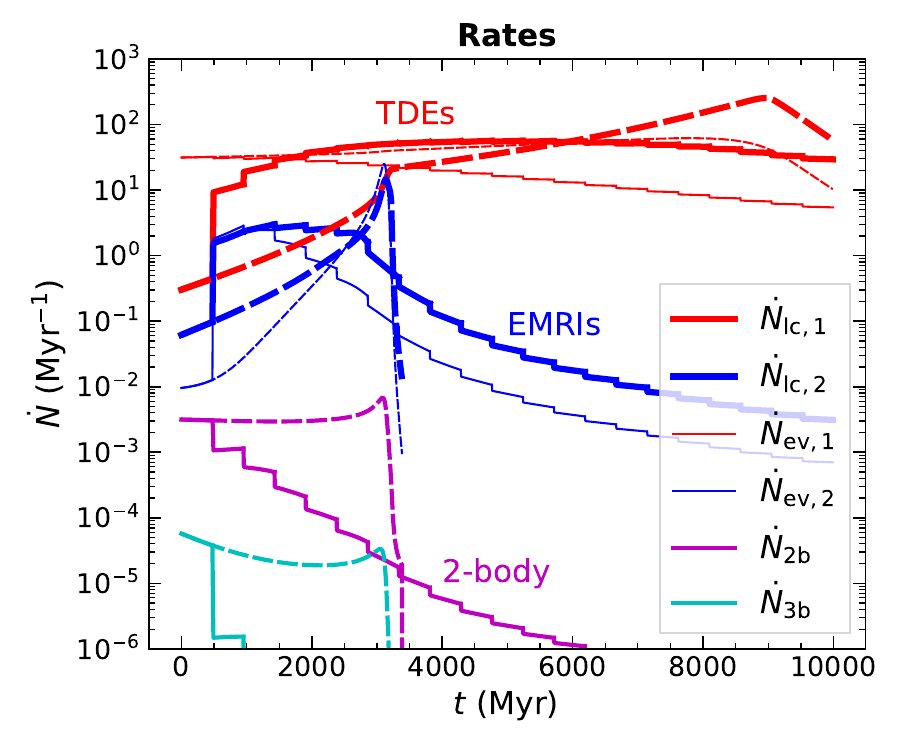}
    \caption{Time evolution of the system shown in Fig.~\ref{fig:initial-snapshot} with (solid) and without (dashed) gas accretion. In the former scenario, 20 gas inflow episodes, each of $8\cdot10^4\,M_\odot$ of gas, are added uniformly over time. {\it Left}: Evolution of the total stellar mass (red), of total mass in stellar-mass BHs (blue), and of the mass of the growing massive BH (black). The vertical green lines correspond to the episodes of gas inflow. {\it Middle}: Evolution of the break radii of the stellar (red-solid) and BH (blue-solid) population. {\it Right}: Loss-cone (evaporation) rates for stars and BHs are the solid-blue (thin solid-blue) and solid-red (thin solid-red) lines, respectively. The magenta and cyan lines correspond to the 2-body and 3-body stellar-mass binary BH formation rates.}
    \label{fig:time-evolution}
\end{figure*}

\subsection{Initial conditions for a typical NSC with an IMBH}
\label{sec:ICs}

We consider an NSC system with $n_1=n_2=10^5\,\rm pc^{-3}$, $R_1=1\,\rm pc$, $R_2=0.2\,\rm pc$, $\alpha_1=-0.5$, $\alpha_2=-1.0$, and $\beta_1=\beta_2=-5.0$. The density, velocity dispersion, and loss-cone/evaporation rate profiles for this system are shown in Fig.~\ref{fig:initial-snapshot} and correspond to a snapshot of the system at $t=0$. 

The BHs consist of a subsystem more compact than the stellar population and dominate the density near the center. Motivated by simulations by~\cite{2019MNRAS.484.3279P}, we chose a stellar cusp shallower than the cusp formed by stellar-mass BH remnants; see also~\cite{Alexander:2008tq}. The total mass of the system is about $10^6\,M_\odot$. The properties have been chosen to resemble the masses and radii of clusters observed at high redshift $z\sim10$~\citep{2024Natur.632..513A}. At the center, a $10^3\,M_\odot$ BH seed is placed. The influence radii for the stellar and BH population are $\approx0.1\,\rm pc$ and $\approx0.04\,\rm pc$, respectively, and correspond to the radii at which the total enclosed mass of each species is $2000\,M_\odot$.
Moreover, the stellar distribution is expected to be cuspy around a massive BH~\citep{1976MNRAS.176..633F,Rozner:2025iec}.

The IMBH seed at the center may have formed from an earlier phase of runaway stellar collisions, past rapid gas accretion, direct collapse of a metal-poor cloud, or may be of primordial origin. We are agnostic on its exact formation mechanism, and here we are concerned with the question of the subsequent mass growth of this intermediate-mass BH into the SMBH regime.
See~\cite{2021ApJ...908...57E,2023MNRAS.522.4224V,Pacucci:2025ojp,2025arXiv250321879R} for simulation examples of the collisional runaway scenario that produce $\sim10^3\,M_\odot$ BH seeds in similar environments.
\cite{2025MNRAS.539.2561C} discuss a direct collapse scenario in metal-enriched environments of a supermassive star that forms an IMBH of at least $\sim10^3\,M_\odot$ and further fragmentation into dense star clusters with high star formation efficiencies is possible~\citep{2024A&A...690A..94P}.

At a given radius within $1\,\rm pc$ stars have a larger velocity dispersion than BHs by a factor of a few. The discontinuity feature in the slope of $\sigma_2(r)$ at $r=R_2=0.2\,\rm pc$ corresponds to an artificial effect caused by the choice of a break in the slope of the power law in the density profile and is not physical. In reality, the velocity dispersion is a smooth function of $r$. The escape velocity from the center takes its maximum value and is about $\sqrt{3}\cdot80\,\rm km\,s^{-1}\simeq140\,\rm km\,s^{-1}$. The three-dimensional root-mean-square velocity dispersion of the central intermediate-mass BH is $\simeq4\,\rm km\,s^{-1}$. As a consequence, the wandering radius of the IMBH is small enough for it to be considered anchored to the center of the system. However, in our analysis, we do take into consideration the small value of the IMBH's velocity dispersion when computing TDE and captured EMRI rates.

In the right panel of Fig.~\ref{fig:initial-snapshot}, we show the loss-cone and evaporation rate radial profiles for stars and BHs. For the parameters chosen in this example, at $t=0$ all BHs are in the full-loss-cone regime while stars transition into the empty-loss-cone regime below the critical radius at $\simeq0.02\,\rm pc$. Every $1\,\rm Myr$ about 30 stars are ejected from the system as a consequence of two-body relaxation, and the stellar evaporation flux peaks at a few$\times0.1\,\rm pc$. The stellar-mass BH evaporation rate is $\approx0.0088\,\rm Myr^{-1}$.
The central IMBH accretes one star every $\approx3\,\rm Myr$ and one stellar-mass BH every $\approx16\,\rm Myr$ from loss-cone effects.

\subsection{Time evolution of the typical NSC}
\label{sec:evolution-of-typical-NSC}

Next, we evolve the system presented above in time by numerically integrating Eqs.~\eqref {eq:dN1dt}, \eqref{eq:dRdt}, and~\eqref{eq:dMBHdt}. The system evaporates with time due to relaxation, and the central IMBH's mass grows by disrupting stars and capturing stellar-mass BHs.
We consider two growth scenarios: one in which the BH grows solely through accretion, and another in which discrete phases of gas accretion are allowed to boost the mass of the BH.

The BH grows in mass initially primarily through capturing stellar-mass BHs. After the complete evaporation of the BH subsystem at $t\approx3000\,\rm Myr$, the growth steadily proceeds through TDEs to finally reach an asymptotic value of $\approx3\times10^5\,M_\odot$ (cf.~Fig.~\ref{fig:time-evolution}, left panel). Even though the TDE rate is higher than the BH capture rate by a factor of a few (cf.~Fig.~\ref{fig:time-evolution}, right panel), the BH mass growth is initially dominated by captured EMRIs because BHs are a factor of 10 heavier than stars. Furthermore, the gravitational-wave recoil kick obtained during the merger of the IMBH with $10\,M_\odot$ BHs is insignificant due to the significant mass asymmetry. Thus, the central BH is retained by the NSC.

The radius of the dry cluster decreases with time and reaches $\sim0.1\,\rm pc$ by the end of the simulation. This is happening because as the BH grows, the central stellar velocity dispersion cusps and is $\propto R_1^{-1/2}$ while the total mass term changes by a smaller fraction.

Next, we consider 20 inflow episodes each of $8\times10^4\,M_\odot$ spread uniformly in time.
The gas reservoir is $1.6\times10^6\,M_\odot$.
We can safely neglect the impact of gas influence on the dynamics as at each moment in time the gas contributes very little to the mass budget of the NSC. If a significant amount of gas inflows into the system and rapidly removed, the result would be a significant expansion of the system~\citep{1980ApJ...235..986H}.
This time, the BH's mass growth is boosted, and by the end of the simulation, the mass has grown to $>10^6\,M_\odot$.
In our model, the accretion rate is higher with higher BH mass, and the feedback is proportional to the accretion rate. Thus, the duty cycle decreases as the BH's mass grows with time. In this example, the AGN lifetime shifts from $\sim20\,\rm Myr$ down to less than $\sim1\,\rm Myr$ throughout the simulation. Given that each inflow occurs every $500\,\rm Myr$, the AGN duty cycle drops from $\sim4\%$ to about 0.1\%.

The radius evolution in the simulation with gas inflow episodes exhibits a tooth-like appearance. This is attributed to adiabatic expansion during slow gas expulsion from the system, which happens each time a gas inflow episode occurs, followed by gas removal due to AGN feedback~\citep{1980ApJ...235..986H}. The gas thus prevents the system from collapsing, as in the dry case, and the break radius does not evolve significantly over $10\,\rm Gyr$.
Moreover, the BH subsystem does not totally evaporate, and a population of $\sim1300$ stellar-mass BHs survives after a Hubble time due to the effect of the gas.

\begin{figure}
    \centering
    \includegraphics[width=1\linewidth]{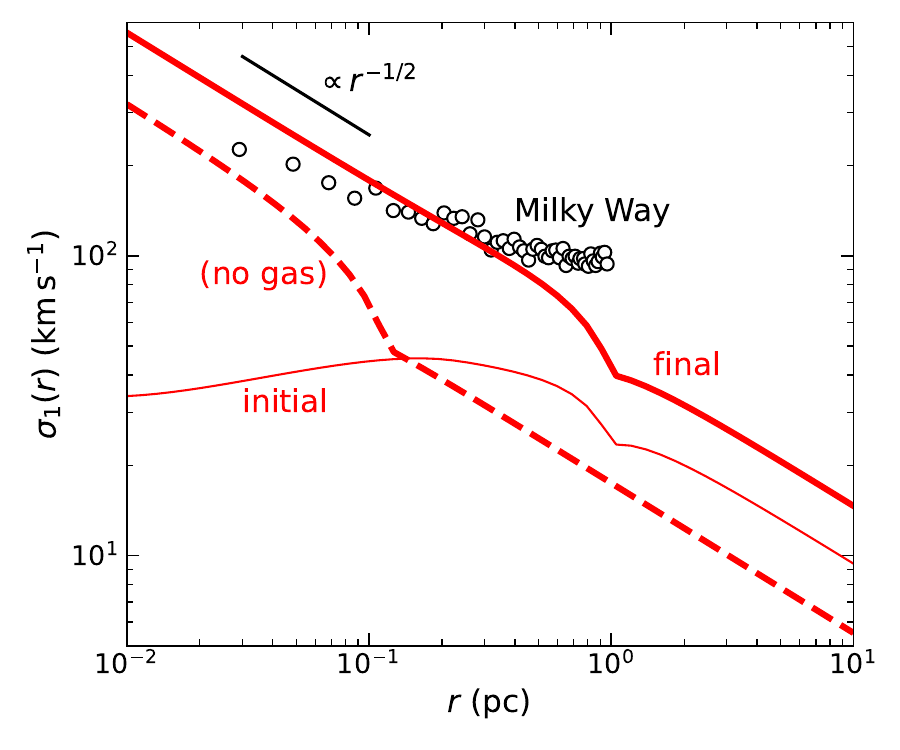}
    \caption{Stellar velocity dispersion profile snapshots at $t=0$ (initial, thin line) and $t=10\,\rm Gyr$ (final) with (thick-solid) and without (thick-dashed) gas accretion episodes. The black line segment shows the radius-dependent Keplerian circular velocity. Finally, the black hollow circles correspond to Milky Way data from~\cite{2009A&A...502...91S}.}
    \label{fig:velo_disp_ini_fin}
\end{figure}

In the initial snapshot, the stellar velocity dispersion is roughly flat. As the central massive BH grows, a stronger velocity cusp develops. At the final snapshot of the simulation at $t=10\,\rm Gyr$, the velocity dispersion increases $>100\,\rm km\,s^{-1}$ for radii $<0.1\,\rm pc$ (see Fig.~\ref{fig:velo_disp_ini_fin}). We verify that within the central region, the stellar velocity dispersion follows the expected $\sigma_1(r)\propto r^{-1/2}$ relation expected from a dominant central point-mass potential. Since the simulation, including gas inflow episodes, yields a more massive final SMBH, the velocity dispersion has a larger central value.
In Fig.~\ref{fig:velo_disp_ini_fin}, we also show for reference the central velocity dispersion profile around Sagittarius A*. We use data from Fig.~12 of~\cite{2009A&A...502...91S} and assume the Earth is at the galactocentric distance of $8\,\rm kpc$. At $r\gtrsim0.5\,\rm pc$ the velocity dispersion in the Milky Way is roughly constant at the value $\approx100\,\rm km\,s^{-1}$, consistent with an isothermal profile ($r\propto r^{-2}$). This differs from our predicted velocity dispersion profile at large distances, which continues to decrease due to a more rapid assumed decline in the mass density profile ($r^{-5}$). Nevertheless, this comparison is only qualitative since the simulated final properties of our NSC do not match those of the NSC at the center of the Milky Way. The Milky Way is perhaps the best example that exhibits a clear Keplerian rise near the center~\citep[chapter~1]{Merritt:2013book}.

The timescale for the duration of a TDE fallback timescale is given by Eq.~(4) in~\cite{Gezari:2021bmb}. For solar stars this TDE flare timescale is $t_{\rm TDE}\approx0.11\,{\rm yr}\, M_{6}^{1/2}$ where $M_{6}$ is the mass of the central BH normalized to $10^6\,M_\odot$. We compute the TDE duty cycle as $t_{\rm TDE}\dot{N}_{\rm lc,1}$. The TDE duty cycle steadily increases from $\sim10^{-8}$ to $\sim10^{-5}$ in the first $5\,\rm Gyr$ and then to a maximum value of $\sim10^{-4}$ at $t\sim9\,\rm Gyr$ after which it drops by a factor of $\sim3$ by the end of the simulation. 
When we add gas inflows, the TDE duty cycle increases at a faster rate and then remains at higher values, approximately a few times $10^{-5}$.

\subsection{An extremely massive and compact NSC example}
\label{sec:evolution-of-extreme-NSC}

\begin{figure*}[ht]
    \centering
    \begin{minipage}[c]{0.39\linewidth}
        \vspace*{0.\textheight}
        \centering
        \includegraphics[width=\linewidth]{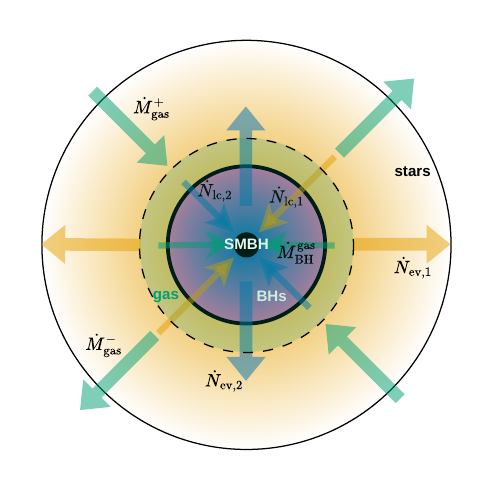}
    \end{minipage}
    \begin{minipage}[c]{0.59\linewidth}
        \includegraphics[width=\linewidth]{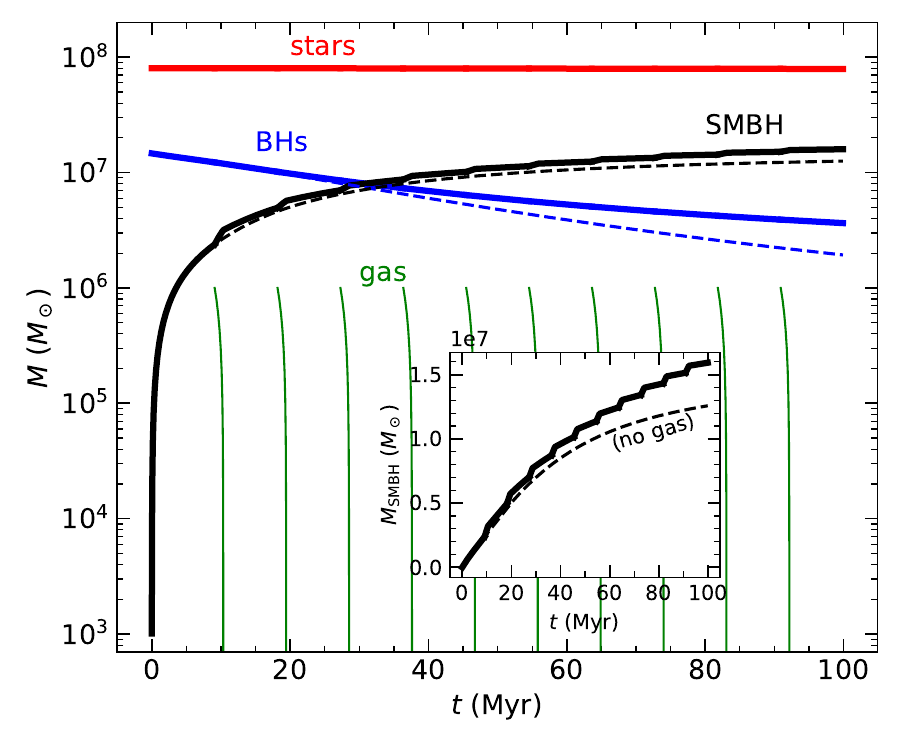}
    \end{minipage}%
    \hfill
    \caption{{\it Left}: Cartoon of the spherical cluster model developed in this work. The outer diffuse {orange region bounded by the solid black line} corresponds to the bulk stellar population, and the inner blue overdensity {bounded by the thick black line} is the subcluster of stellar-mass BH remnants (BHs) in the core surrounding a supermassive BH (SMBH) represented by the central {solid} black sphere. The faint {green dashed} sphere indicates the presence of ionized hydrogen gas in the system. The thin inward (thick outward) arrows show the loss-cone influx (evaporation outflux) of stars $\dot{N}_{\rm lc,1}$ ($\dot{N}_{\rm ev,1}$) and BHs $\dot{N}_{\rm lc,2}$ ($\dot{N}_{\rm ev,2}$). Finally, the green inward and outward green arrows show the influx $\dot{M}_{\rm gas}^{+}$ and outflux $\dot{M}_{\rm gas}^{-}$ of gas from the system, while the inward green arrows $\dot{M}_{\rm BH}^{\rm gas}$ indicate gas accretion into the massive BH. {\it Right}: Time evolution of the total stellar mass (red), total mass of stellar-mass BHs (blue), and mass of the massive BH (black). The dashed lines correspond to no gas inflow episodes, while the solid green lines show the gas inflow episodes. The inset shows a linear scale of the massive BH growth and the gas inflow episodes.}
    \label{fig:SMBH-growth}
\end{figure*}

For our second example, we consider a denser version of the typical system we examined above. In particular, we set $n_1=n_2=10^8\,\rm pc^{-3}$ and choose stronger cusps, with $\alpha_1=-1.0$ and $\alpha_2=-1.5$ (close to a Bahcall-Wolf law with $\alpha=-1.75$). Furthermore, we take $R_1=4R_2=0.4\,\rm pc$. All other parameters are kept the same as in~Sec.~\ref{sec:ICs}. We simulate this system for $100\,\rm Myr$. The total initial mass of this NSC (including $M_1$, $M_2$, and $M_{\rm BH}$) is $\approx10^8\,M_\odot$.
We show a cartoon of our three-dimensional model for the NSC interpretation of LRDs in the left panel of Fig.~\ref{fig:SMBH-growth}.

Observations of very dense clusters~\citep{2024Natur.632..513A}, or very massive/compact galaxies~\citep{2024ApJ...965...98C}, as well as
simulations~\citep{2024MNRAS.531.3770R, 2025ApJ...990..135W, 2025MNRAS.543.1023L}, hint at massive and compact environments like the one we consider in this subsection present already in place at high-redshift ($z\gtrsim10$).

In this case, the growth of the IMBH is more dramatic and a $\sim10^7\,M_\odot$ SMBH emerges within a few$\times10\,\rm Myr$ (right panel of Fig.~\ref{fig:SMBH-growth}). That is, the SMBH formation efficiency for this NSC ---the final SMBH to initial total cluster mass as defined in~\cite{2025arXiv250814260V}--- is $\sim10\%$. Moreover, the population of stellar-mass BHs does not dissolve on the simulated timescale, although $\approx90\%$ of the initial number of BHs is lost either in ejections or loss-cone effects. The corresponding loss fraction for stars is $\approx2\%$ on the same timescale.

\begin{figure*}
    \centering
    \includegraphics[width=0.49\textwidth]{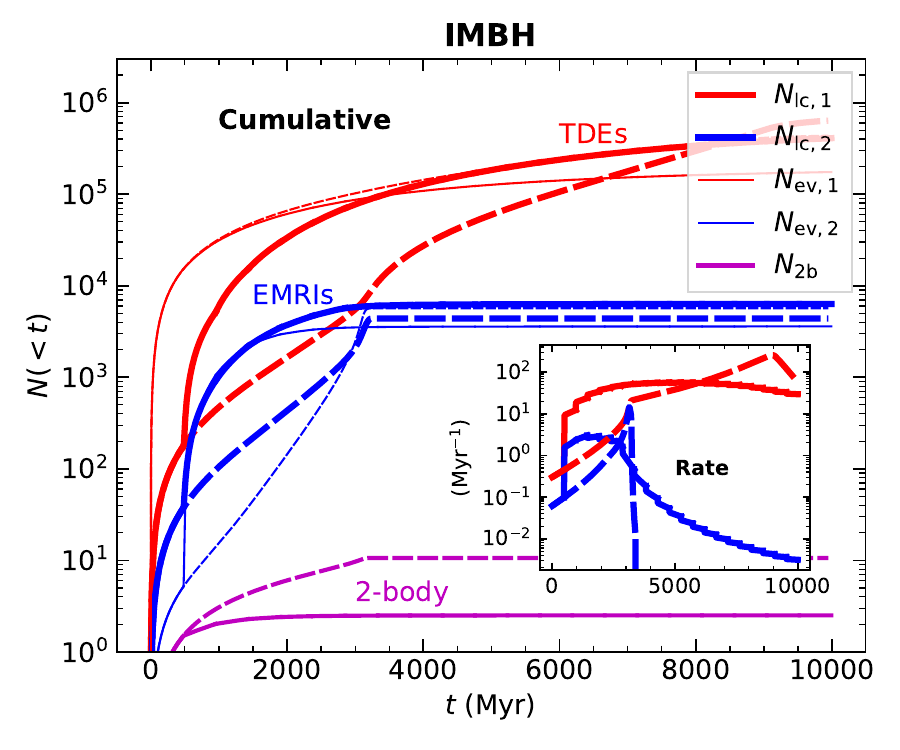}
    \includegraphics[width=0.49\textwidth]{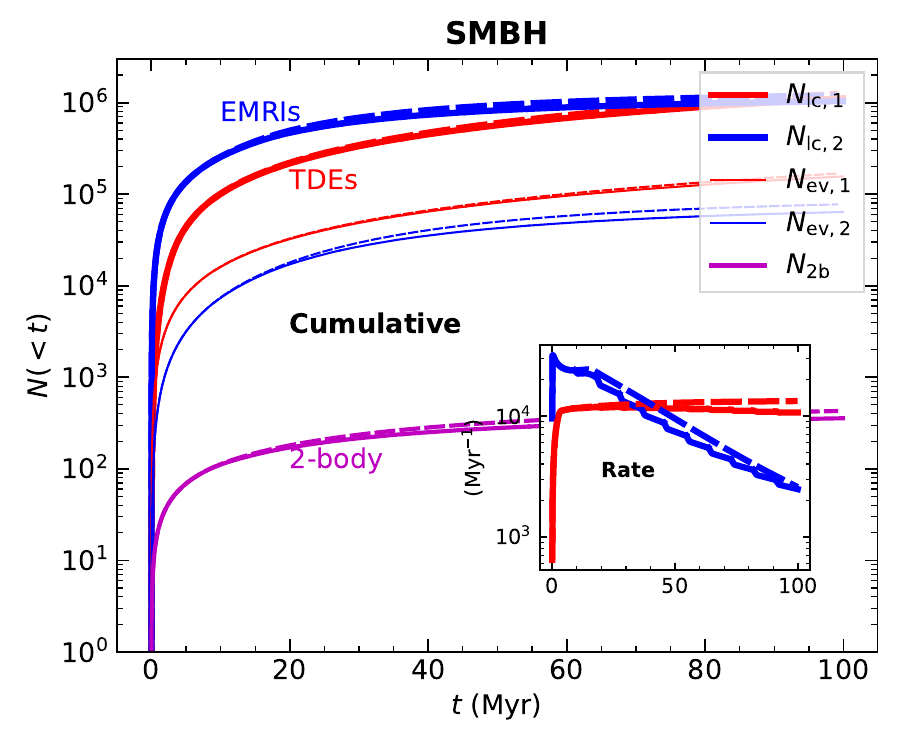}
    \caption{Cumulative number of tidal disruption events (TDEs), captured extreme-mass ratio inspirals (EMRIs), and 2-body binaries (2-body). {\it Left}: The cumulative numbers for the simulations of Fig.~\ref{fig:time-evolution}. The inset shows the rates as a function of time, with the same x-axis as in the main plot. {\it Right}: Correspondingly for the simulations of Fig.~\ref{fig:SMBH-growth}. The cumulative number of 3-body binaries is less than unity in both cases.}
    \label{fig:TDE-capture-rates}
\end{figure*}

Over the $100\,\rm Myr$ simulated, one million TDEs and captured BHs occurred~(cf.~Fig.~\ref{fig:TDE-capture-rates}, right panel).
Hundreds of 2-body binaries are formed between two stellar-mass BHs within the cusp around the SMBH. Most of these binaries are formed in the first $20\,\rm Myr$ with an initial peak rate of $\approx70\,\rm Myr^{-1}$ which continually drops down to a few$\times\rm Myr^{-1}$ at later times. The rates for 3-body binary formation are at least two orders of magnitude smaller than the 2-body rate. Thus, stellar-mass BH-BH merger rates from these 2-body and 3-body channels are expected to be low in environments similar to the one examined here.

We repeat the simulation for this system with the addition of ten gas inflow episodes every $\sim9\,\rm Myr$, each containing $10^6\,M_\odot$ of gas, assuming a gas reservoir of $10^7\,M_\odot$. This time, the AGN duty cycle is $\sim10\%$, higher than in the previous example, and is maintained throughout the $100\,\rm Myr$. With the addition of gas, the BH grows by an extra $\sim30\%$ compared to the growth it would have had via TDEs and captured EMRIs without the gas inflow episodes (see inset in Fig.~\ref{fig:SMBH-growth}).
A plausible scenario for LRDs would be the formation of the IMBH seed through runaway stellar collisions within a few$\,\rm Myr$ and the subsequent growth of the central BH through repeated TDEs, captured EMRIs, and gas accretion. Finally, the TDE duty cycle in this example is higher than in the previous one, rises rapidly to a few percent within $\sim10\,\rm Myr$ as the BH rapidly grows, and there is not much difference between the gas and gas-free cases.

\section{Tidal disruption rates}
\label{sec:Rates}

In the previous section, we have presented the evolution of an NSC and the growth of its central massive BH through the combined effects of gas accretion, TDEs, and captured EMRIs. In this section, we estimate lower limits on cosmological TDE and EMRI rates based on the observed number density of LRDs at $z=4$--$6$ and on properties of high-redshift star clusters detected with the JWST.

Based on the results by~\cite{2025arXiv250403551J}, we assume that the $M_{\rm BH}$-$\sigma_\star$ relation holds up to redshift $z\sim9$, where $\sigma_\star$ is the stellar velocity dispersion of the bulge. We use the relation from~\cite{2020ARA&A..58..257G}. Up to an unknown fudge factor, we determine the break radius of the stellar density profile by the influence radius of the central SMBH, $R_1=GM_{\rm BH}/\sigma_\star^2$. This is motivated by observations of low-redshift cored galaxies; see~\cite{Merritt:2013book}, pg.~20.
Furthermore, based on standard initial mass function assumptions, the total mass in stellar-mass BHs is roughly a fraction of $\approx10\%$ the stellar mass, and assuming energy equipartition among stars and BHs, we estimate $R_2=M_2R_1/M_1\approx0.1R_1$~\citep{Choksi:2018jnq}. Notice that if equipartition is not achieved, then BHs would undergo Spitzer instability, compactify even further, and have a smaller value for $R_2$ than what we estimate here, leading to an increase in captured EMRI rates.

We have used the constraints from Table~6 of~\cite{Matthee:2023utn} on the number density of LRDs in the SMBH mass range $\sim10^7\,M_\odot$--$10^8\,M_\odot$. The measured number densities are provided for three SMBH mass bins $[10^{6.9}, 10^{7.3}]\,M_\odot$, $[10^{7.3}, 10^{7.7}]\,M_\odot$, and $[10^{7.7}, 10^{8.5}]\,M_\odot$, with corresponding number densities $\Phi_{A}$, $\Phi_{B}$, and $\Phi_{C}$, respectively. Correspondingly given the rates per LRD as $\Gamma_{A}$, $\Gamma_{B}$, and $\Gamma_{C}$, we write the source-frame volumetric rate density as $R_{s}=\Phi_{A}\Gamma_{A} + \Phi_{B}\Gamma_{B} + \Phi_{C}\Gamma_{C}$. For simplicity, we assume this to be constant in the redshift range $z=4$--$6$. 
Given the SMBH masses and the stellar velocity dispersions obtained from the $M_{\rm BH}$-$\sigma_\star$ relation, we determine the break radii of our LRDs to be in the range from a few $\rm pc$ up to $130\, \rm pc$, which represents well their observed compactness.

Based on their global properties, the relaxation time of LRDs is greater than a Hubble time~\citep{Escala:2025jlt}; however, presumably, their central NSCs are collisionally relaxed systems.
We do not currently know, at the time of writing, the exact properties of putative NSCs in LRDs. We make some reasonable assumptions on the mass of NSCs when computing rates based on~\cite{2023ApJ...945...53V}, where massive compact clusters with masses $10^{6-7}\,M_\odot$ have been discovered at similar redshifts ($z\sim6$). We thus use $10^6\,M_\odot$ or $10^7\,M_\odot$ as an educated guess for the mass of these systems.

The spin values of the SMBHs in the LRDs are also not known. However, under the assumption that these BHs have undergone significant accretion in the past, their spin is likely high, close to the extremal value, and spun up during the last mass doubling~\citep{Bardeen:1970zz}. Low-redshift observations indicate that a substantial number of active SMBHs are spinning rapidly (dimensionless spin parameter $\gtrsim0.5$), although there are still systematics in the techniques used in determining SMBH spin~\citep{Reynolds:2020jwt}.
Moreover, the spin magnitude of BHs tends to decrease statistically and asymptote to zero through the repeated accretion of stars and BHs with random orientations~\citep{Kritos:2024kpn,Kritos:2025bby}.
Depending on the spin magnitude and orbital inclination of the approaching star, the Hills mass ---the maximum mass of a BH that can tidally disrupt the star--- can be up to several$\times10^8\,M_\odot$~\citep{Mummery:2023meb}.
For simplicity, we assume that all SMBHs in our examined mass range can produce an observable TDE. A more precise understanding of the SMBH spin distribution at $z\sim5$ would enable us to estimate the Hills mass and, thus, the fraction of TDEs that produce an electromagnetic counterpart more accurately.

Once we compute the volumetric source-frame rate densities for TDEs and captured EMRIs, we integrate over comoving volume to obtain the observer-frame intrinsic cumulative yearly event rate of these transients in the redshift range $z=4$--$6$. Under the assumption that $R_{s}$ is constant in that redshift range we write~\citep{Rodriguez:2016kxx,Ricarte:2018mzn}
\begin{align}
    {\cal R}_{o} = \int_{z=4}^{z=6} R_s {1\over1+z} {dV_c}\approx R_s\int_4^6 {dV_c\over dz} {dz\over1+z},
    \label{eq:Ro}
\end{align}
where the factor of $(1+z)^{-1}$ converts time from source-frame to observer-frame. We write ${\cal R}_{o}\approx (1.5\times10^{11}\,{\rm Mpc^3})\cdot R_{s}$.

Since there is a range of predicted rates depending on initial conditions and the number density of LRDs, we report the $5\%$, $50\%$, and $95\%$ percentiles, corresponding to pessimistic, median, and optimistic rates, respectively. We propagate errors in the LRD densities ($\Phi\pm\Delta\Phi$) and scatter in the $M_{\rm BH}$-$\sigma_\star$ relation into our predicted rates for TDEs and captured EMRI rates. We show our results in Table~\ref{tab:TDE-rates} and~\ref{tab:EMRI-rates}, respectively.

With our choice of assumptions, we find a fixed ratio of $R_{\rm TDE}/R_{\rm EMRI}\sim10\,$:$\,1$. Thus, for every captured EMRI, there are about ten stellar tidal disruption events happening at $z=4$--$6$. The corresponding median volumetric source-frame TDE rate density is $R_{\rm TDE}\approx0.003\,\rm yr^{-1}\,Gpc^{-3}$ with an uncertainty of about $1\,\rm dex$. This assumes an NSC mass scale of $10^6\,M_\odot$. If the typical mass of NSCs in those LRDs were $10^7\,M_\odot$, the TDE rate would be higher by a factor of $\sim100$. When integrated over the comoving volume [cf.~Eq.~\eqref{eq:Ro}], it results in a cumulative observer-frame TDE rate of $\approx34\,\rm yr^{-1}$.
Since we assume a fixed merger rate density over $z\in[4,6]$, the same ratio of $\sim10\,$:$\,1$ holds for the volume-integrated observer-frame TDE and captured EMRI rates. This implies that only a few captured EMRIs occur every observer year at those redshifts.

Little Red Dots represent only a fraction of active SMBHs, and the difficulty in observing means that our estimates are only lower limits on the TDE and captured EMRI rates at $z\sim5$. In particular, the number densities of NSCs in each of the three SMBH mass bins we considered, $\Phi_1$, $\Phi_2$, and $\Phi_3$, are only lower limits.
Moreover, under our LRD paradigm, we do not know the properties of the NSCs, and this is the dominant source of uncertainty in our results.

According to~\cite{Babak:2017tow}, depending on the astrophysical model adopted, the intrinsic EMRI rate can be between a few tens and a few tens of thousands at all redshifts. \cite{Kritos:2024sgd} finds a total of a few thousand captured EMRIs up to $z=10$. Moreover, the future Laser Interferometer Space Antenna (LISA) is unlikely to be able to detect EMRIs with primary in the range $10^7\,M_\odot$--$10^8\,M_\odot$ and at the redshifts where the majority of the LRDs are observed~\citep{Babak:2017tow}. However, a gravitational-wave background of EMRIs, i.e., the cumulative gravitational-wave signal from all LRD sources, may still be observable at $\mu$Hz frequencies.

Comparison with previous work highlights both methodological and physical differences. \cite{2024ApJ...977....7R} finds that black holes develop a steady-state density profile scaling as $r^{-4}$ within the radius of influence, which becomes shallower as $r^{-1.75}$ at smaller radii, while stars follow slopes of $r^{-1.5}$ closer to the supermassive black hole. Their framework predicts thousands of detectable extreme mass-ratio inspirals (EMRIs) with LISA. In contrast, in our study the power-law slopes are treated as free parameters (see the Appendix), and if EMRIs are expected to outnumber direct plunges, our estimates may represent a lower limit. Similarly, \cite{2025ApJ...991..146R} derives cusp slopes of 1.75 for black holes and 1.5 for lighter stars based on analytic arguments, whereas our adopted values of 1.5 and 1.0, respectively, imply reduced rates compared to steady-state predictions. Semi-analytic Fokker-Planck estimates of compact object influxes are presented in \cite{2025ApJ...980..150K}, while \cite{2022ApJ...940..101L} incorporates a mass spectrum and shows that, in steady state, each mass group satisfies a zero-energy-flow solution. A comprehensive comparison with these approaches is beyond the scope of this work. Moreover, we do not compute EMRI rates here and instead focus exclusively on direct plunges.

\begin{table}[]
    \centering
    \begin{tabular}{c c c}
    \hline
         & TDE & \\
         & $\rm yr^{-1}\, Mpc^{-3}$ ($\rm yr^{-1}$) & \\
         pessimistic & median & optimistic \\
    \hline
         $6\cdot10^{-13}$  & $3\cdot10^{-12}$  & $2\cdot10^{-11}$  \\
         $6\cdot10^{-11}$ (8) & $2\cdot10^{-10}$ (34) & $9\cdot10^{-10}$ (133) \\
    \hline
    \end{tabular}
    \caption{The high-redshift ($z=4$--$6$) source-frame TDE rate density ($R_s$) from the population of LRDs in units of $\rm yr^{-1}\,Mpc^{-3}$ calculated for the SMBH mass range $10^7\,M_\odot$--$10^8\,M_\odot$ and in parentheses the yearly observer-frame cumulative number of TDEs (${\cal R}_{o}$) in the redshift range $z=4$--$6$. The first (second) row assumes LRDs contain a $10^6\,M_\odot$ ($10^7\,M_\odot$) star cluster. The three columns correspond to the pessimistic, median, and optimistic estimates.}
    \label{tab:TDE-rates}
\end{table}

\begin{table}[]
    \centering
    \begin{tabular}{c c c}
    \hline
         & captured EMRI & \\
         & $\rm yr^{-1}\, Mpc^{-3}$ ($\rm yr^{-1}$) & \\
         pessimistic & median & optimistic \\
    \hline
         $6\cdot10^{-14}$  & $3\cdot10^{-13}$  & $2\cdot10^{-12}$  \\
         $8\cdot10^{-12}$ (1) & $4\cdot10^{-11}$ (6) & $2\cdot10^{-10}$ (31) \\
    \hline
    \end{tabular}
    \caption{Same as Table~\ref{tab:TDE-rates} but for the captured EMRI rates.}
    \label{tab:EMRI-rates}
\end{table}

\section{Why are the x-ray luminosities low?}
\label{sec:x-rays}

\begin{figure}
    \centering
    \includegraphics[width=\linewidth]{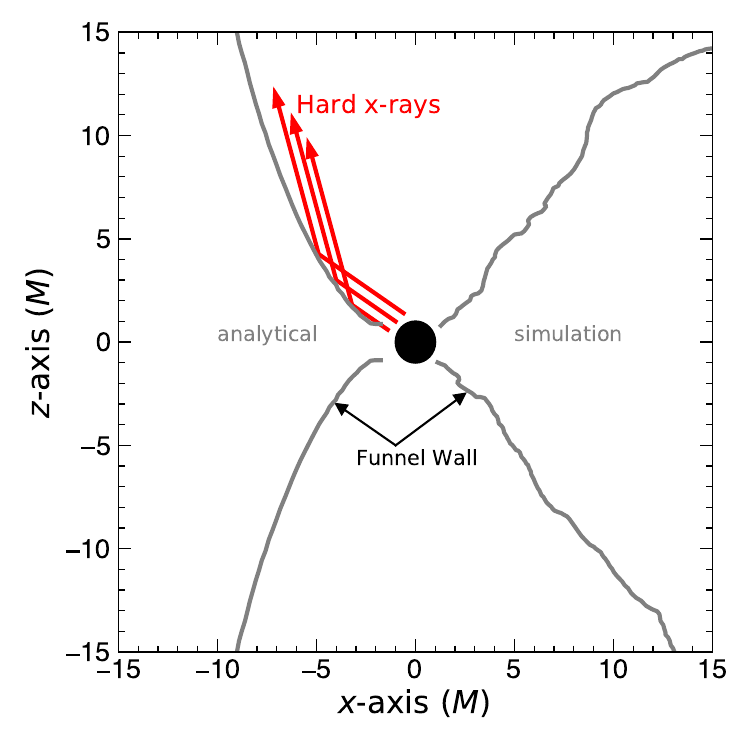}
    \caption{Hard x-rays from the BH hot inner corona scattered on Polish doughnut funnel walls. Funnel model and thick accretion disk are from analytical (left) and numerical simulations (right)~\citep[the values of the envelope are digitized from Fig.~1 in][with units in gravitational radii]{Lei:2008ui}. Scattering of hard x-rays as depicted for proxy model of Cygnus X-3~\citep{2024A&A...688L..27V}.}
    \label{fig:disk-and-x-rays}
\end{figure}

In general, LRDs are subluminous in x-rays~\citep{Ananna:2024jug,Sacchi:2025xjk}. This superficially is a problem, especially because of the phases we invoke of super-Eddington accretion onto the central BH.
Explanations in the literature include an inhomogeneous broad emission line region with high covering factor~\citep{Maiolino:2024uon}. \cite{Pacucci:2024tws} consider mildly Eddington accretion rates which results in lower SMBH masses [see also~\cite{Lambrides:2024ugh}]. 
An analogy with ULXs which have highly beamed x-ray emission from super-Eddington accretion onto the central BH provides yet another interpretation~\citep{King:2024oqb}. 
\cite{Madau:2024fdv} discuss x-ray scattering from the funnel walls, a model that we develop further below.

We can use Cygnus X-3 as a template for the observability, or lack thereof, of x-rays from the  SMBH. Cygnus X-3 is a microquasar produced by a super-Eddington x-ray binary. The central x-ray source is hidden behind the optically thick accretion disk. The narrow disk funnel scatters hard x-rays from the central BH at source inclination of $\sim$30~degrees, as measured by the unexpectedly high~(12\%) x-ray polarization as measured by IXPE~\citep{2024A&A...688L..27V}. Due to transient funnel filling phases of variable accretion, the resulting luminous ultrasoft x-ray flux ionizes the accretion disk corona.

We argue that a very similar phenomenon applies in the super-Eddington phase of the LRD~\citep{Lei:2008ui}. 
There is a deep funnel surrounding the BH in the super-Eddington accretion-induced phase (see Fig.~\ref{fig:disk-and-x-rays}). This is the so-called Polish doughnut~\citep{2025arXiv250109854M}, which alternates with a slim disk as the accretion rate varies from high to low. The super-Eddington phases are intermittent and brief. They result in intense hard x-rays that are, however, scattered in the funnel. One consequence is that they are viewed only at an oblique angle. Hence, strong x-ray sources associated with super-Eddington accretion are not directly observed in the direction of the LRD in this phase. The soft x-ray component escapes directly and plays a role in ionizing the accretion disk corona. Observational consequences would include unassociated x-ray sources and high x-ray polarization as high as 10\%--20\% by analogy with Cygnus X-3~\citep{2024A&A...688L..27V}.

\section{Conclusions}
\label{sec:Conclusions}

Regardless of interpretation, LRDs must inevitably host a massive BH; even in the stars-only scenario, the inferred stellar densities are so high that a massive object would rapidly form through runaway stellar collisions~\citep{Pacucci:2025ojp,Escala:2025jlt}.

If a significant amount of gas is present in the center of LRDs~\citep{2025arXiv250920455J}, EMRIs may be accelerated by the enhanced dynamical friction from the gas environment. As demonstrated by~\cite{Speri:2022upm}, the associated gravitational-wave dephasing imprinted in the gravitational waveform signal may be detectable by LISA.

We found that a significant mass of the SMBH in LRDs could be assembled through the accretion of stars and stellar-mass BHs from the surrounding dense NSC environment. See also~\cite{Stone:2016ryd,2021MNRAS.500.3944P,Rizzuto:2022fdp,2024A&A...689A.204P} for similar conclusions. 
High stellar densities increase the rates of TDEs and EMRIs~\citep{Zhang:2025jmm}.
The gas further boosts the growth.
In the densest system simulated, the IMBH seed rapidly grew through TDEs, EMRIs, and gas accretion within a few tens of Myr. This allows for enough time for the BH to grow and the system to resemble the properties of LRDs by $z\sim5$.

In the densest systems, an IMBH seed would rapidly grow through TDEs, EMRIs, and gas accretion. Simulations we carried out for NSCs with a central density of $10^8\,\rm pc^{-3}$ resulted in the rapid growth of a $10^3\,M_\odot$ IMBH seed into the SMBH regime (final mass $\simeq2\times10^7\,M_\odot$) within a few$\times10\,\rm Myr$, giving rise to the $z\sim5$ population of LRDs. Again, a significant fraction of the SMBH's mass ($\sim80\%$) is assembled through TDEs and captured EMRIs. Note that this fraction is dependent on the gas mass reservoir, which is taken in this example to be $10^7\,M_\odot$ and supplied in ten inflow episodes over a $100\,\rm Myr$ period. Moreover, there is a residual population of stellar-mass BHs around the SMBH, leading to subsequent high-energy phenomena such as BH mergers and EMRIs.

While the central massive BH is in the active accretion phase, such a gas-rich NSC with an accreting SMBH in its center would observationally resemble an LRD~\citep{2025ApJ...980L..27I}. After the residual gas is removed due to feedback and gas accretion shuts off, the BH becomes quiet. During the inactive phase, though, the BH continues to grow via successive TDEs and EMRIs.  
\cite{Bellovary:2025aeg} finds a high-redshift TDE rate of $\sim10^{-4}\,\rm yr^{-1}$ by fitting the observed abundance of LRDs to the theoretical number density of SMBH seeds under the assumption that the emission of LRDs is exclusively powered by TDEs. While this rate estimate is five orders of magnitude lower than ours, it cannot explain the necessary growth rate of the BHs.
In this paper we envision that the emission of LRDs is dominated by the AGN and gas accretion plays an important role in the growth of BHs.

Using JWST data,~\cite{Karmen:2025xhz} claims a candidate for high-redshift TDE at $z\approx5$.
Moreover, \cite{Graham:2025tla} has identified the superluminous optical flare J2245+3743, most likely corresponding to the tidal disruption of a massive star ($\gtrsim10\,M_\odot$) by a $\sim3\times10^8\,M_\odot$ SMBH at redshift $z\approx2$. Based on this single observation, \cite{Graham:2025tla} infers the rate of high-redshift ($z\sim2$) TDEs to be $\sim3\times10^{-13}\,{\rm yr}^{-1}\,{\rm Mpc}^{-3}$. This is contrasted with our predicted range $6\times10^{-13}$--$2\times10^{-11}\,{\rm yr}^{-1}\,{\rm Mpc}^{-3}$ for the $z=4$--$6$ TDE rate density. For comparison, according to \cite{vanVelzen:2014dna} the local TDE rate density is $\sim3\times10^{-7}\,{\rm yr}^{-1}\,{\rm Mpc}^{-3}$, which is a factor of $\sim10$ less from what theoretical models predict for $z\sim0$~\citep{Stone:2014wxa}.
At the time of writing, other five TDE candidates have been identified at $z>1$~\citep{Gu:2024nzu}.
We expect that future experiments, such as UVEX~\citep{Kulkarni:2021tit}, will utilize imaging and spectroscopy to identify many more TDE candidates. Other diagnostics that are currently inconclusive but are expected to rapidly improve include radio detectability~\citep{2025A&A...693L...2P,Latif:2025xyo} and variability~\citep{2025ApJ...991..137Z,2025A&A...698A.227F,2025ApJ...983L..26T}. 

As the gas is being ejected away from the episodic accretion feedback or gets converted into stars in subsequent starburst episodes, the galaxy builds up its stellar population and eventually starves of gas, limiting SMBH growth through accretion~\citep{Silk:2024rsf}.
Thus, some SMBHs may have grown faster than the stellar population around them~\citep{2025arXiv250821748J}, and therefore the $M_{\rm BH}/M_\star$ ratios are larger at higher redshifts~\citep{2025arXiv250622147G,2025arXiv251007376J}.

The population of stellar-mass BHs around the SMBH in LRDs can lead to other electromagnetic counterparts, such as micro-TDEs~\citep{Perets:2016pwr,Kremer:2019zql,Rastello:2025foo}. These involve the tidal disruption of a main-sequence star by a stellar-mass BH.
If a population of stellar-mass BHs is present around SMBHs in LRDs, a fraction of stars will be tidally disrupted by BH remnants. This is very likely due to the evolution of massive stars ($\gtrsim20\,M_\odot$) into BHs in the dense NSC environment.
Moreover, there might be other types of high-energy phenomena such as white dwarf tidal disruptions, mergers involving neutron stars, and stellar collisions, all of which produce multimessenger signals.

Preferential gas accretion onto the secondary in a SMBH merger is argued to inevitably boost SMBH merger rate~\citep{Krause:2025gkq}, reconciling the Pulsar Timing Array signal with the predicted merger rate at low redshift~\citep{2025A&A...700A.135T}, and providing potentially even more significant boosts in the merger rate predicted in the gas-rich regime that is characteristic of the high redshift regime which LISA is expected to probe~\citep{Comerford:2025tjl}.

Given the uncertainties of LRD number densities and the unknown properties of their putative central NSCs, we make predictions of intrinsic TDE and captured EMRI rates at $z=4$--$6$ with median values of $\sim500\,\rm yr^{-1}$ and $\sim4\,\rm yr^{-1}$, respectively, depending on the properties of the LRD's NSCs.
The number density values of LRDs we used from~\cite{Matthee:2023utn} are lower bounds of the number density of SMBHs. Together with selection effects, the fact that lower-mass SMBHs are more challenging to observe, we conclude that our estimates can only be a lower bound.
We may not be able to see individual EMRIs from LRDs at those high redshifts, but we may be able to identify a background of EMRIs instead~\citep{Bonetti:2020jku}.

\begin{acknowledgments}
We thank Mitch Begelman, Jillian Bellovary, Emanuele Berti, Francesco Bollati, Mitchell Karmen, Piero Madau, Colin Norman, Massimo Ricotti, and Marta Volonteri for discussions.
We also thank Daniel D'Orazio, Brenna Mockler, Connar Rowan, Johan Samsing, Mattia Sormani, James Stone, Christofer Tiede, and Alessandro A. Trani for discussions that took place during the NBIA Workshop on Open Problems in Astrophysical Dynamics.
Finally, we thank the anonymous referee for their comments.
K.K. is supported by NSF Grants No.~AST-2307146, PHY-2513337, PHY-090003, and PHY-20043, by NASA Grant No.~21-ATP21-0010, by John Templeton Foundation Grant No.~62840, by the Simons Foundation [MPS-SIP-00001698, E.B.], by the Simons Foundation International, by Italian Ministry of Foreign Affairs and International Cooperation Grant No.~PGR01167, and by the Onassis Foundation - Scholarship ID: F ZT 041-1/2023-2024.
\end{acknowledgments}

\appendix
\section*{Dependence on the cusp slope}

\begin{figure}
    \centering
    \includegraphics[width=0.24\linewidth]{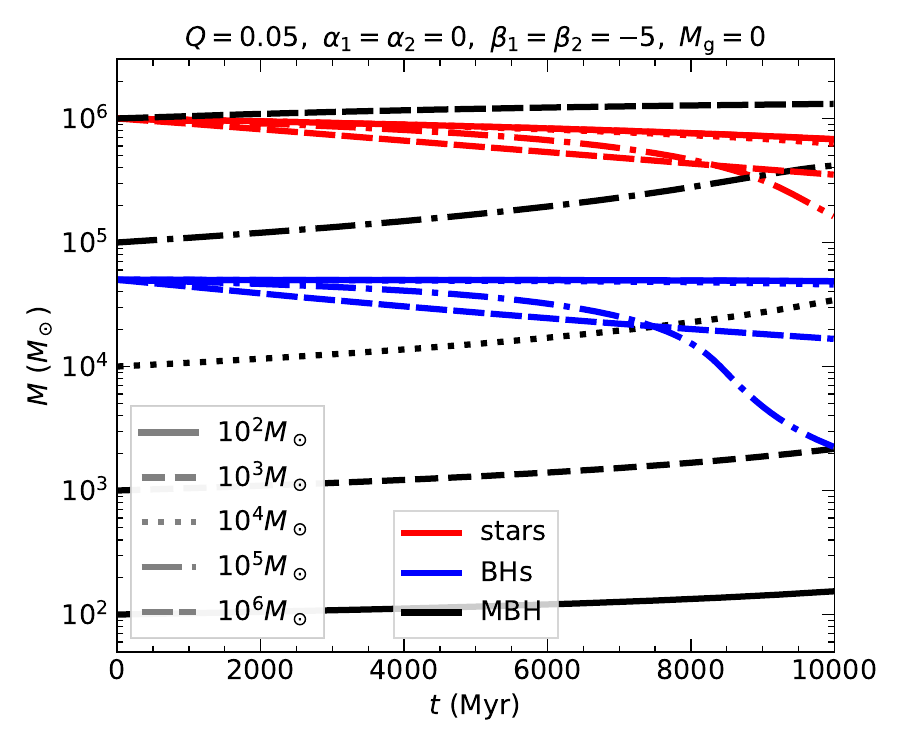}
    \includegraphics[width=0.24\linewidth]{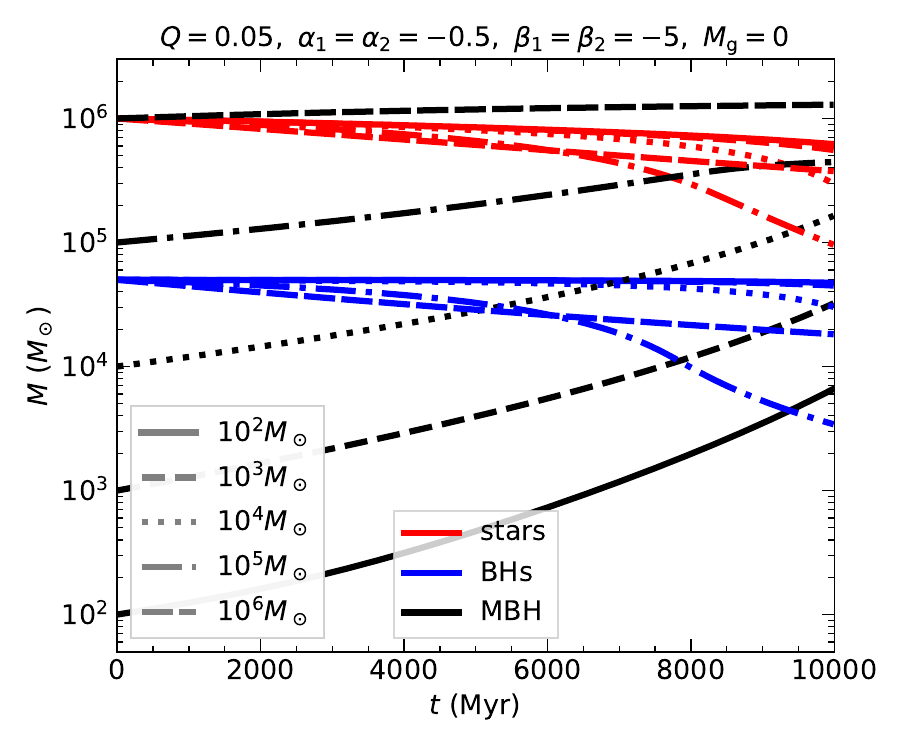}
    \includegraphics[width=0.24\linewidth]{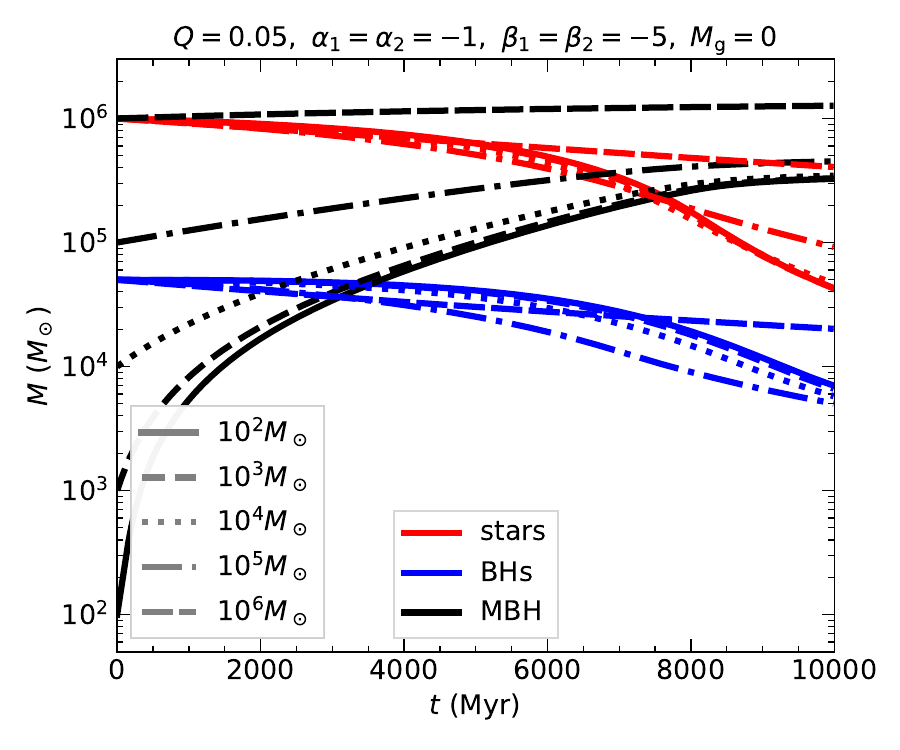}
    \includegraphics[width=0.24\linewidth]{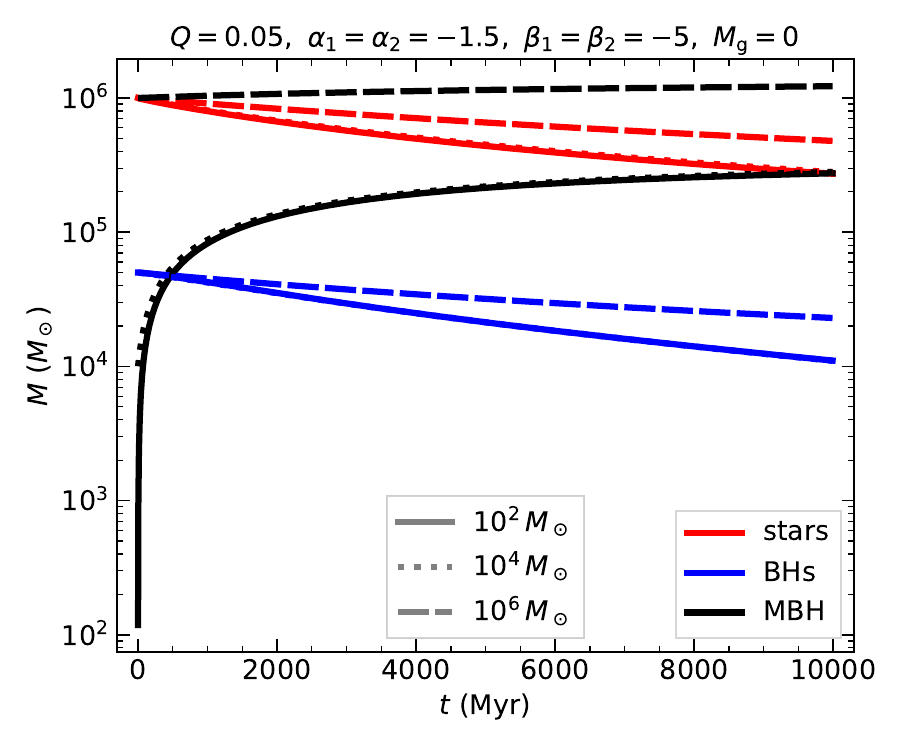}
    \includegraphics[width=0.24\linewidth]{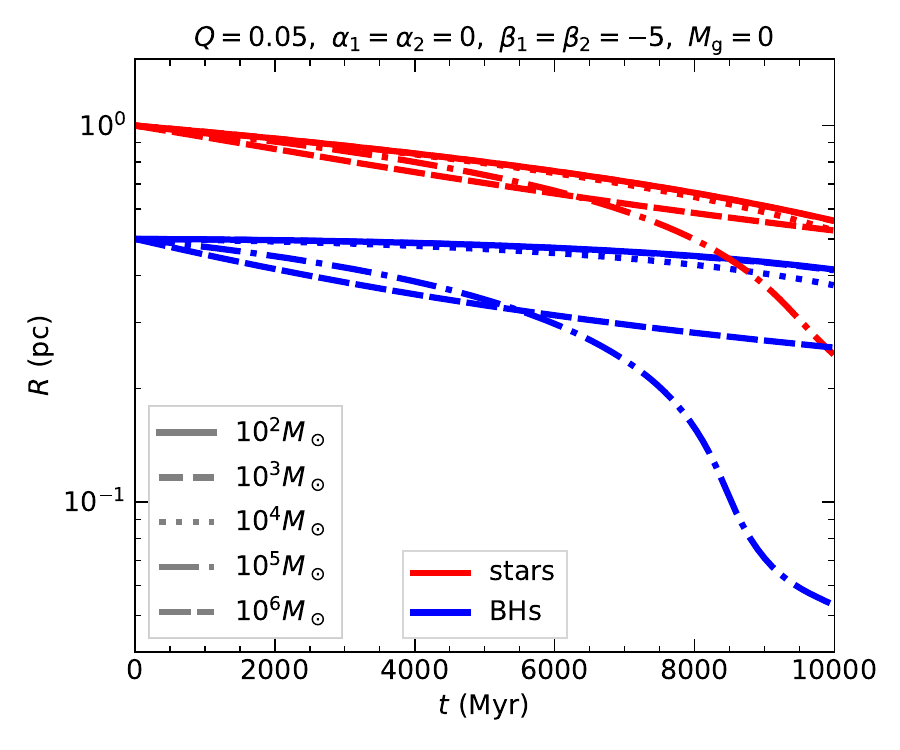}
    \includegraphics[width=0.24\linewidth]{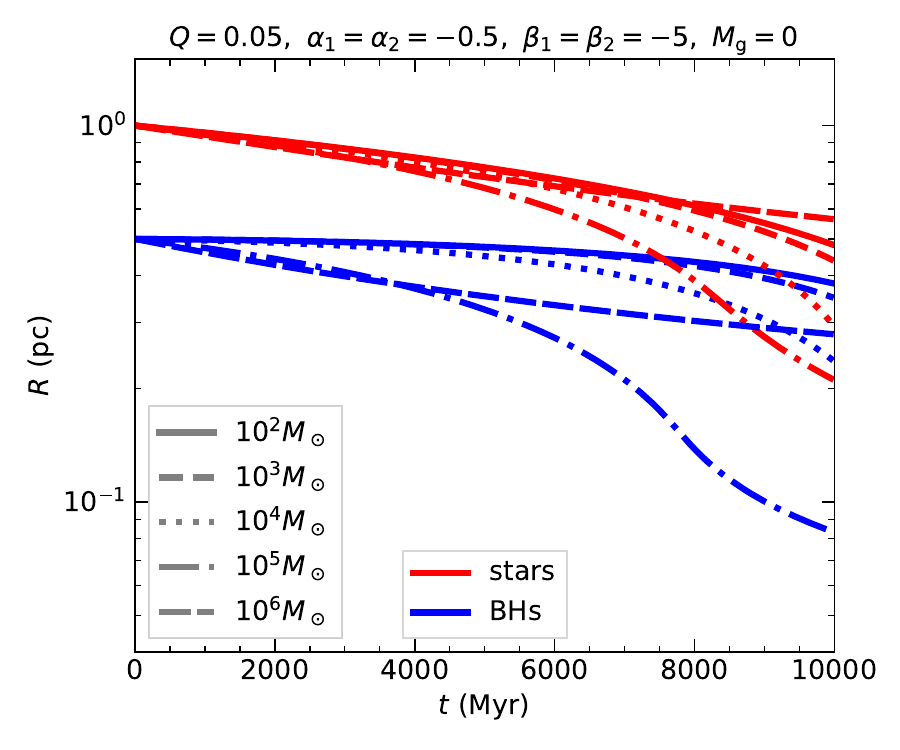}
    \includegraphics[width=0.24\linewidth]{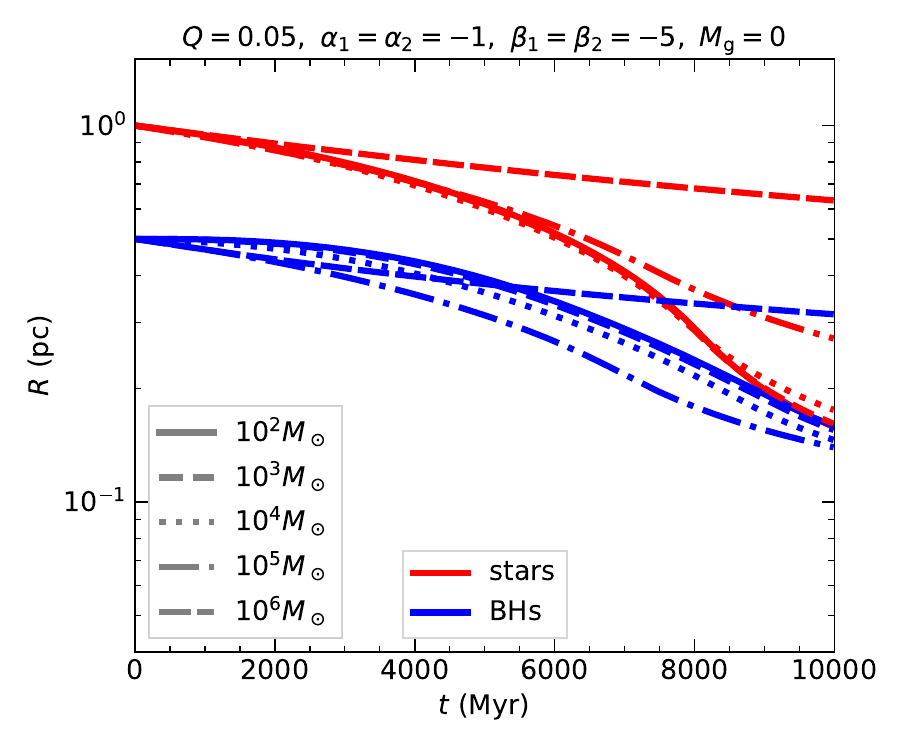}
    \includegraphics[width=0.24\linewidth]{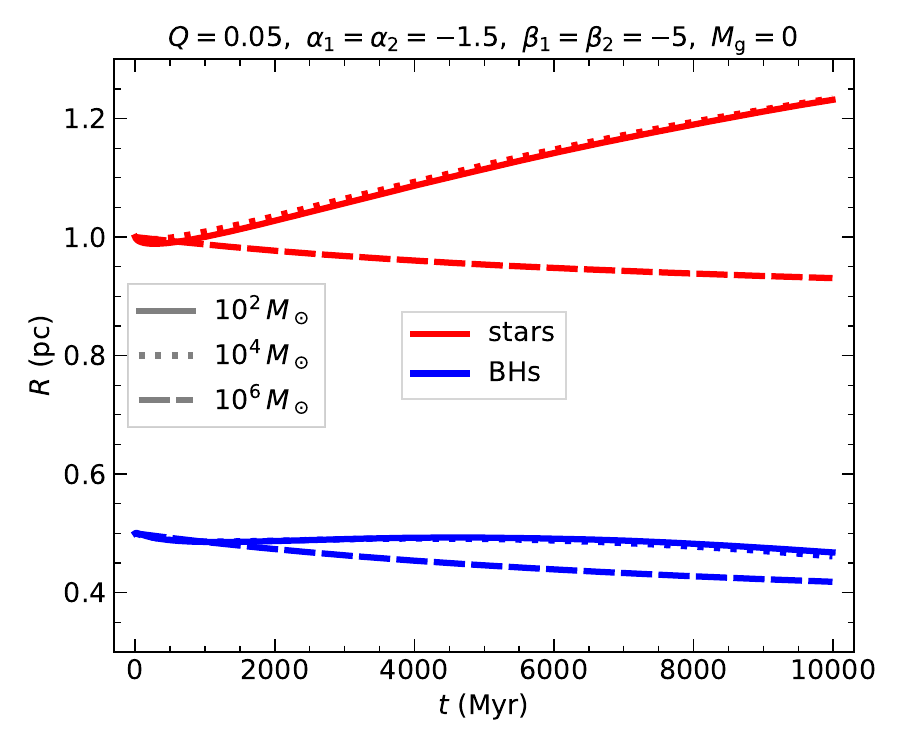}
    \includegraphics[width=0.24\linewidth]{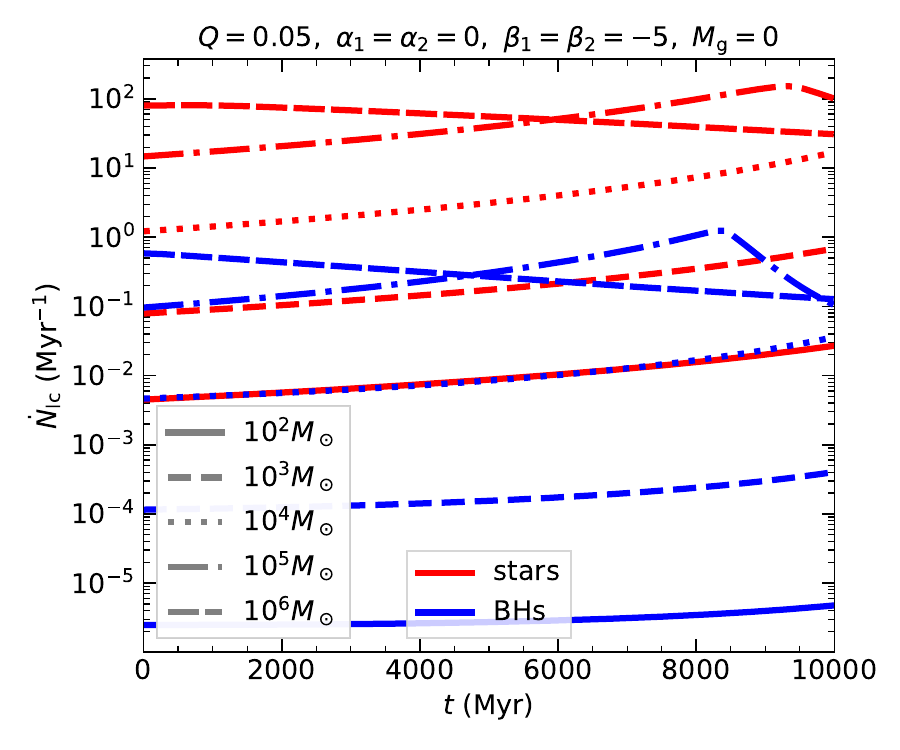}
    \includegraphics[width=0.24\linewidth]{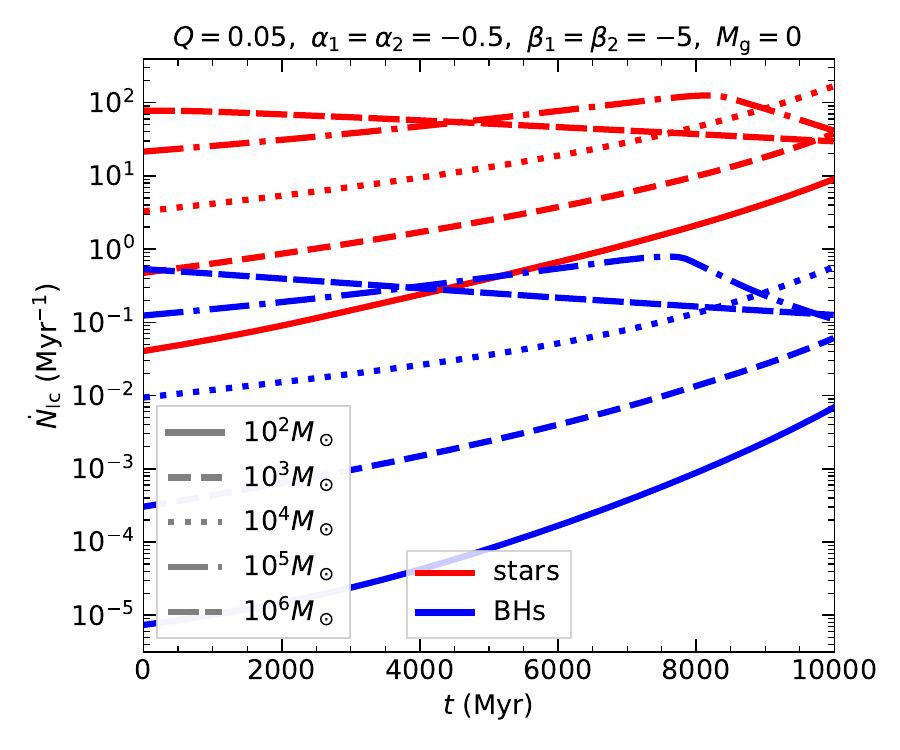}
    \includegraphics[width=0.24\linewidth]{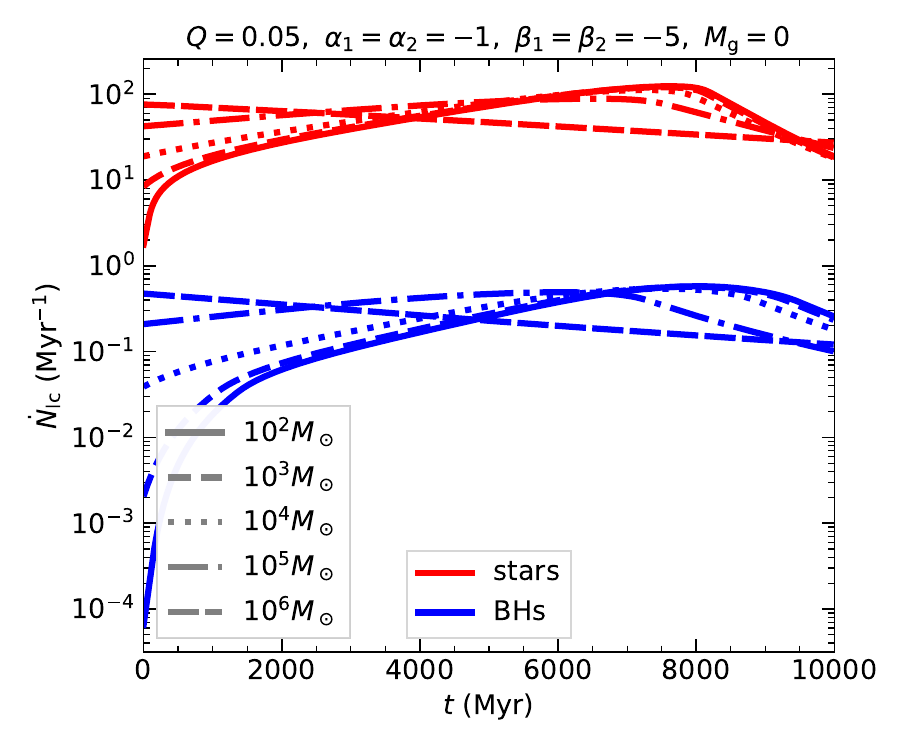}
    \includegraphics[width=0.24\linewidth]{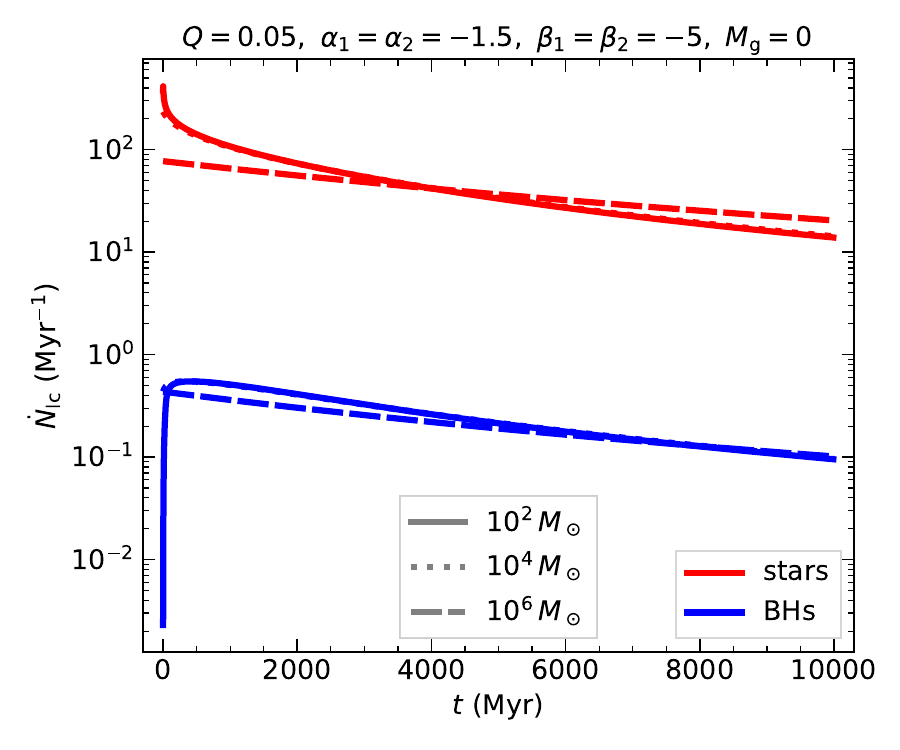}
    \includegraphics[width=0.24\linewidth]{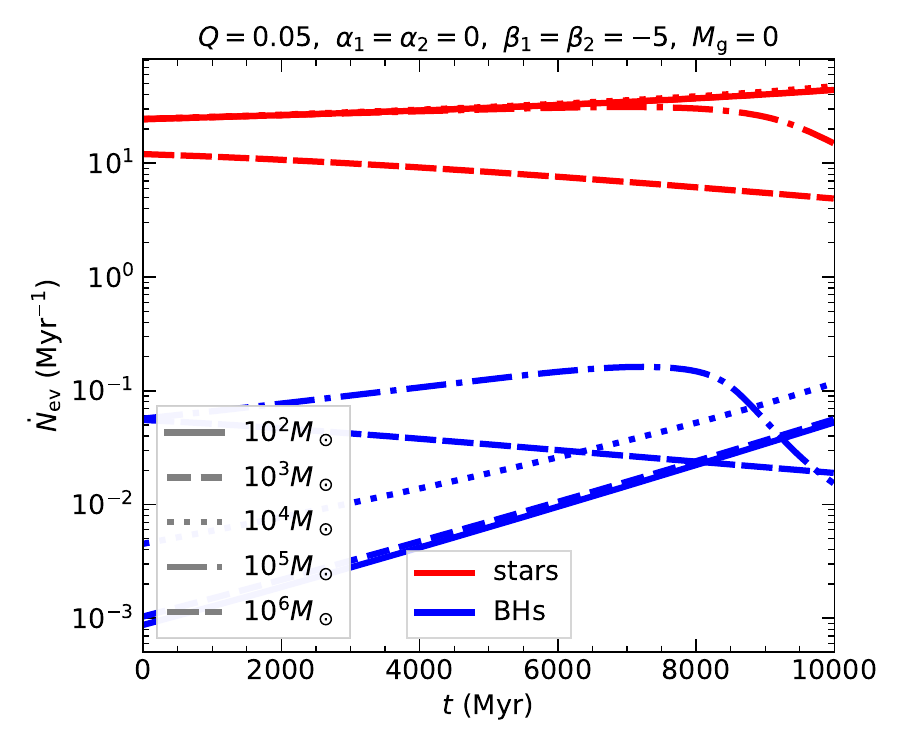}
    \includegraphics[width=0.24\linewidth]{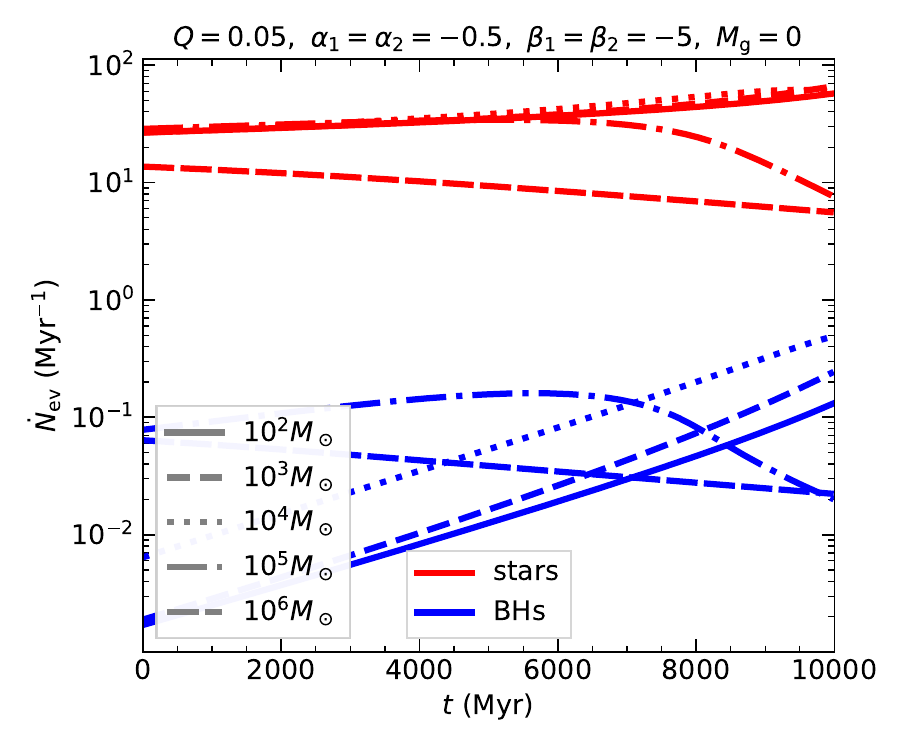}
    \includegraphics[width=0.24\linewidth]{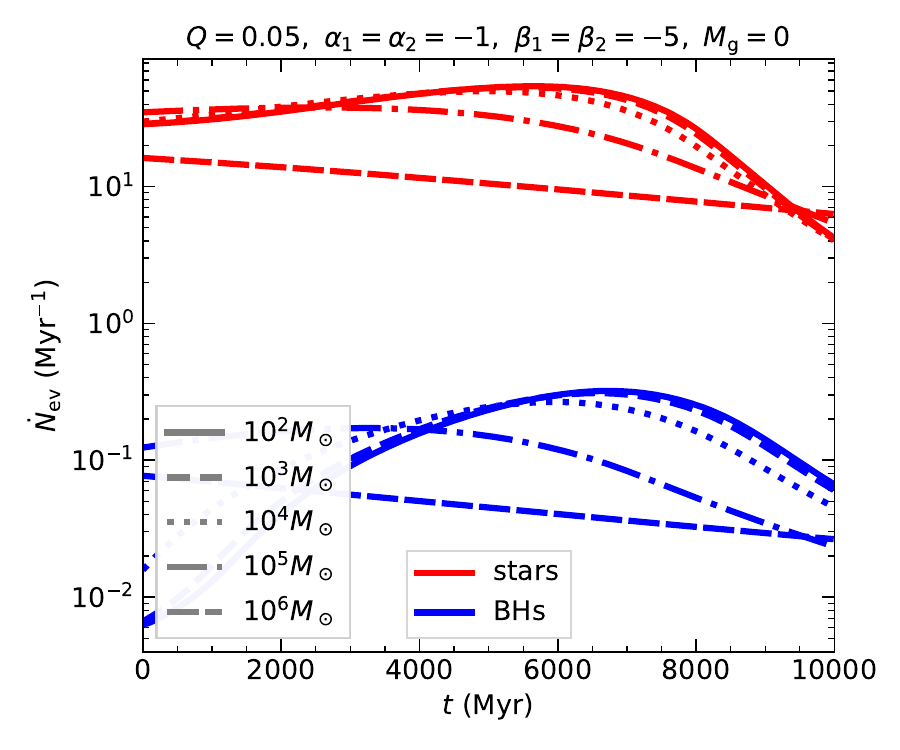}
    \includegraphics[width=0.24\linewidth]{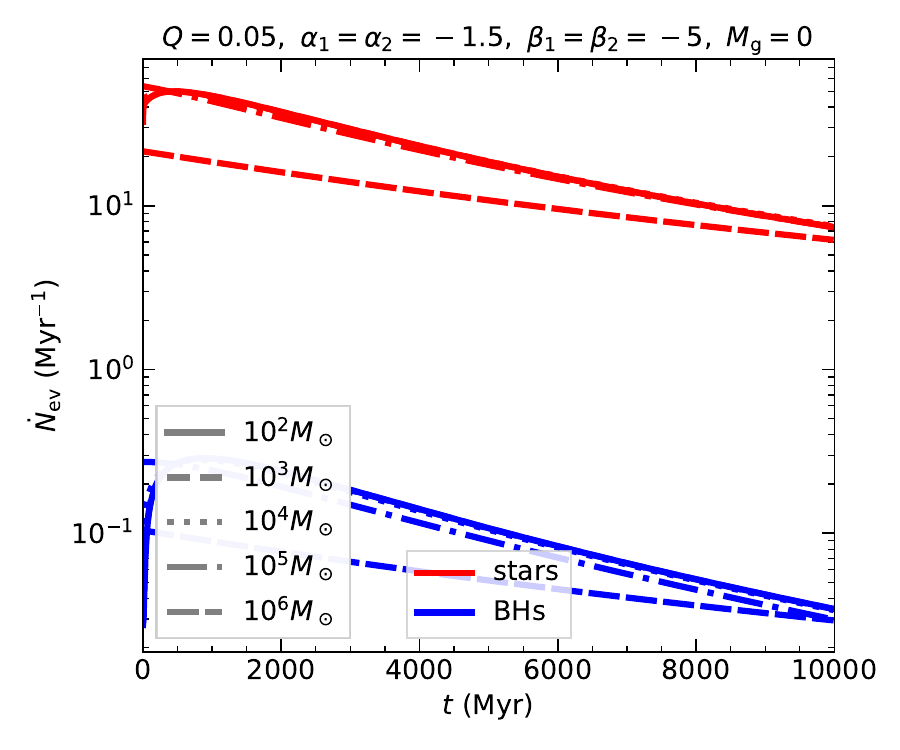}
    \caption{Temporal evolution of the masses ($M$, top row), radii ($R$, second row), loss-cone ($\dot{N}_{\rm lc}$, third row), and evaporation rates ($\dot{N}_{\rm ev}$, bottom row) while varying the seed mass of the central growing BH (black lines, top row). We assume stars and stellar-mass BHs have the same inner cusp slopes $\alpha_1=\alpha_2$ and different columns consider different index values, $\alpha_1=0$ (left column), $\alpha_1=-0.5$ (second column), $\alpha_1=-1.0$ (third column), and $\alpha_1=-1.5$ (right column). All other parameters of the model have been fixed. Here $Q$ is the ratio of total mass in stellar-mass BHs ($m_2=10\,M_\odot$) to the total mass in low-mass stars ($m_1=1\,M_\odot$).}
    \label{fig:dependence-of-cusp}
\end{figure}

In this Appendix, we explore the dependence of our results on the choice of cusp indices $\alpha_{1,2}$. To isolate its effect, we fix all other parameters taking $\beta_1=\beta_2=-5$, $M_1(t=0)=20M_2(t=0)=10^6\,M_\odot$, $R_1(t=0)=2R_2(t=0)=1\,\rm pc$. We vary only the indices for which we further assume $\alpha_1=\alpha_2\in\{0.0, -0.5, -1.0, -1.5\}$ and consider five seed BH masses $M_{\rm BH}\in\{10^2,10^3,10^4,10^5,10^6\}\,M_\odot$; we do not attempt a comprehensive survey of the parameter space in this work. The dependence of the results on the slope of the inner cusp is shown in Fig.~\ref{fig:dependence-of-cusp}.

\newpage
\bibliographystyle{aasjournal_titles}
\bibliography{LRD_model}

@ARTICLE{2025arXiv251007376J,
       author = {{Jones}, Brenda L. and {Kocevski}, Dale D. and {Pacucci}, Fabio and {Taylor}, Anthony J. and {Finkelstein}, Steven L. and {Buchner}, Johannes and {Trump}, Jonathan R. and {Somerville}, Rachel S. and {Hirschmann}, Michaela and {Yung}, L.~Y. Aaron and {Barro}, Guillermo and {Bell}, Eric F. and {Bisigello}, Laura and {Calabro}, Antonello and {Cleri}, Nikko J. and {Dekel}, Avishai and {Dickinson}, Mark and {Gandolfi}, Giovanni and {Giavalisco}, Mauro and {Grogin}, Norman A. and {Inayoshi}, Kohei and {Kartaltepe}, Jeyhan S. and {Koekemoer}, Anton M. and {Napolitano}, Lorenzo and {Onoue}, Masafusa and {Ravindranath}, Swara and {Rodighiero}, Giulia and {Wilkins}, Stephen M.},
        title = "{The $M_{\rm BH}-M_{*}$ Relationship at $3<z<7$: Big Black Holes in Little Red Dots}",
      journal = {arXiv e-prints},
         year = 2025,
        month = oct,
          eid = {arXiv:2510.07376},
        pages = {arXiv:2510.07376},
          doi = {10.48550/arXiv.2510.07376},
archivePrefix = {arXiv},
       eprint = {2510.07376},
 primaryClass = {astro-ph.GA},
       adsurl = {https://ui.adsabs.harvard.edu/abs/2025arXiv251007376J}
}

@article{Lin:2025pnq,
    author = "Lin, Xiaojing and others",
    title = "{The Discovery of Little Red Dots in the Local Universe: Signatures of Cool Gas Envelopes}",
    eprint = "2507.10659",
    archivePrefix = "arXiv",
    primaryClass = "astro-ph.GA",
    month = "7",
    year = "2025"
}

@article{Stone:2014wxa,
    author = "Stone, Nicholas C. and Metzger, Brian D.",
    title = "{Rates of Stellar Tidal Disruption as Probes of the Supermassive Black Hole Mass Function}",
    eprint = "1410.7772",
    archivePrefix = "arXiv",
    primaryClass = "astro-ph.HE",
    doi = "10.1093/mnras/stv2281",
    journal = "Mon. Not. Roy. Astron. Soc.",
    volume = "455",
    number = "1",
    pages = "859--883",
    year = "2016"
}

@article{Gu:2024nzu,
    author = "Gu, Ying and Zhang, Xue-Guang and Chen, Xing-Qian and Yang, Xing and Liang, En-Wei",
    title = "{A candidate of high-z central tidal disruption event in quasar SDSS J000118.70+003314.0}",
    eprint = "2412.17046",
    archivePrefix = "arXiv",
    primaryClass = "astro-ph.GA",
    doi = "10.1093/mnras/stae2816",
    journal = "Mon. Not. Roy. Astron. Soc.",
    volume = "537",
    number = "1",
    pages = "84--96",
    year = "2025"
}

@article{vanVelzen:2014dna,
    author = "van Velzen, Sjoert and Farrar, Glennys R.",
    title = "{Measurement of the rate of stellar tidal disruption flares}",
    eprint = "1407.6425",
    archivePrefix = "arXiv",
    primaryClass = "astro-ph.GA",
    doi = "10.1088/0004-637X/792/1/53",
    journal = "Astrophys. J.",
    volume = "792",
    pages = "53",
    year = "2014"
}

@article{Graham:2025tla,
    author = "Graham, Matthew J. and others",
    title = "{An Extremely Luminous Flare Recorded from a Supermassive Black Hole}",
    eprint = "2511.02178",
    archivePrefix = "arXiv",
    primaryClass = "astro-ph.GA",
    doi = "10.1038/s41550-025-02699-0",
    month = "11",
    year = "2025"
}

@ARTICLE{2025arXiv250612141L,
       author = {{Loiacono}, Federica and {Gilli}, Roberto and {Mignoli}, Marco and {Mazzolari}, Giovanni and {Decarli}, Roberto and {Brusa}, Marcella and {Calura}, Francesco and {Chiaberge}, Marco and {Comastri}, Andrea and {D'Amato}, Quirino and {Iwasawa}, Kazushi and {Juod{\v{z}}balis}, Ignas and {Lanzuisi}, Giorgio and {Maiolino}, Roberto and {Marchesi}, Stefano and {Norman}, Colin and {Peca}, Alessandro and {Prandoni}, Isabella and {Sapori}, Matteo and {Signorini}, Matilde and {Tozzi}, Paolo and {Vanzella}, Eros and {Vignali}, Cristian and {Vito}, Fabio and {Zamorani}, Gianni},
        title = "{A big red dot at cosmic noon}",
      journal = {arXiv e-prints},
         year = 2025,
        month = jun,
          eid = {arXiv:2506.12141},
        pages = {arXiv:2506.12141},
          doi = {10.48550/arXiv.2506.12141},
archivePrefix = {arXiv},
       eprint = {2506.12141},
 primaryClass = {astro-ph.GA},
       adsurl = {https://ui.adsabs.harvard.edu/abs/2025arXiv250612141L}
}

@article{Sacchi:2025xjk,
    author = "Sacchi, Andrea and Bogdan, Akos",
    title = "{Chandra Rules Out Super-Eddington Accretion Models for Little Red Dots}",
    eprint = "2505.09669",
    archivePrefix = "arXiv",
    primaryClass = "astro-ph.GA",
    doi = "10.3847/2041-8213/adf5c8",
    journal = "Astrophys. J. Lett.",
    volume = "989",
    number = "2",
    pages = "L30",
    year = "2025"
}

@article{Bellovary:2025aeg,
    author = "Bellovary, Jillian",
    title = "{Little Red Dots Are Tidal Disruption Events in Runaway-collapsing Clusters}",
    eprint = "2501.03309",
    archivePrefix = "arXiv",
    primaryClass = "astro-ph.GA",
    doi = "10.3847/2041-8213/adce6c",
    journal = "Astrophys. J. Lett.",
    volume = "984",
    number = "2",
    pages = "L55",
    year = "2025"
}

@article{Kulkarni:2021tit,
    author = "Kulkarni, S. R. and others",
    title = "{Science with the Ultraviolet Explorer (UVEX)}",
    eprint = "2111.15608",
    archivePrefix = "arXiv",
    primaryClass = "astro-ph.GA",
    month = "11",
    year = "2021"
}

@ARTICLE{2025ApJ...991..137Z,
       author = {{Zhou}, Shuying and {Sun}, Mouyuan and {Zhang}, Zijian and {Chen}, Jie and {Ho}, Luis C.},
        title = "{On the Variability Features of Active Galactic Nuclei in Little Red Dots}",
      journal = {\apj},
         year = 2025,
        month = oct,
       volume = {991},
       number = {2},
          eid = {137},
        pages = {137},
          doi = {10.3847/1538-4357/adfd5f},
archivePrefix = {arXiv},
       eprint = {2508.16795},
 primaryClass = {astro-ph.GA},
       adsurl = {https://ui.adsabs.harvard.edu/abs/2025ApJ...991..137Z}
}

@ARTICLE{2025A&A...698A.227F,
       author = {{Furtak}, Lukas J. and {Secunda}, Amy R. and {Greene}, Jenny E. and {Zitrin}, Adi and {Labb{\'e}}, Ivo and {Golubchik}, Miriam and {Bezanson}, Rachel and {Kokorev}, Vasily and {Atek}, Hakim and {Brammer}, Gabriel B. and {Chemerynska}, Iryna and {Cutler}, Sam E. and {Dayal}, Pratika and {Feldmann}, Robert and {Fujimoto}, Seiji and {Glazebrook}, Karl and {Leja}, Joel and {Ma}, Yilun and {Matthee}, Jorryt and {Naidu}, Rohan P. and {Nelson}, Erica J. and {Oesch}, Pascal A. and {Pan}, Richard and {Price}, Sedona H. and {Suess}, Katherine A. and {Wang}, Bingjie and {Weaver}, John R. and {Whitaker}, Katherine E.},
        title = "{Investigating photometric and spectroscopic variability in the multiply imaged little red dot A2744-QSO1}",
      journal = {\aap},
         year = 2025,
        month = jun,
       volume = {698},
          eid = {A227},
        pages = {A227},
          doi = {10.1051/0004-6361/202554110},
archivePrefix = {arXiv},
       eprint = {2502.07875},
 primaryClass = {astro-ph.GA},
       adsurl = {https://ui.adsabs.harvard.edu/abs/2025A&A...698A.227F}
}

@ARTICLE{2025ApJ...983L..26T,
       author = {{Tee}, Wei Leong and {Fan}, Xiaohui and {Wang}, Feige and {Yang}, Jinyi},
        title = "{Lack of Rest-frame Ultraviolet Variability in Little Red Dots Based on HST and JWST Observations}",
      journal = {\apjl},
         year = 2025,
        month = apr,
       volume = {983},
       number = {1},
          eid = {L26},
        pages = {L26},
          doi = {10.3847/2041-8213/adc5e3},
archivePrefix = {arXiv},
       eprint = {2412.05242},
 primaryClass = {astro-ph.GA},
       adsurl = {https://ui.adsabs.harvard.edu/abs/2025ApJ...983L..26T}
}

@article{Latif:2025xyo,
    author = "Latif, Muhammad A. and Aftab, Ammara and Whalen, Daniel J. and Mezcua, Mar",
    title = "{Radio emission from little red dots may reveal their true nature}",
    eprint = "2502.03742",
    archivePrefix = "arXiv",
    primaryClass = "astro-ph.GA",
    doi = "10.1051/0004-6361/202453194",
    journal = "Astron. Astrophys.",
    volume = "694",
    pages = "L14",
    year = "2025"
}

@ARTICLE{2025A&A...693L...2P,
       author = {{Perger}, K. and {Fogasy}, J. and {Frey}, S. and {Gab{\'a}nyi}, K. {\'E}.},
        title = "{Deep silence: Radio properties of little red dots}",
      journal = {\aap},
         year = 2025,
        month = jan,
       volume = {693},
          eid = {L2},
        pages = {L2},
          doi = {10.1051/0004-6361/202452422},
archivePrefix = {arXiv},
       eprint = {2411.19518},
 primaryClass = {astro-ph.GA},
       adsurl = {https://ui.adsabs.harvard.edu/abs/2025A&A...693L...2P}
}

@article{Naidu:2025rpo,
    author = "Naidu, Rohan P. and others",
    title = "{A ''Black Hole Star'' Reveals the Remarkable Gas-Enshrouded Hearts of the Little Red Dots}",
    eprint = "2503.16596",
    archivePrefix = "arXiv",
    primaryClass = "astro-ph.GA",
    month = "3",
    year = "2025"
}

@ARTICLE{2019MNRAS.484.3279P,
       author = {{Panamarev}, Taras and {Just}, Andreas and {Spurzem}, Rainer and {Berczik}, Peter and {Wang}, Long and {Arca Sedda}, Manuel},
        title = "{Direct N-body simulation of the Galactic centre}",
      journal = {\mnras},
         year = 2019,
        month = apr,
       volume = {484},
       number = {3},
        pages = {3279-3290},
          doi = {10.1093/mnras/stz208},
archivePrefix = {arXiv},
       eprint = {1805.02153},
 primaryClass = {astro-ph.GA},
       adsurl = {https://ui.adsabs.harvard.edu/abs/2019MNRAS.484.3279P}
}

@ARTICLE{1995AJ....110.2622L,
       author = {{Lauer}, T.~R. and {Ajhar}, E.~A. and {Byun}, Y. -I. and {Dressler}, A. and {Faber}, S.~M. and {Grillmair}, C. and {Kormendy}, J. and {Richstone}, D. and {Tremaine}, S.},
        title = "{The Centers of Early-Type Galaxies with HST.I.An Observational Survey}",
      journal = {\aj},
         year = 1995,
        month = dec,
       volume = {110},
        pages = {2622},
          doi = {10.1086/117719},
       adsurl = {https://ui.adsabs.harvard.edu/abs/1995AJ....110.2622L}
}

@ARTICLE{2025arXiv250316600D,
       author = {{de Graaff}, Anna and {Rix}, Hans-Walter and {Naidu}, Rohan P. and {Labbe}, Ivo and {Wang}, Bingjie and {Leja}, Joel and {Matthee}, Jorryt and {Katz}, Harley and {Greene}, Jenny E. and {Hviding}, Raphael E. and {Baggen}, Josephine and {Bezanson}, Rachel and {Boogaard}, Leindert A. and {Brammer}, Gabriel and {Dayal}, Pratika and {van Dokkum}, Pieter and {Goulding}, Andy D. and {Hirschmann}, Michaela and {Maseda}, Michael V. and {McConachie}, Ian and {Miller}, Tim B. and {Nelson}, Erica and {Oesch}, Pascal A. and {Setton}, David J. and {Shivaei}, Irene and {Weibel}, Andrea and {Whitaker}, Katherine E. and {Williams}, Christina C.},
        title = "{A remarkable Ruby: Absorption in dense gas, rather than evolved stars, drives the extreme Balmer break of a Little Red Dot at $z=3.5$}",
      journal = {arXiv e-prints},
         year = 2025,
        month = mar,
          eid = {arXiv:2503.16600},
        pages = {arXiv:2503.16600},
          doi = {10.48550/arXiv.2503.16600},
archivePrefix = {arXiv},
       eprint = {2503.16600},
 primaryClass = {astro-ph.GA},
       adsurl = {https://ui.adsabs.harvard.edu/abs/2025arXiv250316600D}
}

@ARTICLE{2025ApJ...980L..27I,
       author = {{Inayoshi}, Kohei and {Maiolino}, Roberto},
        title = "{Extremely Dense Gas around Little Red Dots and High-redshift Active Galactic Nuclei: A Nonstellar Origin of the Balmer Break and Absorption Features}",
      journal = {\apjl},
         year = 2025,
        month = feb,
       volume = {980},
       number = {2},
          eid = {L27},
        pages = {L27},
          doi = {10.3847/2041-8213/adaebd},
archivePrefix = {arXiv},
       eprint = {2409.07805},
 primaryClass = {astro-ph.GA},
       adsurl = {https://ui.adsabs.harvard.edu/abs/2025ApJ...980L..27I}
}

@ARTICLE{2025arXiv250113082J,
       author = {{Ji}, Xihan and {Maiolino}, Roberto and {{\"U}bler}, Hannah and {Scholtz}, Jan and {D'Eugenio}, Francesco and {Sun}, Fengwu and {Perna}, Michele and {Turner}, Hannah and {Arribas}, Santiago and {Bennett}, Jake S. and {Bunker}, Andrew and {Carniani}, Stefano and {Charlot}, St{\'e}phane and {Cresci}, Giovanni and {Curti}, Mirko and {Egami}, Eiichi and {Fabian}, Andy and {Inayoshi}, Kohei and {Isobe}, Yuki and {Jones}, Gareth and {Juod{\v{z}}balis}, Ignas and {Kumari}, Nimisha and {Lyu}, Jianwei and {Mazzolari}, Giovanni and {Parlanti}, Eleonora and {Robertson}, Brant and {Rodr{\'\i}guez Del Pino}, Bruno and {Schneider}, Raffaella and {Sijacki}, Debora and {Tacchella}, Sandro and {Trinca}, Alessandro and {Valiante}, Rosa and {Venturi}, Giacomo and {Volonteri}, Marta and {Willott}, Chris and {Witten}, Callum and {Witstok}, Joris},
        title = "{BlackTHUNDER -- A non-stellar Balmer break in a black hole-dominated little red dot at $z=7.04$}",
      journal = {arXiv e-prints},
         year = 2025,
        month = jan,
          eid = {arXiv:2501.13082},
        pages = {arXiv:2501.13082},
          doi = {10.48550/arXiv.2501.13082},
archivePrefix = {arXiv},
       eprint = {2501.13082},
 primaryClass = {astro-ph.GA},
       adsurl = {https://ui.adsabs.harvard.edu/abs/2025arXiv250113082J}
}

@ARTICLE{2025arXiv250321879R,
       author = {{Rantala}, Antti and {Naab}, Thorsten},
        title = "{A rapid channel for the collisional formation and gravitational wave driven mergers of supermassive black hole seeds at high redshift}",
      journal = {arXiv e-prints},
         year = 2025,
        month = mar,
          eid = {arXiv:2503.21879},
        pages = {arXiv:2503.21879},
archivePrefix = {arXiv},
       eprint = {2503.21879},
 primaryClass = {astro-ph.GA},
       adsurl = {https://ui.adsabs.harvard.edu/abs/2025arXiv250321879R}
}

@article{Inayoshi:2019fun,
    author = "Inayoshi, Kohei and Visbal, Eli and Haiman, Zolt\'an",
    title = "{The Assembly of the First Massive Black Holes}",
    eprint = "1911.05791",
    archivePrefix = "arXiv",
    primaryClass = "astro-ph.GA",
    doi = "10.1146/annurev-astro-120419-014455",
    journal = "Ann. Rev. Astron. Astrophys.",
    volume = "58",
    pages = "27--97",
    year = "2020"
}

@article{Kritos:2024upo,
    author = "Kritos, Konstantinos and Berti, Emanuele and Silk, Joseph",
    title = "{Supermassive black holes from runaway mergers and accretion in nuclear star clusters}",
    eprint = "2404.11676",
    archivePrefix = "arXiv",
    primaryClass = "astro-ph.HE",
    doi = "10.1093/mnras/stae1145",
    journal = "Mon. Not. Roy. Astron. Soc.",
    volume = "531",
    number = "1",
    pages = "133--136",
    year = "2024"
}

@article{Madau:2024fdv,
    author = "Madau, Piero and Haardt, Francesco",
    title = "{X-Ray Weak Active Galactic Nuclei from Super-Eddington Accretion onto Infant Black Holes}",
    eprint = "2410.00417",
    archivePrefix = "arXiv",
    primaryClass = "astro-ph.GA",
    doi = "10.3847/2041-8213/ad90e1",
    journal = "Astrophys. J. Lett.",
    volume = "976",
    number = "2",
    pages = "L24",
    year = "2024"
}

@article{Pacucci:2024tws,
    author = "Pacucci, Fabio and Narayan, Ramesh",
    title = "{Mildly Super-Eddington Accretion onto Slowly Spinning Black Holes Explains the X-Ray Weakness of the Little Red Dots}",
    eprint = "2407.15915",
    archivePrefix = "arXiv",
    primaryClass = "astro-ph.HE",
    doi = "10.3847/1538-4357/ad84f7",
    journal = "Astrophys. J.",
    volume = "976",
    number = "1",
    pages = "96",
    year = "2024"
}

@INPROCEEDINGS{2023ASPC..534...83H,
       author = {{Henshaw}, J.~D. and {Barnes}, A.~T. and {Battersby}, C. and {Ginsburg}, A. and {Sormani}, M.~C. and {Walker}, D.~L.},
        title = "{Star Formation in the Central Molecular Zone of the Milky Way}",
    booktitle = {Protostars and Planets VII},
         year = 2023,
       editor = {{Inutsuka}, S. and {Aikawa}, Y. and {Muto}, T. and {Tomida}, K. and {Tamura}, M.},
       series = {Astronomical Society of the Pacific Conference Series},
       volume = {534},
        month = jul,
        pages = {83},
          doi = {10.48550/arXiv.2203.11223},
archivePrefix = {arXiv},
       eprint = {2203.11223},
 primaryClass = {astro-ph.GA},
       adsurl = {https://ui.adsabs.harvard.edu/abs/2023ASPC..534...83H}
}

@article{Lei:2008ui,
    author = "Lei, Q. and Abramowicz, M. A. and Fragile, P. C. and Horak, J. and Machida, M. and Straub, O.",
    title = "{The Polish doughnuts revisited I. The angular momentum distribution and equipressure surfaces}",
    eprint = "0812.2467",
    archivePrefix = "arXiv",
    primaryClass = "astro-ph",
    doi = "10.1051/0004-6361/200811518",
    journal = "Astron. Astrophys.",
    volume = "498",
    pages = "471--477",
    year = "2009"
}

@ARTICLE{2024A&A...688L..27V,
       author = {{Veledina}, Alexandra and {Poutanen}, Juri and {Bocharova}, Anastasiia and {Di Marco}, Alessandro and {Forsblom}, Sofia V. and {La Monaca}, Fabio and {Podgorn{\'y}}, Jakub and {Tsygankov}, Sergey S. and {Zdziarski}, Andrzej A. and {Ahlberg}, Varpu and {Green}, David A. and {Muleri}, Fabio and {Rhodes}, Lauren and {Bianchi}, Stefano and {Costa}, Enrico and {Dov{\v{c}}iak}, Michal and {Loktev}, Vladislav and {McCollough}, Michael and {Soffitta}, Paolo and {Sunyaev}, Rashid},
        title = "{Ultrasoft state of microquasar Cygnus X-3: X-ray polarimetry reveals the geometry of the astronomical puzzle}",
      journal = {\aap},
         year = 2024,
        month = aug,
       volume = {688},
          eid = {L27},
        pages = {L27},
          doi = {10.1051/0004-6361/202451356},
archivePrefix = {arXiv},
       eprint = {2407.02655},
 primaryClass = {astro-ph.HE},
       adsurl = {https://ui.adsabs.harvard.edu/abs/2024A&A...688L..27V}
}

@ARTICLE{2025arXiv250920455J,
       author = {{Jones}, Gareth C. and {{\"U}bler}, Hannah and {Maiolino}, Roberto and {Ji}, Xihan and {Marconi}, Alessandro and {D'Eugenio}, Francesco and {Arribas}, Santiago and {Bunker}, Andrew J. and {Carniani}, Stefano and {Charlot}, St{\'e}phane and {Cresci}, Giovanni and {Inayoshi}, Kohei and {Isobe}, Yuki and {Juod{\v{z}}balis}, Ignas and {Mazzolari}, Giovanni and {P{\'e}rez-Gonz{\'a}lez}, Pablo G. and {Perna}, Michele and {Schneider}, Raffaella and {Scholtz}, Jan and {Tacchella}, Sandro},
        title = "{BlackTHUNDER: Shedding light on a dormant and extreme little red dot at z=8.50}",
      journal = {arXiv e-prints},
         year = 2025,
        month = sep,
          eid = {arXiv:2509.20455},
        pages = {arXiv:2509.20455},
          doi = {10.48550/arXiv.2509.20455},
archivePrefix = {arXiv},
       eprint = {2509.20455},
 primaryClass = {astro-ph.GA},
       adsurl = {https://ui.adsabs.harvard.edu/abs/2025arXiv250920455J}
}

@article{Escala:2025jlt,
    author = "Escala, Andres and Zimmermann, Lucas and Valdebenito, Sebastian and Vergara, Marcelo C. and Schleicher, Dominik R. G. and Liempi, Matias",
    title = "{On the Fate of Little Red Dots}",
    eprint = "2509.20453",
    archivePrefix = "arXiv",
    primaryClass = "astro-ph.GA",
    month = "9",
    year = "2025"
}

@article{Kremer:2019zql,
    author = "Kremer, Kyle and Lu, Wenbin and Rodriguez, Carl L. and Lachat, Mitchell and Rasio, Frederic",
    title = "{Tidal Disruptions of Stars by Black Hole Remnants in Dense Star Clusters}",
    eprint = "1904.06353",
    archivePrefix = "arXiv",
    primaryClass = "astro-ph.HE",
    doi = "10.3847/1538-4357/ab2e0c",
    month = "4",
    year = "2019"
}

@article{Speri:2022upm,
    author = "Speri, Lorenzo and Antonelli, Andrea and Sberna, Laura and Babak, Stanislav and Barausse, Enrico and Gair, Jonathan R. and Katz, Michael L.",
    title = "{Probing Accretion Physics with Gravitational Waves}",
    eprint = "2207.10086",
    archivePrefix = "arXiv",
    primaryClass = "gr-qc",
    doi = "10.1103/PhysRevX.13.021035",
    journal = "Phys. Rev. X",
    volume = "13",
    number = "2",
    pages = "021035",
    year = "2023"
}

@ARTICLE{2025arXiv250919422I,
       author = {{Inayoshi}, Kohei and {Murase}, Kohta and {Kashiyama}, Kazumi},
        title = "{Spectral Uniformity of Little Red Dots: A Natural Outcome of Coevolving Seed Black Holes and Nascent Starbursts}",
      journal = {arXiv e-prints},
         year = 2025,
        month = sep,
          eid = {arXiv:2509.19422},
        pages = {arXiv:2509.19422},
archivePrefix = {arXiv},
       eprint = {2509.19422},
 primaryClass = {astro-ph.GA},
       adsurl = {https://ui.adsabs.harvard.edu/abs/2025arXiv250919422I}
}

@article{Maiolino:2023zdu,
    author = "Maiolino, Roberto and others",
    title = "{A small and vigorous black hole in the early Universe}",
    eprint = "2305.12492",
    archivePrefix = "arXiv",
    primaryClass = "astro-ph.GA",
    doi = "10.1038/s41586-024-07494-x",
    journal = "Nature",
    volume = "627",
    number = "8002",
    pages = "59--63",
    year = "2024",
    note = "[Erratum: Nature 630, E2 (2024)]"
}

@article{Perets:2016pwr,
    author = "Perets, Hagai B. and Li, Zhuo and Lombardi, James C. and Milcarek, Stephen R.",
    title = "{Micro - tidal disruption events by stellar compact objects and the production of ultra-long GRBs}",
    eprint = "1602.07698",
    archivePrefix = "arXiv",
    primaryClass = "astro-ph.HE",
    doi = "10.3847/0004-637X/823/2/113",
    journal = "Astrophys. J.",
    volume = "823",
    number = "2",
    pages = "113",
    year = "2016"
}

@article{Rodriguez:2016kxx,
    author = "Rodriguez, Carl L. and Chatterjee, Sourav and Rasio, Frederic A.",
    title = "{Binary Black Hole Mergers from Globular Clusters: Masses, Merger Rates, and the Impact of Stellar Evolution}",
    eprint = "1602.02444",
    archivePrefix = "arXiv",
    primaryClass = "astro-ph.HE",
    doi = "10.1103/PhysRevD.93.084029",
    journal = "Phys. Rev. D",
    volume = "93",
    number = "8",
    pages = "084029",
    year = "2016"
}

@article{Babak:2017tow,
    author = "Babak, Stanislav and Gair, Jonathan and Sesana, Alberto and Barausse, Enrico and Sopuerta, Carlos F. and Berry, Christopher P. L. and Berti, Emanuele and Amaro-Seoane, Pau and Petiteau, Antoine and Klein, Antoine",
    title = "{Science with the space-based interferometer LISA. V: Extreme mass-ratio inspirals}",
    eprint = "1703.09722",
    archivePrefix = "arXiv",
    primaryClass = "gr-qc",
    doi = "10.1103/PhysRevD.95.103012",
    journal = "Phys. Rev. D",
    volume = "95",
    number = "10",
    pages = "103012",
    year = "2017"
}

@article{Bardeen:1970zz,
    author = "Bardeen, James M.",
    title = "{Kerr Metric Black Holes}",
    doi = "10.1038/226064a0",
    journal = "Nature",
    volume = "226",
    pages = "64--65",
    year = "1970"
}

@article{Reynolds:2020jwt,
    author = "Reynolds, Christopher S.",
    title = "{Observational Constraints on Black Hole Spin}",
    eprint = "2011.08948",
    archivePrefix = "arXiv",
    primaryClass = "astro-ph.HE",
    doi = "10.1146/annurev-astro-112420-035022",
    journal = "Ann. Rev. Astron. Astrophys.",
    volume = "59",
    pages = "117--154",
    year = "2021"
}

@article{Choksi:2018jnq,
    author = "Choksi, Nick and Volonteri, Marta and Colpi, Monica and Gnedin, Oleg Y. and Li, Hui",
    title = "{The star clusters that make black hole binaries across cosmic time}",
    eprint = "1809.01164",
    archivePrefix = "arXiv",
    primaryClass = "astro-ph.GA",
    doi = "10.3847/1538-4357/aaffde",
    journal = "Astrophys. J.",
    volume = "873",
    number = "1",
    pages = "100",
    year = "2019"
}

@article{Silk:2024rsf,
    author = "Silk, Joseph and Begelman, Mitchell C. and Norman, Colin and Nusser, Adi and Wyse, Rosemary F. G.",
    title = "{Which Came First: Supermassive Black Holes or Galaxies? Insights from JWST}",
    eprint = "2401.02482",
    archivePrefix = "arXiv",
    primaryClass = "astro-ph.GA",
    doi = "10.3847/2041-8213/ad1bf0",
    journal = "Astrophys. J. Lett.",
    volume = "961",
    number = "2",
    pages = "L39",
    year = "2024"
}

@article{Mummery:2023meb,
    author = "Mummery, Andrew",
    title = "{The maximum mass of a black hole which can tidally disrupt a star: measuring black hole spins with tidal disruption events}",
    eprint = "2312.00557",
    archivePrefix = "arXiv",
    primaryClass = "gr-qc",
    doi = "10.1093/mnras/stad3636",
    journal = "Mon. Not. Roy. Astron. Soc.",
    volume = "527",
    number = "3",
    pages = "6233--6252",
    year = "2023"
}

@ARTICLE{2020ARA&A..58..257G,
       author = {{Greene}, Jenny E. and {Strader}, Jay and {Ho}, Luis C.},
        title = "{Intermediate-Mass Black Holes}",
      journal = {\araa},
         year = 2020,
        month = aug,
       volume = {58},
        pages = {257-312},
          doi = {10.1146/annurev-astro-032620-021835},
archivePrefix = {arXiv},
       eprint = {1911.09678},
 primaryClass = {astro-ph.GA},
       adsurl = {https://ui.adsabs.harvard.edu/abs/2020ARA&A..58..257G}
}

@article{Rizzuto:2022fdp,
    author = "Rizzuto, Francesco Paolo and Naab, Thorsten and Rantala, Antti and Johansson, Peter H. and Ostriker, Jeremiah P. and Stone, Nicholas C. and Liao, Shihong and Irodotou, Dimitrios",
    title = "{The growth of intermediate mass black holes through tidal captures and tidal disruption events}",
    eprint = "2211.13320",
    archivePrefix = "arXiv",
    primaryClass = "astro-ph.GA",
    doi = "10.1093/mnras/stad734",
    journal = "Mon. Not. Roy. Astron. Soc.",
    volume = "521",
    number = "2",
    pages = "2930--2948",
    year = "2023"
}

@article{Stone:2016ryd,
    author = {Stone, Nicholas C. and K{\"u}pper, Andreas H. W. and Ostriker, Jeremiah P.},
    title = "{Formation of Massive Black Holes in Galactic Nuclei: Runaway Tidal Encounters}",
    eprint = "1606.01909",
    archivePrefix = "arXiv",
    primaryClass = "astro-ph.GA",
    doi = "10.1093/mnras/stx097",
    journal = "Mon. Not. Roy. Astron. Soc.",
    volume = "467",
    number = "4",
    pages = "4180--4199",
    year = "2017"
}

@ARTICLE{2024MNRAS.529.4104V,
       author = {{van Donkelaar}, Floor and {Mayer}, Lucio and {Capelo}, Pedro R. and {Tamfal}, Tomas and {Quinn}, Thomas R. and {Madau}, Piero},
        title = "{Stellar cluster formation in a Milky Way-sized galaxy at z > 4 - II. A hybrid formation scenario for the nuclear star cluster and its connection to the nuclear stellar ring}",
      journal = {\mnras},
         year = 2024,
        month = apr,
       volume = {529},
       number = {4},
        pages = {4104-4116},
          doi = {10.1093/mnras/stae804},
archivePrefix = {arXiv},
       eprint = {2303.12828},
 primaryClass = {astro-ph.GA},
       adsurl = {https://ui.adsabs.harvard.edu/abs/2024MNRAS.529.4104V}
}

@ARTICLE{2024Natur.636..332M,
       author = {{Mowla}, Lamiya and {Iyer}, Kartheik and {Asada}, Yoshihisa and {Desprez}, Guillaume and {Tan}, Vivian Yun Yan and {Martis}, Nicholas and {Sarrouh}, Ghassan and {Strait}, Victoria and {Abraham}, Roberto and {Brada{\v{c}}}, Maru{\v{s}}a and {Brammer}, Gabriel and {Muzzin}, Adam and {Pacifici}, Camilla and {Ravindranath}, Swara and {Sawicki}, Marcin and {Willott}, Chris and {Estrada-Carpenter}, Vince and {Jahan}, Nusrath and {Noirot}, Ga{\"e}l and {Matharu}, Jasleen and {Rihtar{\v{s}}i{\v{c}}}, Gregor and {Zabl}, Johannes},
        title = "{Formation of a low-mass galaxy from star clusters in a 600-million-year-old Universe}",
      journal = {\nat},
         year = 2024,
        month = dec,
       volume = {636},
       number = {8042},
        pages = {332-336},
          doi = {10.1038/s41586-024-08293-0},
archivePrefix = {arXiv},
       eprint = {2402.08696},
 primaryClass = {astro-ph.GA},
       adsurl = {https://ui.adsabs.harvard.edu/abs/2024Natur.636..332M}
}

@ARTICLE{2024Natur.636..594J,
       author = {{Juod{\v{z}}balis}, Ignas and {Maiolino}, Roberto and {Baker}, William M. and {Tacchella}, Sandro and {Scholtz}, Jan and {D'Eugenio}, Francesco and {Witstok}, Joris and {Schneider}, Raffaella and {Trinca}, Alessandro and {Valiante}, Rosa and {DeCoursey}, Christa and {Curti}, Mirko and {Carniani}, Stefano and {Chevallard}, Jacopo and {de Graaff}, Anna and {Arribas}, Santiago and {Bennett}, Jake S. and {Bourne}, Martin A. and {Bunker}, Andrew J. and {Charlot}, St{\'e}phane and {Jiang}, Brian and {Koudmani}, Sophie and {Perna}, Michele and {Robertson}, Brant and {Sijacki}, Debora and {{\"U}bler}, Hannah and {Williams}, Christina C. and {Willott}, Chris},
        title = "{A dormant overmassive black hole in the early Universe}",
      journal = {\nat},
         year = 2024,
        month = dec,
       volume = {636},
       number = {8043},
        pages = {594-597},
          doi = {10.1038/s41586-024-08210-5},
archivePrefix = {arXiv},
       eprint = {2403.03872},
 primaryClass = {astro-ph.GA},
       adsurl = {https://ui.adsabs.harvard.edu/abs/2024Natur.636..594J}
}

@article{Kovacs:2024zfh,
    author = "Kovacs, Orsolya E. and others",
    title = "{A Candidate Supermassive Black Hole in a Gravitationally Lensed Galaxy at Z {\ensuremath{\approx}} 10}",
    eprint = "2403.14745",
    archivePrefix = "arXiv",
    primaryClass = "astro-ph.GA",
    doi = "10.3847/2041-8213/ad391f",
    journal = "Astrophys. J. Lett.",
    volume = "965",
    number = "2",
    pages = "L21",
    year = "2024"
}

@article{Lambrides:2024ugh,
    author = "Lambrides, Erini and others",
    title = "{The Case for Super-Eddington Accretion: Connecting Weak X-ray and UV Line Emission in JWST Broad-Line AGN During the First Gyr of Cosmic Time}",
    eprint = "2409.13047",
    archivePrefix = "arXiv",
    primaryClass = "astro-ph.HE",
    month = "9",
    year = "2024"
}

@article{Krause:2025gkq,
    author = "Krause, Martin G. H. and others",
    title = "{Evidence for Supermassive Black Hole Binaries}",
    eprint = "2510.07534",
    archivePrefix = "arXiv",
    primaryClass = "astro-ph.HE",
    month = "10",
    year = "2025"
}

@ARTICLE{2025A&A...700A.135T,
       author = {{Toubiana}, A. and {Sberna}, L. and {Volonteri}, M. and {Barausse}, E. and {Babak}, S. and {Enficiaud}, R. and {Izquierdo{\textendash}Villalba}, D. and {Gair}, J.~R. and {Greene}, J.~E. and {Quelquejay Leclere}, H.},
        title = "{Reconciling PTA and JWST, and preparing for LISA with POMPOCO: a Parametrisation Of the Massive black hole POpulation for Comparison to Observations}",
      journal = {\aap},
         year = 2025,
        month = aug,
       volume = {700},
          eid = {A135},
        pages = {A135},
          doi = {10.1051/0004-6361/202453027},
archivePrefix = {arXiv},
       eprint = {2410.17916},
 primaryClass = {astro-ph.GA},
       adsurl = {https://ui.adsabs.harvard.edu/abs/2025A&A...700A.135T}
}

@ARTICLE{2024A&A...689A.204P,
       author = {{Polkas}, M. and {Bonoli}, S. and {Bortolas}, E. and {Izquierdo-Villalba}, D. and {Sesana}, A. and {Broggi}, L. and {Hoyer}, N. and {Spinoso}, D.},
        title = "{Demographics of tidal disruption events with L-Galaxies: I. Volumetric TDE rates and the abundance of nuclear star clusters}",
      journal = {\aap},
         year = 2024,
        month = sep,
       volume = {689},
          eid = {A204},
        pages = {A204},
          doi = {10.1051/0004-6361/202449470},
archivePrefix = {arXiv},
       eprint = {2312.13242},
 primaryClass = {astro-ph.HE},
       adsurl = {https://ui.adsabs.harvard.edu/abs/2024A&A...689A.204P}
}

@ARTICLE{2021MNRAS.500.3944P,
       author = {{Pfister}, Hugo and {Dai}, Jane Lixin and {Volonteri}, Marta and {Auchettl}, Katie and {Trebitsch}, Maxime and {Ramirez-Ruiz}, Enrico},
        title = "{Tidal disruption events in the first billion years of a galaxy}",
      journal = {\mnras},
         year = 2021,
        month = jan,
       volume = {500},
       number = {3},
        pages = {3944-3956},
          doi = {10.1093/mnras/staa3471},
archivePrefix = {arXiv},
       eprint = {2006.06565},
 primaryClass = {astro-ph.GA},
       adsurl = {https://ui.adsabs.harvard.edu/abs/2021MNRAS.500.3944P}
}

@article{Kritos:2024kpn,
    author = "Kritos, Konstantinos and Reali, Luca and Gerosa, Davide and Berti, Emanuele",
    title = "{Minimum gas mass accreted by spinning intermediate-mass black holes in stellar clusters}",
    eprint = "2409.15439",
    archivePrefix = "arXiv",
    primaryClass = "astro-ph.HE",
    doi = "10.1103/PhysRevD.110.123017",
    journal = "Phys. Rev. D",
    volume = "110",
    number = "12",
    pages = "123017",
    year = "2024"
}

@article{Kritos:2025bby,
    author = "Kritos, Konstantinos and Reali, Luca and Ng, Ken K. Y. and Antonini, Fabio and Berti, Emanuele",
    title = "{Gravitational wave inference of star cluster properties from intermediate-mass black hole mergers}",
    eprint = "2501.16422",
    archivePrefix = "arXiv",
    primaryClass = "astro-ph.HE",
    doi = "10.1103/PhysRevD.111.063056",
    journal = "Phys. Rev. D",
    volume = "111",
    number = "6",
    pages = "063056",
    year = "2025"
}

@article{Comerford:2025tjl,
    author = "Comerford, Julia M. and Simon, Joseph",
    title = "{Preferential Accretion onto the Secondary Black Hole Strengthens Gravitational Wave Signals}",
    eprint = "2510.06325",
    archivePrefix = "arXiv",
    primaryClass = "astro-ph.GA",
    month = "10",
    year = "2025"
}

@ARTICLE{2025arXiv250109854M,
       author = {{Madau}, Piero},
        title = "{Chasing the Light: Shadowing, Collimation, and the Super-Eddington Growth of Infant Black Holes in JWST-Discovered AGNs}",
      journal = {arXiv e-prints},
         year = 2025,
        month = jan,
          eid = {arXiv:2501.09854},
        pages = {arXiv:2501.09854},
          doi = {10.48550/arXiv.2501.09854},
archivePrefix = {arXiv},
       eprint = {2501.09854},
 primaryClass = {astro-ph.HE},
       adsurl = {https://ui.adsabs.harvard.edu/abs/2025arXiv250109854M}
}

@article{Inayoshi:2025isg,
    author = "Inayoshi, Kohei",
    title = "{Little Red Dots as the Very First Activity of Black Hole Growth}",
    doi = "10.3847/2041-8213/adea66",
    journal = "Astrophys. J. Lett.",
    volume = "988",
    number = "1",
    pages = "L22",
    year = "2025"
}

@article{Zhang:2025jmm,
    author = "Zhang, Fupeng and Seoane, Pau Amaro",
    title = "{Co-evolution of Nuclear Star Clusters and Massive Black Holes: Extreme Mass-Ratio Inspirals}",
    eprint = "2510.10821",
    archivePrefix = "arXiv",
    primaryClass = "astro-ph.GA",
    month = "10",
    year = "2025"
}

@ARTICLE{2025arXiv250818358W,
       author = {{Wang}, Bingjie and {Leja}, Joel and {Katz}, Harley and {Inayoshi}, Kohei and {Cleri}, Nikko J. and {de Graaff}, Anna and {Hviding}, Raphael E. and {van Dokkum}, Pieter and {Greene}, Jenny E. and {Labb{\'e}}, Ivo and {Matthee}, Jorryt and {McConachie}, Ian and {Naidu}, Rohan P. and {Nelson}, Erica J.},
        title = "{The Missing Hard Photons of Little Red Dots: Their Incident Ionizing Spectra Resemble Massive Stars}",
      journal = {arXiv e-prints},
         year = 2025,
        month = aug,
          eid = {arXiv:2508.18358},
        pages = {arXiv:2508.18358},
          doi = {10.48550/arXiv.2508.18358},
archivePrefix = {arXiv},
       eprint = {2508.18358},
 primaryClass = {astro-ph.GA},
       adsurl = {https://ui.adsabs.harvard.edu/abs/2025arXiv250818358W}
}

@article{Ricarte:2018mzn,
    author = "Ricarte, Angelo and Natarajan, Priyamvada",
    title = "{The Observational Signatures of Supermassive Black Hole Seeds}",
    eprint = "1809.01177",
    archivePrefix = "arXiv",
    primaryClass = "astro-ph.GA",
    doi = "10.1093/mnras/sty2448",
    journal = "Mon. Not. Roy. Astron. Soc.",
    volume = "481",
    number = "3",
    pages = "3278--3292",
    year = "2018"
}

@article{Kritos:2022non,
    author = "Kritos, Konstantinos and Berti, Emanuele and Silk, Joseph",
    title = "{Massive black hole assembly in nuclear star clusters}",
    eprint = "2212.06845",
    archivePrefix = "arXiv",
    primaryClass = "astro-ph.HE",
    doi = "10.1103/PhysRevD.108.083012",
    journal = "Phys. Rev. D",
    volume = "108",
    number = "8",
    pages = "083012",
    year = "2023"
}

@article{Kritos:2024sgd,
    author = "Kritos, Konstantinos and Beckmann, Ricarda S. and Silk, Joseph and Berti, Emanuele and Yi, Sophia and Volonteri, Marta and Dubois, Yohan and Devriendt, Julien",
    title = "{Supermassive black hole growth in hierarchically merging nuclear star clusters}",
    eprint = "2412.15334",
    archivePrefix = "arXiv",
    primaryClass = "astro-ph.GA",
    month = "12",
    year = "2024"
}

@article{Matthee:2023utn,
    author = "Matthee, Jorryt and others",
    title = "{Little Red Dots: An Abundant Population of Faint Active Galactic Nuclei at z \ensuremath{\sim} 5 Revealed by the EIGER and FRESCO JWST Surveys}",
    eprint = "2306.05448",
    archivePrefix = "arXiv",
    primaryClass = "astro-ph.GA",
    doi = "10.3847/1538-4357/ad2345",
    journal = "Astrophys. J.",
    volume = "963",
    number = "2",
    pages = "129",
    year = "2024"
}

@article{Greene:2024phl,
    author = "Greene, Jenny E. and others",
    title = "{UNCOVER Spectroscopy Confirms the Surprising Ubiquity of Active Galactic Nuclei in Red Sources at z \ensuremath{>} 5}",
    doi = "10.3847/1538-4357/ad1e5f",
    journal = "Astrophys. J.",
    volume = "964",
    number = "1",
    pages = "39",
    year = "2024"
}

@article{Ananna:2024jug,
    author = "Ananna, Tonima Tasnim and Bogd\'an, \'Akos and Kov\'acs, Orsolya E. and Natarajan, Priyamvada and Hickox, Ryan C.",
    title = "{X-Ray View of Little Red Dots: Do They Host Supermassive Black Holes?}",
    eprint = "2404.19010",
    archivePrefix = "arXiv",
    primaryClass = "astro-ph.GA",
    doi = "10.3847/2041-8213/ad5669",
    journal = "Astrophys. J. Lett.",
    volume = "969",
    number = "1",
    pages = "L18",
    year = "2024"
}

@article{Maiolino:2023bpi,
    author = "Maiolino, Roberto and others",
    title = "{JADES - The diverse population of infant black holes at 4 \ensuremath{<} z \ensuremath{<} 11: Merging, tiny, poor, but mighty}",
    eprint = "2308.01230",
    archivePrefix = "arXiv",
    primaryClass = "astro-ph.GA",
    doi = "10.1051/0004-6361/202347640",
    journal = "Astron. Astrophys.",
    volume = "691",
    pages = "A145",
    year = "2024"
}

@article{Mouri:2002mc,
    author = "Mouri, Hideaki and Taniguchi, Yoshiaki",
    title = "{Runaway merging of black holes: analytical constraint on the timescale}",
    eprint = "astro-ph/0201102",
    archivePrefix = "arXiv",
    doi = "10.1086/339472",
    journal = "Astrophys. J. Lett.",
    volume = "566",
    pages = "L17--L20",
    year = "2002"
}

@book{Merritt:2013book,
    author = {{Merritt}, D.},
    title = "{Dynamics and evolution of galactic nuclei}",
    publisher = {Princeton University Press},
    address = {Princeton, NJ},
    year = 2013
}

@ARTICLE{1988Natur.333..523R,
       author = {{Rees}, Martin J.},
        title = "{Tidal disruption of stars by black holes of {}10$^{6}$-{}10$^{8}$ solar masses in nearby galaxies}",
      journal = {\nat},
         year = 1988,
        month = jun,
       volume = {333},
       number = {6173},
        pages = {523-528},
          doi = {10.1038/333523a0},
       adsurl = {https://ui.adsabs.harvard.edu/abs/1988Natur.333..523R}
}

@article{Perets:2006bz,
    author = "Perets, Hagai B. and Hopman, Clovis and Alexander, Tal",
    title = "{Massive perturber-driven interactions of stars with a massive black hole}",
    eprint = "astro-ph/0606443",
    archivePrefix = "arXiv",
    doi = "10.1086/510377",
    journal = "Astrophys. J.",
    volume = "656",
    pages = "709--720",
    year = "2007"
}

@article{Bonetti:2020jku,
    author = "Bonetti, Matteo and Sesana, Alberto",
    title = "{Gravitational wave background from extreme mass ratio inspirals}",
    eprint = "2007.14403",
    archivePrefix = "arXiv",
    primaryClass = "astro-ph.GA",
    doi = "10.1103/PhysRevD.102.103023",
    journal = "Phys. Rev. D",
    volume = "102",
    number = "10",
    pages = "103023",
    year = "2020"
}

@article{Poutanen:2006uc,
    author = "Poutanen, Juri and Fabrika, Sergei and Butkevich, Alexey G. and Abolmasov, Pavel",
    title = "{Supercritically accreting stellar mass black holes as ultraluminous X-ray sources}",
    eprint = "astro-ph/0609274",
    archivePrefix = "arXiv",
    reportNumber = "SLAC-PUB-12118",
    doi = "10.1111/j.1365-2966.2007.11668.x",
    journal = "Mon. Not. Roy. Astron. Soc.",
    volume = "377",
    pages = "1187--1194",
    year = "2007"
}

@article{Begelman:2025upi,
    author = "Begelman, Mitchell C. and Dexter, Jason",
    title = "{Little Red Dots As Late-stage Quasi-stars}",
    eprint = "2507.09085",
    archivePrefix = "arXiv",
    primaryClass = "astro-ph.GA",
    month = "7",
    year = "2025"
}

@article{Maiolino:2025tih,
    author = "Maiolino, Roberto and others",
    title = "{A black hole in a near-pristine galaxy 700 million years after the Big Bang}",
    eprint = "2505.22567",
    archivePrefix = "arXiv",
    primaryClass = "astro-ph.GA",
    month = "5",
    year = "2025"
}

@article{OLeary:2008myb,
    author = "O'Leary, Ryan M. and Kocsis, Bence and Loeb, Abraham",
    title = "{Gravitational waves from scattering of stellar-mass black holes in galactic nuclei}",
    eprint = "0807.2638",
    archivePrefix = "arXiv",
    primaryClass = "astro-ph",
    doi = "10.1111/j.1365-2966.2009.14653.x",
    journal = "Mon. Not. Roy. Astron. Soc.",
    volume = "395",
    number = "4",
    pages = "2127--2146",
    year = "2009"
}

@article{Gezari:2021bmb,
    author = "Gezari, Suvi",
    title = "{Tidal Disruption Events}",
    eprint = "2104.14580",
    archivePrefix = "arXiv",
    primaryClass = "astro-ph.HE",
    doi = "10.1146/annurev-astro-111720-030029",
    journal = "Ann. Rev. Astron. Astrophys.",
    volume = "59",
    pages = "21--58",
    year = "2021"
}

@ARTICLE{2009A&A...502...91S,
       author = {{Sch{\"o}del}, R. and {Merritt}, D. and {Eckart}, A.},
        title = "{The nuclear star cluster of the Milky Way: proper motions and mass}",
      journal = {\aap},
         year = 2009,
        month = jul,
       volume = {502},
       number = {1},
        pages = {91-111},
          doi = {10.1051/0004-6361/200810922},
archivePrefix = {arXiv},
       eprint = {0902.3892},
 primaryClass = {astro-ph.GA},
       adsurl = {https://ui.adsabs.harvard.edu/abs/2009A&A...502...91S}
}

@ARTICLE{2024MNRAS.531.3770R,
       author = {{Rantala}, Antti and {Naab}, Thorsten and {Lah{\'e}n}, Natalia},
        title = "{FROST-CLUSTERS - I. Hierarchical star cluster assembly boosts intermediate-mass black hole formation}",
      journal = {\mnras},
         year = 2024,
        month = jul,
       volume = {531},
       number = {3},
        pages = {3770-3799},
          doi = {10.1093/mnras/stae1413},
archivePrefix = {arXiv},
       eprint = {2403.10602},
 primaryClass = {astro-ph.GA},
       adsurl = {https://ui.adsabs.harvard.edu/abs/2024MNRAS.531.3770R}
}

@ARTICLE{2025MNRAS.543.1023L,
       author = {{Lah{\'e}n}, Natalia and {Naab}, Thorsten and {Rantala}, Antti and {Partmann}, Christian},
        title = "{Mergers all the way down: stellar collisions and kinematics of a dense hierarchically forming massive star cluster in a dwarf starburst}",
      journal = {\mnras},
         year = 2025,
        month = oct,
       volume = {543},
       number = {2},
        pages = {1023-1038},
          doi = {10.1093/mnras/staf1546},
archivePrefix = {arXiv},
       eprint = {2504.18620},
 primaryClass = {astro-ph.GA},
       adsurl = {https://ui.adsabs.harvard.edu/abs/2025MNRAS.543.1023L}
}

@ARTICLE{2025ApJ...990..135W,
       author = {{Williams}, Claire E. and {Naoz}, Smadar and {Lake}, William and {Burkhart}, Blakesley and {Marinacci}, Federico and {Vogelsberger}, Mark and {Yoshida}, Naoki and {Menon}, Shyam H. and {Chen}, Avi and {Adamo}, Angela},
        title = "{{\ensuremath{\Lambda}}CDM Star Clusters at Cosmic Dawn: Stellar Densities, Environment, and Equilibrium}",
      journal = {\apj},
         year = 2025,
        month = sep,
       volume = {990},
       number = {2},
          eid = {135},
        pages = {135},
          doi = {10.3847/1538-4357/adf19d},
archivePrefix = {arXiv},
       eprint = {2502.17561},
 primaryClass = {astro-ph.GA},
       adsurl = {https://ui.adsabs.harvard.edu/abs/2025ApJ...990..135W}
}

@ARTICLE{2024ApJ...965...98C,
       author = {{Casey}, Caitlin M. and {Akins}, Hollis B. and {Shuntov}, Marko and {Ilbert}, Olivier and {Paquereau}, Louise and {Franco}, Maximilien and {Hayward}, Christopher C. and {Finkelstein}, Steven L. and {Boylan-Kolchin}, Michael and {Robertson}, Brant E. and {Allen}, Natalie and {Brinch}, Malte and {Cooper}, Olivia R. and {Ding}, Xuheng and {Drakos}, Nicole E. and {Faisst}, Andreas L. and {Fujimoto}, Seiji and {Gillman}, Steven and {Harish}, Santosh and {Hirschmann}, Michaela and {Jin}, Shuowen and {Kartaltepe}, Jeyhan S. and {Koekemoer}, Anton M. and {Kokorev}, Vasily and {Liu}, Daizhong and {Long}, Arianna S. and {Magdis}, Georgios and {Maraston}, Claudia and {Martin}, Crystal L. and {McCracken}, Henry Joy and {McKinney}, Jed and {Mobasher}, Bahram and {Rhodes}, Jason and {Rich}, R. Michael and {Sanders}, David B. and {Silverman}, John D. and {Toft}, Sune and {Vijayan}, Aswin P. and {Weaver}, John R. and {Wilkins}, Stephen M. and {Yang}, Lilan and {Zavala}, Jorge A.},
        title = "{COSMOS-Web: Intrinsically Luminous z {\ensuremath{\gtrsim}} 10 Galaxy Candidates Test Early Stellar Mass Assembly}",
      journal = {\apj},
         year = 2024,
        month = apr,
       volume = {965},
       number = {1},
          eid = {98},
        pages = {98},
          doi = {10.3847/1538-4357/ad2075},
archivePrefix = {arXiv},
       eprint = {2308.10932},
 primaryClass = {astro-ph.GA},
       adsurl = {https://ui.adsabs.harvard.edu/abs/2024ApJ...965...98C}
}

@ARTICLE{2022ApJ...940..101L,
       author = {{Linial}, Itai and {Sari}, Re'em},
        title = "{Stellar Distributions around a Supermassive Black Hole: Strong-segregation Regime Revisited}",
      journal = {\apj},
         year = 2022,
        month = dec,
       volume = {940},
       number = {2},
          eid = {101},
        pages = {101},
          doi = {10.3847/1538-4357/ac9bfd},
archivePrefix = {arXiv},
       eprint = {2206.14817},
 primaryClass = {astro-ph.GA},
       adsurl = {https://ui.adsabs.harvard.edu/abs/2022ApJ...940..101L}
}

@ARTICLE{2025ApJ...980..150K,
       author = {{Kaur}, Karamveer and {Rom}, Barak and {Sari}, Re'em},
        title = "{Semianalytical Fokker-Planck Models for Nuclear Star Clusters}",
      journal = {\apj},
         year = 2025,
        month = feb,
       volume = {980},
       number = {1},
          eid = {150},
        pages = {150},
          doi = {10.3847/1538-4357/ada8a8},
archivePrefix = {arXiv},
       eprint = {2406.07627},
 primaryClass = {astro-ph.HE},
       adsurl = {https://ui.adsabs.harvard.edu/abs/2025ApJ...980..150K}
}

@ARTICLE{2025ApJ...991..146R,
       author = {{Rom}, Barak and {Sari}, Re'em},
        title = "{Mass Segregation and Transient Formation in Nuclear Stellar Clusters}",
      journal = {\apj},
         year = 2025,
        month = oct,
       volume = {991},
       number = {2},
          eid = {146},
        pages = {146},
          doi = {10.3847/1538-4357/adfb6c},
archivePrefix = {arXiv},
       eprint = {2502.13209},
 primaryClass = {astro-ph.HE},
       adsurl = {https://ui.adsabs.harvard.edu/abs/2025ApJ...991..146R}
}

@ARTICLE{2024ApJ...977....7R,
       author = {{Rom}, Barak and {Linial}, Itai and {Kaur}, Karamveer and {Sari}, Re'em},
        title = "{Dynamics Around Supermassive Black Holes: Extreme-mass-ratio Inspirals as Gravitational-wave Sources}",
      journal = {\apj},
         year = 2024,
        month = dec,
       volume = {977},
       number = {1},
          eid = {7},
        pages = {7},
          doi = {10.3847/1538-4357/ad8b1d},
archivePrefix = {arXiv},
       eprint = {2406.19443},
 primaryClass = {astro-ph.GA},
       adsurl = {https://ui.adsabs.harvard.edu/abs/2024ApJ...977....7R}
}

@ARTICLE{2023ApJ...955...30R,
       author = {{Rose}, Sanaea C. and {Naoz}, Smadar and {Sari}, Re'em and {Linial}, Itai},
        title = "{Stellar Collisions in the Galactic Center: Massive Stars, Collision Remnants, and Missing Red Giants}",
      journal = {\apj},
         year = 2023,
        month = sep,
       volume = {955},
       number = {1},
          eid = {30},
        pages = {30},
          doi = {10.3847/1538-4357/acee75},
archivePrefix = {arXiv},
       eprint = {2304.10569},
 primaryClass = {astro-ph.GA},
       adsurl = {https://ui.adsabs.harvard.edu/abs/2023ApJ...955...30R}
}

@ARTICLE{2024ApJ...963L..17R,
       author = {{Rose}, Sanaea C. and {MacLeod}, Morgan},
        title = "{Collisional Shaping of Nuclear Star Cluster Density Profiles}",
      journal = {\apjl},
         year = 2024,
        month = mar,
       volume = {963},
       number = {1},
          eid = {L17},
        pages = {L17},
          doi = {10.3847/2041-8213/ad251f},
archivePrefix = {arXiv},
       eprint = {2310.19912},
 primaryClass = {astro-ph.GA},
       adsurl = {https://ui.adsabs.harvard.edu/abs/2024ApJ...963L..17R}
}

@ARTICLE{2021ApJ...908...57E,
       author = {{Escala}, Andr{\'e}s},
        title = "{Observational Support for Massive Black Hole Formation Driven by Runaway Stellar Collisions in Galactic Nuclei}",
      journal = {\apj},
         year = 2021,
        month = feb,
       volume = {908},
       number = {1},
          eid = {57},
        pages = {57},
          doi = {10.3847/1538-4357/abd93c},
archivePrefix = {arXiv},
       eprint = {2006.01826},
 primaryClass = {astro-ph.GA},
       adsurl = {https://ui.adsabs.harvard.edu/abs/2021ApJ...908...57E}
}

@ARTICLE{2023MNRAS.522.4224V,
       author = {{Vergara}, M.~C. and {Escala}, A. and {Schleicher}, D.~R.~G. and {Reinoso}, B.},
        title = "{Global instability by runaway collisions in nuclear stellar clusters: numerical tests of a route for massive black hole formation}",
      journal = {\mnras},
         year = 2023,
        month = jul,
       volume = {522},
       number = {3},
        pages = {4224-4237},
          doi = {10.1093/mnras/stad1253},
archivePrefix = {arXiv},
       eprint = {2209.15066},
 primaryClass = {astro-ph.GA},
       adsurl = {https://ui.adsabs.harvard.edu/abs/2023MNRAS.522.4224V}
}

@article{Alexander:2008tq,
    author = "Alexander, Tal and Hopman, Clovis",
    title = "{Strong mass segregation around a massive black hole}",
    eprint = "0808.3150",
    archivePrefix = "arXiv",
    primaryClass = "astro-ph",
    doi = "10.1088/0004-637X/697/2/1861",
    journal = "Astrophys. J.",
    volume = "697",
    pages = "1861--1869",
    year = "2009"
}

@article{Rozner:2025iec,
    author = "Rozner, Mor and Ramirez-Ruiz, Enrico",
    title = "{Stellar Distributions around Supermassive Black Holes in Gas-rich Nuclear Star Clusters}",
    eprint = "2506.04229",
    archivePrefix = "arXiv",
    primaryClass = "astro-ph.GA",
    doi = "10.3847/2041-8213/adeca7",
    journal = "Astrophys. J. Lett.",
    volume = "988",
    number = "1",
    pages = "L21",
    year = "2025"
}

@ARTICLE{1980ApJ...235..986H,
       author = {{Hills}, J.~G.},
        title = "{The effect of mass loss on the dynamical evolution of a stellar system - Analytic approximations}",
      journal = {\apj},
         year = 1980,
        month = feb,
       volume = {235},
        pages = {986-991},
          doi = {10.1086/157703},
       adsurl = {https://ui.adsabs.harvard.edu/abs/1980ApJ...235..986H}
}

@ARTICLE{1977ApJ...217..281S,
       author = {{Shapiro}, S.~L.},
        title = "{The dissolution of globular clusters containing massive black holes.}",
      journal = {\apj},
         year = 1977,
        month = oct,
       volume = {217},
        pages = {281-286},
          doi = {10.1086/155577},
       adsurl = {https://ui.adsabs.harvard.edu/abs/1977ApJ...217..281S}
}

@ARTICLE{1990Ap&SS.168..233Y,
       author = {{Zhou}, Yuan and {Zhong}, Xie Guang},
        title = "{The Core Evolution of a Globular Cluster Containing Massive Black-Holes}",
      journal = {\apss},
         year = 1990,
        month = jun,
       volume = {168},
       number = {2},
        pages = {233-241},
          doi = {10.1007/BF00636869},
       adsurl = {https://ui.adsabs.harvard.edu/abs/1990Ap&SS.168..233Y}
}

@ARTICLE{1991MNRAS.251..564D,
       author = {{Dokuchaev}, Viacheslav I.},
        title = "{Joint evolution of a galactic nucleus and central massive black hole.}",
      journal = {\mnras},
         year = 1991,
        month = aug,
       volume = {251},
        pages = {564-574},
          doi = {10.1093/mnras/251.4.564},
       adsurl = {https://ui.adsabs.harvard.edu/abs/1991MNRAS.251..564D}
}

@ARTICLE{1976MNRAS.176..633F,
       author = {{Frank}, J. and {Rees}, M.~J.},
        title = "{Effects of massive black holes on dense stellar systems.}",
      journal = {\mnras},
         year = 1976,
        month = sep,
       volume = {176},
        pages = {633-647},
          doi = {10.1093/mnras/176.3.633},
       adsurl = {https://ui.adsabs.harvard.edu/abs/1976MNRAS.176..633F}
}

@article{King:2024oqb,
    author = "King, Andrew",
    title = "{The black hole masses of high-redshift QSOs}",
    eprint = "2404.16832",
    archivePrefix = "arXiv",
    primaryClass = "astro-ph.GA",
    doi = "10.1093/mnras/stae1171",
    journal = "Mon. Not. Roy. Astron. Soc.",
    volume = "531",
    number = "1",
    pages = "550--553",
    year = "2024"
}

@article{Maiolino:2024uon,
    author = "Maiolino, Roberto and others",
    title = "{JWST meets Chandra: a large population of Compton thick, feedback-free, and intrinsically X-ray weak AGN, with a sprinkle of SNe}",
    eprint = "2405.00504",
    archivePrefix = "arXiv",
    primaryClass = "astro-ph.GA",
    doi = "10.1093/mnras/staf359",
    journal = "Mon. Not. Roy. Astron. Soc.",
    volume = "538",
    number = "3",
    pages = "1921--1943",
    year = "2025"
}

@article{Zwick:2025eik,
    author = "Zwick, Lorenz and Tiede, Christopher and Mayer, Lucio",
    title = "{Little Red Dots as self-gravitating discs accreting on supermassive stars: Spectral appearance and formation pathway of the progenitors to direct collapse black holes}",
    eprint = "2507.22014",
    archivePrefix = "arXiv",
    primaryClass = "astro-ph.GA",
    month = "7",
    year = "2025"
}

@article{Karmen:2025xhz,
    author = "Karmen, Mitchell and others",
    title = "{JWST Discovery of a High-redshift Tidal Disruption Event Candidate in COSMOS-Web}",
    eprint = "2504.13248",
    archivePrefix = "arXiv",
    primaryClass = "astro-ph.HE",
    doi = "10.3847/1538-4357/adf216",
    journal = "Astrophys. J.",
    volume = "990",
    number = "2",
    pages = "149",
    year = "2025"
}

@article{Kaur:2024ofj,
    author = "Kaur, Karamveer and Rom, Barak and Sari, Re'em",
    title = "{Semianalytical Fokker{\textendash}Planck Models for Nuclear Star Clusters}",
    eprint = "2406.07627",
    archivePrefix = "arXiv",
    primaryClass = "astro-ph.HE",
    doi = "10.3847/1538-4357/ada8a8",
    journal = "Astrophys. J.",
    volume = "980",
    number = "1",
    pages = "150",
    year = "2025"
}

@article{Zana:2025uuk,
    author = "Zana, Tommaso and Capelo, Pedro R. and Boresta, Mairo and Schneider, Raffaella and Lupi, Alessandro and Trinca, Alessandro and Mayer, Lucio and Valiante, Rosa and Graziani, Luca",
    title = "{Super-Eddington accretion in protogalactic cores}",
    eprint = "2508.21114",
    archivePrefix = "arXiv",
    primaryClass = "astro-ph.GA",
    month = "8",
    year = "2025"
}

@ARTICLE{2025arXiv250912325B,
       author = {{Bonoli}, Silvia and {Izquierdo-Villalba}, David and {Spinoso}, Daniele and {Colpi}, Monica and {Sesana}, Alberto and {Polkas}, Markos and {Springel}, Volker},
        title = "{Constraints on the early growth of massive black holes from PTA and JWST with L-GalaxiesBH}",
      journal = {arXiv e-prints},
         year = 2025,
        month = sep,
          eid = {arXiv:2509.12325},
        pages = {arXiv:2509.12325},
archivePrefix = {arXiv},
       eprint = {2509.12325},
 primaryClass = {astro-ph.GA},
       adsurl = {https://ui.adsabs.harvard.edu/abs/2025arXiv250912325B}
}

@article{Rastello:2025foo,
    author = "Rastello, Sara and Iorio, Giuliano and Gieles, Mark and Wang, Long",
    title = "{Micro-Tidal Disruption Events in Young Star Clusters}",
    eprint = "2509.07067",
    archivePrefix = "arXiv",
    primaryClass = "astro-ph.HE",
    month = "9",
    year = "2025"
}

@ARTICLE{2025arXiv250403551J,
       author = {{Juod{\v{z}}balis}, Ignas and {Maiolino}, Roberto and {Baker}, William M. and {Lake}, Emma Curtis and {Scholtz}, Jan and {D'Eugenio}, Francesco and {Trefoloni}, Bartolomeo and {Isobe}, Yuki and {Tacchella}, Sandro and {Bunker}, Andrew J. and {Carniani}, Stefano and {Charlot}, St{\'e}phane and {Jones}, Gareth C. and {Parlanti}, Eleonora and {Perna}, Michele and {Rinaldi}, Pierluigi and {Robertson}, Brant and {{\"U}bler}, Hannah and {Venturi}, Giacomo and {Willott}, Chris},
        title = "{JADES: comprehensive census of broad-line AGN from Reionization to Cosmic Noon revealed by JWST}",
      journal = {arXiv e-prints},
         year = 2025,
        month = apr,
          eid = {arXiv:2504.03551},
        pages = {arXiv:2504.03551},
          doi = {10.48550/arXiv.2504.03551},
archivePrefix = {arXiv},
       eprint = {2504.03551},
 primaryClass = {astro-ph.GA},
       adsurl = {https://ui.adsabs.harvard.edu/abs/2025arXiv250403551J}
}

@ARTICLE{2023ApJ...945...53V,
       author = {{Vanzella}, Eros and {Claeyssens}, Ad{\'e}la{\"\i}de and {Welch}, Brian and {Adamo}, Angela and {Coe}, Dan and {Diego}, Jose M. and {Mahler}, Guillaume and {Khullar}, Gourav and {Kokorev}, Vasily and {Oguri}, Masamune and {Ravindranath}, Swara and {Furtak}, Lukas J. and {Hsiao}, Tiger Yu-Yang and {Abdurro'uf} and {Mandelker}, Nir and {Brammer}, Gabriel and {Bradley}, Larry D. and {Brada{\v{c}}}, Maru{\v{s}}a and {Conselice}, Christopher J. and {Dayal}, Pratika and {Nonino}, Mario and {Andrade-Santos}, Felipe and {Windhorst}, Rogier A. and {Pirzkal}, Nor and {Sharon}, Keren and {de Mink}, S.~E. and {Fujimoto}, Seiji and {Zitrin}, Adi and {Eldridge}, Jan J. and {Norman}, Colin},
        title = "{JWST/NIRCam Probes Young Star Clusters in the Reionization Era Sunrise Arc}",
      journal = {\apj},
         year = 2023,
        month = mar,
       volume = {945},
       number = {1},
          eid = {53},
        pages = {53},
          doi = {10.3847/1538-4357/acb59a},
archivePrefix = {arXiv},
       eprint = {2211.09839},
 primaryClass = {astro-ph.GA},
       adsurl = {https://ui.adsabs.harvard.edu/abs/2023ApJ...945...53V}
}

@ARTICLE{2025arXiv250613852M,
       author = {{McClymont}, William and {Tacchella}, Sandro and {Ji}, Xihan and {Kannan}, Rahul and {Maiolino}, Roberto and {Simmonds}, Charlotte and {Smith}, Aaron and {Puchwein}, Ewald and {Garaldi}, Enrico and {Vogelsberger}, Mark and {D'Eugenio}, Francesco and {Keating}, Laura and {Shen}, Xuejian and {Trefoloni}, Bartolomeo and {Zier}, Oliver},
        title = "{Overmassive black holes in the early Universe can be explained by gas-rich, dark matter-dominated galaxies}",
      journal = {arXiv e-prints},
         year = 2025,
        month = jun,
          eid = {arXiv:2506.13852},
        pages = {arXiv:2506.13852},
          doi = {10.48550/arXiv.2506.13852},
archivePrefix = {arXiv},
       eprint = {2506.13852},
 primaryClass = {astro-ph.GA},
       adsurl = {https://ui.adsabs.harvard.edu/abs/2025arXiv250613852M}
}

@ARTICLE{2025arXiv250622147G,
       author = {{Geris}, Sophia and {Maiolino}, Roberto and {Isobe}, Yuki and {Scholtz}, Jan and {D'Eugenio}, Francesco and {Ji}, Xihan and {Juodzbalis}, Ignas and {Simmonds}, Charlotte and {Dayal}, Pratika and {Trinca}, Alessandro and {Schneider}, Raffaella and {Arribas}, Santiago and {Bhatawdekar}, Rachana and {Bunker}, Andrew J. and {Carniani}, Stefano and {Charlot}, Stephane and {Chevallard}, Jacopo and {Curtis-Lake}, Emma and {Johnson}, Benjamin D. and {Parlanti}, Eleonora and {Rinaldi}, Pierluigi and {Robertson}, Brant and {Tacchella}, Sandro and {Uebler}, Hannah and {Venturi}, Giacomo and {Williams}, Christina C. and {Witstok}, Joris},
        title = "{JADES reveals a large population of low mass black holes at high redshift}",
      journal = {arXiv e-prints},
         year = 2025,
        month = jun,
          eid = {arXiv:2506.22147},
        pages = {arXiv:2506.22147},
          doi = {10.48550/arXiv.2506.22147},
archivePrefix = {arXiv},
       eprint = {2506.22147},
 primaryClass = {astro-ph.GA},
       adsurl = {https://ui.adsabs.harvard.edu/abs/2025arXiv250622147G}
}

@ARTICLE{2014A&A...566A..47S,
       author = {{Sch{\"o}del}, R. and {Feldmeier}, A. and {Kunneriath}, D. and {Stolovy}, S. and {Neumayer}, N. and {Amaro-Seoane}, P. and {Nishiyama}, S.},
        title = "{Surface brightness profile of the Milky Way's nuclear star cluster}",
      journal = {\aap},
         year = 2014,
        month = jun,
       volume = {566},
          eid = {A47},
        pages = {A47},
          doi = {10.1051/0004-6361/201423481},
archivePrefix = {arXiv},
       eprint = {1403.6657},
 primaryClass = {astro-ph.GA},
       adsurl = {https://ui.adsabs.harvard.edu/abs/2014A&A...566A..47S}
}

@ARTICLE{2025arXiv250821748J,
       author = {{Juod{\v{z}}balis}, Ignas and {Marconcini}, Cosimo and {D'Eugenio}, Francesco and {Maiolino}, Roberto and {Marconi}, Alessandro and {{\"U}bler}, Hannah and {Scholtz}, Jan and {Ji}, Xihan and {Arribas}, Santiago and {Bennett}, Jake S. and {Bromm}, Volker and {Bunker}, Andrew J. and {Carniani}, Stefano and {Charlot}, St{\'e}phane and {Cresci}, Giovanni and {Egami}, Eiichi and {Fabian}, Andrew and {Inayoshi}, Kohei and {Isobe}, Yuki and {Ivey}, Lucy and {Jones}, Gareth C. and {Koudmani}, Sophie and {Laporte}, Nicolas and {Liu}, Boyuan and {Lyu}, Jianwei and {Mazzolari}, Giovanni and {Monty}, Stephanie and {Parlanti}, Eleonora and {P{\'e}rez-Gonz{\'a}lez}, Pablo G. and {Perna}, Michele and {Robertson}, Brant and {Schneider}, Raffaella and {Sijacki}, Debora and {Tacchella}, Sandro and {Trinca}, Alessandro and {Valiante}, Rosa and {Volonteri}, Marta and {Witstok}, Joris and {Zhang}, Saiyang},
        title = "{A direct black hole mass measurement in a Little Red Dot at the Epoch of Reionization}",
      journal = {arXiv e-prints},
         year = 2025,
        month = aug,
          eid = {arXiv:2508.21748},
        pages = {arXiv:2508.21748},
          doi = {10.48550/arXiv.2508.21748},
archivePrefix = {arXiv},
       eprint = {2508.21748},
 primaryClass = {astro-ph.GA},
       adsurl = {https://ui.adsabs.harvard.edu/abs/2025arXiv250821748J}
}

@article{Pacucci:2025ojp,
    author = "Pacucci, Fabio and Hernquist, Lars and Fujii, Michiko",
    title = "{Little Red Dots Are Nurseries of Massive Black Holes}",
    eprint = "2509.02664",
    archivePrefix = "arXiv",
    primaryClass = "astro-ph.GA",
    month = "9",
    year = "2025"
}

@ARTICLE{2025arXiv250814260V,
       author = {{Vergara}, M.~C. and {Askar}, A. and {Flammini Dotti}, F. and {Schleicher}, D.~R.~G. and {Escala}, A. and {Spurzem}, R. and {Giersz}, M. and {Hurley}, J. and {Arca Sedda}, M. and {Neumayer}, N.},
        title = "{Efficient black hole seed formation in low metallicity and dense stellar clusters with implications for JWST sources}",
      journal = {arXiv e-prints},
         year = 2025,
        month = aug,
          eid = {arXiv:2508.14260},
        pages = {arXiv:2508.14260},
          doi = {10.48550/arXiv.2508.14260},
archivePrefix = {arXiv},
       eprint = {2508.14260},
 primaryClass = {astro-ph.GA},
       adsurl = {https://ui.adsabs.harvard.edu/abs/2025arXiv250814260V}
}

@ARTICLE{2025MNRAS.539.2561C,
       author = {{Chon}, Sunmyon and {Omukai}, Kazuyuki},
        title = "{Formation of supermassive stars and dense star clusters in metal-poor clouds exposed to strong FUV radiation}",
      journal = {\mnras},
         year = 2025,
        month = may,
       volume = {539},
       number = {3},
        pages = {2561-2582},
          doi = {10.1093/mnras/staf598},
archivePrefix = {arXiv},
       eprint = {2412.14900},
 primaryClass = {astro-ph.GA},
       adsurl = {https://ui.adsabs.harvard.edu/abs/2025MNRAS.539.2561C}
}

@ARTICLE{2024ApJ...977L..13B,
       author = {{Baggen}, Josephine F.~W. and {van Dokkum}, Pieter and {Brammer}, Gabriel and {de Graaff}, Anna and {Franx}, Marijn and {Greene}, Jenny and {Labb{\'e}}, Ivo and {Leja}, Joel and {Maseda}, Michael V. and {Nelson}, Erica J. and {Rix}, Hans-Walter and {Wang}, Bingjie and {Weibel}, Andrea},
        title = "{The Small Sizes and High Implied Densities of ``Little Red Dots'' with Balmer Breaks Could Explain Their Broad Emission Lines without an Active Galactic Nucleus}",
      journal = {\apjl},
         year = 2024,
        month = dec,
       volume = {977},
       number = {1},
          eid = {L13},
        pages = {L13},
          doi = {10.3847/2041-8213/ad90b8},
archivePrefix = {arXiv},
       eprint = {2408.07745},
 primaryClass = {astro-ph.GA},
       adsurl = {https://ui.adsabs.harvard.edu/abs/2024ApJ...977L..13B}
}

@ARTICLE{2023Natur.616..266L,
       author = {{Labb{\'e}}, Ivo and {van Dokkum}, Pieter and {Nelson}, Erica and {Bezanson}, Rachel and {Suess}, Katherine A. and {Leja}, Joel and {Brammer}, Gabriel and {Whitaker}, Katherine and {Mathews}, Elijah and {Stefanon}, Mauro and {Wang}, Bingjie},
        title = "{A population of red candidate massive galaxies  600 Myr after the Big Bang}",
      journal = {\nat},
         year = 2023,
        month = apr,
       volume = {616},
       number = {7956},
        pages = {266-269},
          doi = {10.1038/s41586-023-05786-2},
archivePrefix = {arXiv},
       eprint = {2207.12446},
 primaryClass = {astro-ph.GA},
       adsurl = {https://ui.adsabs.harvard.edu/abs/2023Natur.616..266L}
}

@ARTICLE{2024Natur.632..513A,
       author = {{Adamo}, Angela and {Bradley}, Larry D. and {Vanzella}, Eros and {Claeyssens}, Ad{\'e}la{\"\i}de and {Welch}, Brian and {Diego}, Jose M. and {Mahler}, Guillaume and {Oguri}, Masamune and {Sharon}, Keren and {Abdurro'uf} and {Hsiao}, Tiger Yu-Yang and {Xu}, Xinfeng and {Messa}, Matteo and {Lassen}, Augusto E. and {Zackrisson}, Erik and {Brammer}, Gabriel and {Coe}, Dan and {Kokorev}, Vasily and {Ricotti}, Massimo and {Zitrin}, Adi and {Fujimoto}, Seiji and {Inoue}, Akio K. and {Resseguier}, Tom and {Rigby}, Jane R. and {Jim{\'e}nez-Teja}, Yolanda and {Windhorst}, Rogier A. and {Hashimoto}, Takuya and {Tamura}, Yoichi},
        title = "{Bound star clusters observed in a lensed galaxy 460 Myr after the Big Bang}",
      journal = {\nat},
         year = 2024,
        month = aug,
       volume = {632},
       number = {8025},
        pages = {513-516},
          doi = {10.1038/s41586-024-07703-7},
archivePrefix = {arXiv},
       eprint = {2401.03224},
 primaryClass = {astro-ph.GA},
       adsurl = {https://ui.adsabs.harvard.edu/abs/2024Natur.632..513A}
}

@ARTICLE{2024A&A...690A..94P,
       author = {{Polak}, Brooke and {Mac Low}, Mordecai-Mark and {Klessen}, Ralf S. and {Wei Teh}, Jia and {Cournoyer-Cloutier}, Claude and {Andersson}, Eric P. and {Appel}, Sabrina M. and {Tran}, Aaron and {Lewis}, Sean C. and {Wilhelm}, Maite J.~C. and {Portegies Zwart}, Simon and {Glover}, Simon C.~O. and {Rieder}, Steven and {Wang}, Long and {McMillan}, Stephen L.~W.},
        title = "{Massive star cluster formation: I. High star formation efficiency while resolving feedback of individual stars}",
      journal = {\aap},
         year = 2024,
        month = oct,
       volume = {690},
          eid = {A94},
        pages = {A94},
          doi = {10.1051/0004-6361/202348840},
archivePrefix = {arXiv},
       eprint = {2312.06509},
 primaryClass = {astro-ph.GA},
       adsurl = {https://ui.adsabs.harvard.edu/abs/2024A&A...690A..94P}
}

@ARTICLE{2025ApJ...981L..28M,
       author = {{Mayer}, Lucio and {van Donkelaar}, Floor and {Messa}, Matteo and {Capelo}, Pedro R. and {Adamo}, Angela},
        title = "{In Situ Formation of Star Clusters at z > 7 via Galactic Disk Fragmentation: Shedding Light on Ultracompact Clusters and Overmassive Black Holes Seen by JWST}",
      journal = {\apjl},
         year = 2025,
        month = mar,
       volume = {981},
       number = {2},
          eid = {L28},
        pages = {L28},
          doi = {10.3847/2041-8213/adadfe},
archivePrefix = {arXiv},
       eprint = {2411.00670},
 primaryClass = {astro-ph.GA},
       adsurl = {https://ui.adsabs.harvard.edu/abs/2025ApJ...981L..28M}
}

@ARTICLE{2025arXiv250418620L,
       author = {{Lah{\'e}n}, Natalia and {Naab}, Thorsten and {Rantala}, Antti and {Partmann}, Christian},
        title = "{Mergers all the way down: stellar collisions and kinematics of a dense hierarchically forming massive star cluster in a dwarf starburst}",
      journal = {arXiv e-prints},
         year = 2025,
        month = apr,
          eid = {arXiv:2504.18620},
        pages = {arXiv:2504.18620},
          doi = {10.48550/arXiv.2504.18620},
archivePrefix = {arXiv},
       eprint = {2504.18620},
 primaryClass = {astro-ph.GA},
       adsurl = {https://ui.adsabs.harvard.edu/abs/2025arXiv250418620L}
}

@ARTICLE{2025arXiv250308779G,
       author = {{Garcia}, Fred Angelo Batan and {Ricotti}, Massimo and {Sugimura}, Kazuyuki},
        title = "{Seeding Cores: A Pathway for Nuclear Star Clusters from Bound Star Clusters in the First Billion Years}",
      journal = {arXiv e-prints},
         year = 2025,
        month = mar,
          eid = {arXiv:2503.08779},
        pages = {arXiv:2503.08779},
          doi = {10.48550/arXiv.2503.08779},
archivePrefix = {arXiv},
       eprint = {2503.08779},
 primaryClass = {astro-ph.GA},
       adsurl = {https://ui.adsabs.harvard.edu/abs/2025arXiv250308779G}
}

@ARTICLE{2022A&A...667A.101F,
       author = {{Fahrion}, Katja and {Bulichi}, Teodora-Elena and {Hilker}, Michael and {Leaman}, Ryan and {Lyubenova}, Mariya and {M{\"u}ller}, Oliver and {Neumayer}, Nadine and {Pinna}, Francesca and {Rejkuba}, Marina and {van de Ven}, Glenn},
        title = "{Nuclear star cluster formation in star-forming dwarf galaxies}",
      journal = {\aap},
         year = 2022,
        month = nov,
       volume = {667},
          eid = {A101},
        pages = {A101},
          doi = {10.1051/0004-6361/202244932},
archivePrefix = {arXiv},
       eprint = {2210.01556},
 primaryClass = {astro-ph.GA},
       adsurl = {https://ui.adsabs.harvard.edu/abs/2022A&A...667A.101F}
}

@article{Neumayer:2020gno,
    author = "Neumayer, Nadine and Seth, Anil and Boeker, Torsten",
    title = "{Nuclear star clusters}",
    eprint = "2001.03626",
    archivePrefix = "arXiv",
    primaryClass = "astro-ph.GA",
    doi = "10.1007/s00159-020-00125-0",
    journal = "Astron. Astrophys. Rev.",
    volume = "28",
    number = "1",
    pages = "4",
    year = "2020"
}

@ARTICLE{1993ApJ...403..271G,
       author = {{Goodman}, Jeremy and {Hut}, Piet},
        title = "{Binary--Single-Star Scattering. V. Steady State Binary Distribution in a Homogeneous Static Background of Single Stars}",
      journal = {\apj},
         year = 1993,
        month = jan,
       volume = {403},
        pages = {271},
          doi = {10.1086/172200},
       adsurl = {https://ui.adsabs.harvard.edu/abs/1993ApJ...403..271G}
}

@ARTICLE{2023MNRAS.523.2918S,
       author = {{Sormani}, Mattia C. and {Barnes}, Ashley T. and {Sun}, Jiayi and {Stuber}, Sophia K. and {Schinnerer}, Eva and {Emsellem}, Eric and {Leroy}, Adam K. and {Glover}, Simon C.~O. and {Henshaw}, Jonathan D. and {Meidt}, Sharon E. and {Neumann}, Justus and {Querejeta}, Miguel and {Williams}, Thomas G. and {Bigiel}, Frank and {Eibensteiner}, Cosima and {Fragkoudi}, Francesca and {Levy}, Rebecca C. and {Grasha}, Kathryn and {Klessen}, Ralf S. and {Kruijssen}, J.~M. Diederik and {Neumayer}, Nadine and {Pinna}, Francesca and {Rosolowsky}, Erik W. and {Smith}, Rowan J. and {Teng}, Yu-Hsuan and {Tress}, Robin G. and {Watkins}, Elizabeth J.},
        title = "{Fuelling the nuclear ring of NGC 1097}",
      journal = {\mnras},
         year = 2023,
        month = aug,
       volume = {523},
       number = {2},
        pages = {2918-2927},
          doi = {10.1093/mnras/stad1554},
archivePrefix = {arXiv},
       eprint = {2305.14437},
 primaryClass = {astro-ph.GA},
       adsurl = {https://ui.adsabs.harvard.edu/abs/2023MNRAS.523.2918S}
}

@article{Syer:1998xh,
    author = "Syer, D. and Ulmer, A.",
    title = "{Tidal disruption rates of stars in observed galaxies}",
    eprint = "astro-ph/9812389",
    archivePrefix = "arXiv",
    doi = "10.1046/j.1365-8711.1999.02445.x",
    journal = "Mon. Not. Roy. Astron. Soc.",
    volume = "306",
    pages = "35",
    year = "1999"
}

@ARTICLE{1989ApJ...343..725Q,
       author = {{Quinlan}, Gerald D. and {Shapiro}, Stuart L.},
        title = "{Dynamical Evolution of Dense Clusters of Compact Stars}",
      journal = {\apj},
         year = 1989,
        month = aug,
       volume = {343},
        pages = {725},
          doi = {10.1086/167745},
       adsurl = {https://ui.adsabs.harvard.edu/abs/1989ApJ...343..725Q}
}

@ARTICLE{1990Natur.345..679S,
       author = {{Shlosman}, Isaac and {Begelman}, Mitchell C. and {Frank}, Julian},
        title = "{The fuelling of active galactic nuclei}",
      journal = {\nat},
         year = 1990,
        month = jun,
       volume = {345},
       number = {6277},
        pages = {679-686},
          doi = {10.1038/345679a0},
       adsurl = {https://ui.adsabs.harvard.edu/abs/1990Natur.345..679S}
}

@ARTICLE{1989Natur.338...45S,
       author = {{Shlosman}, Isaac and {Frank}, Juhan and {Begelman}, Mitchell C.},
        title = "{Bars within bars: a mechanism for fuelling active galactic nuclei}",
      journal = {\nat},
         year = 1989,
        month = mar,
       volume = {338},
       number = {6210},
        pages = {45-47},
          doi = {10.1038/338045a0},
       adsurl = {https://ui.adsabs.harvard.edu/abs/1989Natur.338...45S}
}

@article{Guo:2025sjb,
    author = "Guo, Minghao and Stone, James M. and Quataert, Eliot and Springel, Volker",
    title = "{Cyclic Zoom: Multi-scale GRMHD Modeling of Black Hole Accretion and Feedback}",
    eprint = "2504.16802",
    archivePrefix = "arXiv",
    primaryClass = "astro-ph.HE",
    month = "4",
    year = "2025"
}

@article{Zhang:2025uug,
    author = "Zhang, Lizhong and Stone, James M. and Mullen, Patrick D. and Davis, Shane W. and Jiang, Yan-Fei and White, Christopher J.",
    title = "{Radiation GRMHD Models of Accretion onto Stellar-Mass Black Holes: I. Survey of Eddington Ratios}",
    eprint = "2506.02289",
    archivePrefix = "arXiv",
    primaryClass = "astro-ph.HE",
    month = "6",
    year = "2025"
}

\end{document}